\documentclass{article}

\newif \ifarxiv 
\arxivtrue

\usepackage[doublespacing]{setspace}
\usepackage[utf8]{inputenc}
\usepackage[top=0.9in, bottom=0.9in, lmargin=1in, rmargin=1in]{geometry}
\usepackage[authoryear,semicolon]{natbib}
\usepackage{authblk}
\usepackage{parskip} 
\usepackage{amsmath, amssymb, mathtools, bm} 
\usepackage{graphicx, xcolor}
\usepackage{algorithm, algpseudocode}
\usepackage{scalerel} 
\usepackage{url}
\usepackage{soul} 
\usepackage{caption}
\usepackage{subcaption}
\usepackage{array, booktabs, makecell, multirow} 
\usepackage{xr-hyper} 
\usepackage[hypertexnames=false]{hyperref}
\usepackage{comment}

\ifarxiv
\else 
    
    \makeatletter
    \newcommand*{\addFileDependency}[1]{
    \typeout{(#1)}
    \@addtofilelist{#1}
    \IfFileExists{#1}{}{\typeout{No file #1.}}
    }\makeatother
    
    \newcommand*{\myexternaldocument}[1]{%
    \externaldocument{#1}%
    \addFileDependency{#1.tex}%
    \addFileDependency{#1.aux}%
    }
    
    \myexternaldocument{supplement}
    \myexternaldocument{response_to_reviewers}

\fi 

\providecommand{\keywords}[1]{\textbf{\textit{Keywords: }} #1} 

\makeatletter
\def\thanks#1{\protected@xdef\@thanks{\@thanks
        \protect\footnotetext{#1}}}
\makeatother

\renewcommand{\d}{\, \mathrm{d}}
\newcommand{\A}{\mathbf{A}}

\newcommand{\C}{\mathbf{C}}

\newcommand{\I}{\mathbf{I}}
\newcommand{\J}{\mathbf{J}}

\renewcommand{\L}{\mathbf{L}}
\newcommand{\M}{\mathbf{M}}

\newcommand{\V}{\mathbf{V}}

\newcommand{\f}{\mathbf{f}}

\renewcommand{\r}{\mathbf{r}}

\newcommand{\y}{\mathbf{y}}

\newcommand{\x}{\mathbf{x}}
\newcommand{\z}{\mathbf{z}}

\newcommand{\0}{\mathbf{0}}

\newcommand{\btheta}{\bm{\theta}}
\newcommand{\boldeta}{\bm{\eta}}

\newcommand{\bxi}{\bm{\xi}}
\newcommand{\bepsilon}{\bm{\varepsilon}}
\newcommand{\bmu}{\bm{\mu}}

\newcommand{\bTheta}{\mathbf{\Theta}}
\newcommand{\bSigma}{\mathbf{\Sigma}}

\newcommand{\bPhi}{\mathbf{\Phi}}
\newcommand{\bPsi}{\mathbf{\Psi}}

\newcommand{\Gau}{\mathrm{Gau}}

\newcommand{\Exp}{\mathbb{E}}
\newcommand{\Var}{\operatorname{Var}}

\newcommand{\Tr}{\operatorname{Tr}}
\newcommand{\diag}{\operatorname{diag}}

\renewcommand{\vec}{\operatorname{vec}}

\newcommand{\red}{\textcolor{red}}

\DeclareMathOperator*{\argmin}{arg\,min}

\DeclarePairedDelimiter\abs{\lvert}{\rvert}%
\DeclarePairedDelimiter\norm{\lVert}{\rVert}%

\DeclarePairedDelimiter\floor{\lfloor}{\rfloor}%
\DeclarePairedDelimiter\ceil{\lceil}{\rceil}%

\DeclarePairedDelimiterX{\divergence}[2]{(}{)}{%
  #1\;\delimsize\|\;#2%
}

\makeatletter
\let\oldabs\abs
\def\abs{\@ifstar{\oldabs}{\oldabs*}}
\let\oldnorm\norm
\def\norm{\@ifstar{\oldnorm}{\oldnorm*}}
\makeatother

\usepackage{clipboard}
\newclipboard{output-myclipboard} 

\title{\vspace{-1cm}
Recursive variational Gaussian approximation with the Whittle likelihood for linear non-Gaussian state space models}
\author[1,2,*]{Bao Anh Vu\thanks{* Corresponding author \\ E-mail address: bavu@uow.edu.au}}
\author[1]{David Gunawan}
\author[1,2]{Andrew Zammit-Mangion}

\affil[1]{School of Mathematics and Applied Statistics, University of Wollongong, Wollongong, New South Wales, Australia}
\affil[2] {Securing Antarctica’s Environmental Future, University of Wollongong, Wollongong, New South Wales, Australia}
\date{\today}

\begin{document}

\maketitle

\begin{abstract}
    Parameter inference for linear and non-Gaussian state space models is challenging because the likelihood function contains an intractable integral over the latent state variables. While Markov chain Monte Carlo (MCMC) methods provide exact samples from the posterior distribution as the number of samples goes to infinity, they tend to have high computational cost, particularly for observations of a long time series. 
    When inference with MCMC methods is computationally expensive, variational Bayes (VB) methods are a useful alternative. VB methods approximate the posterior density of the parameters with a simple and tractable distribution found through optimisation. This work proposes a novel sequential VB approach that makes use of the Whittle likelihood for computationally efficient parameter inference in linear, non-Gaussian state space models. Our algorithm, called Recursive Variational Gaussian Approximation with the Whittle Likelihood (R-VGA-Whittle), updates the variational parameters by processing data in the frequency domain. At each iteration, R-VGA-Whittle requires the gradient and Hessian of the Whittle log-likelihood, which are available in closed form. Through several examples involving a linear Gaussian state space model; a univariate/bivariate stochastic volatility model; and a state space model with Student's t measurement error, where the latent states follow an autoregressive fractionally integrated moving average (ARFIMA) model, we show that R-VGA-Whittle provides good approximations to posterior distributions of the parameters, and that it is very computationally efficient when compared to asymptotically exact methods such as Hamiltonian Monte Carlo.

\end{abstract}


\ifarxiv
\else 
MOS subject classification: 62-08; 62F15; 62L12; 62M05; 62M15.
\fi

\section{Introduction}




State space models (SSMs) are a widely used class of hierarchical time series models employed in a diverse range of applications in fields such as epidemiology~\citep[e.g.,][]{dukic2012tracking}, economics~\cite[e.g.,][]{chan2023bayesian}, finance~\cite[e.g.,][]{kim1998stochastic}, and engineering~\cite[e.g.,][]{gordon1993novel}. In an SSM, the observed data depend on unobserved, time-evolving latent state variables and unknown parameters. The observations are assumed to be conditionally independent given the latent states, the dynamics of the latent states are described through a Markov process, and often a simplifying assumption of linearity of the state variables is made. When the state variables are non-Gaussian, Bayesian inference for the parameters of these SSMs is challenging since the likelihood function involves an intractable integral over the latent states. 




A popular class of inference methods for non-Gaussian SSMs is Markov chain Monte Carlo (MCMC). 
While MCMC methods can provide samples from the posterior distribution that are asymptotically exact, they tend to be computationally expensive. 
An early application of MCMC to nonlinear, non-Gaussian state-space models (SSMs) is the work of \cite{meyer2003stochastic}, which combines the Laplace approximation~\citep{laplace1986memoir} and the extended Kalman filter~\citep{harvey1990forecasting} to approximate the likelihood function. This approximation is then used within a Metropolis-Hastings algorithm to estimate parameters in a nonlinear, non-Gaussian stochastic volatility model. While this likelihood approximation is analytical, it is only exact for linear, Gaussian SSMs. An alternative approach is the pseudo-marginal Metropolis-Hastings algorithm developed by~\cite{andrieu2009pseudo}, which employs an unbiased likelihood estimate obtained through a particle filter. However, particle filters are computationally intensive and prone to particle degeneracy, especially when dealing with large datasets (see~\cite{kantas2009overview} for a detailed discussion of this issue). Both approaches typically require many iterations, each involving a full data pass with complexity $\textrm{O}(T)$, where $T$ is the number of observations, making them costly for long data sequences. Additionally, other alternatives, such as the Hamiltonian Monte Carlo (HMC) method ~\citep{duane1987hybrid,neal2011mcmc,hoffman2014no}, target the joint posterior distribution of states and parameters and typically require many iterations to achieve parameter convergence when $T$ is large and the state dimension is high.

Variational Bayes (VB) methods are becoming increasingly popular for Bayesian inference for state space models \citep{zammit-mangion2012, gunawan2021variational,frazier2023variational}. VB approximates the posterior distribution of model parameters with a simpler and more tractable class of distributions constructed using so-called variational parameters, which are unknown and estimated via optimisation techniques. While VB methods tend to be more computationally efficient than MCMC methods, they provide approximate posterior distributions. For a review of VB methods, see~\cite{blei2017variational}.


There are two main classes of VB methods: batch and sequential. A batch VB framework \citep[e.g., ][]{tran2017variational, tan2018gaussian, loaiza2022fast,gunawan2023flexible} requires an optimisation algorithm to repeatedly process the entire dataset until variational parameters converge to a fixed point. A sequential VB framework, on the other hand, updates variational parameters using one observation at a time. Sequential methods only pass through the entire dataset once, and are thus ideal for use with large datasets. 
Sequential VB methods are available for statistical models for independent data \citep{lambert2022recursive}, for classical autoregressive models for time-series data \citep{tomasetti2022updating}, for generalised linear mixed models \citep{vu2024r}, and for SSMs where the state evolution model is linear and Gaussian~(\citealp[][Chapter 5]{beal2003thesis}; \citealp{zammitmangion2011a}).
\cite{campbell2021online} proposed a sequential VB approach that is applicable to an SSM with non-linear state dynamics and non-Gaussian measurement errors, but this approach can be computationally expensive as it requires solving an optimisation problem and a regression task at each time step.


Our article proposes a sequential VB approach to estimate parameters in linear non-Gaussian state space models. This approach builds on the Recursive Variational Gaussian Approximation  (R-VGA) algorithm of \cite{lambert2022recursive}. We first transform time-domain data into frequency-domain data and then use the Whittle likelihood~\citep{whittle1953estimation} in a sequential VB scheme. The logarithm of the Whittle likelihood function is a sum of the individual log-likelihood components in Fourier space, and is thus well-suited for use in sequential updating. We call our algorithm the Recursive Variational Gaussian Approximation with the Whittle likelihood (R-VGA-Whittle) algorithm.
At each iteration, R-VGA-Whittle sequentially updates the variational parameters by processing one frequency or a block of frequencies at a time. Although sequential in nature, and requiring only one pass through the frequencies, we note that the R-VGA-Whittle algorithm requires the full dataset to be available in order to construct the periodogram, from which we obtain the frequency-domain data. 


We demonstrate the performance of R-VGA-Whittle in simulation studies involving a linear Gaussian state space model, univariate and bivariate stochastic volatility models, and a state space model with measurement errors that follow a Student's $t$ distribution and latent states that follow an autoregressive fractionally integrated moving average model (ARFIMA; \cite{granger1980introduction}). We show that R-VGA-Whittle is accurate and is often much faster than both HMC with the exact likelihood and HMC based on the Whittle likelihood; hereafter, we refer to these two methods as HMC-exact and HMC-Whittle, respectively. We subsequently apply R-VGA-Whittle to analyse exchange rate data.

The rest of the article is organised as follows. Section~\ref{sec:methodology} gives an overview of SSMs and the univariate and multivariate Whittle likelihood functions, and then introduces the R-VGA-Whittle and HMC-Whittle algorithms. Section~\ref{sec:applications} compares the performance of R-VGA-Whittle, HMC-Whittle and HMC with the exact likelihood in several examples involving linear Gaussian and non-Gaussian SSMs. Section~\ref{sec:conclusion} summarises and discusses our main findings. The article also has an online supplement with additional technical details and empirical results.

\section{Methodology}
\label{sec:methodology}

We give an overview of SSMs in Section~\ref{sec:SSM}, and the univariate and multivariate Whittle likelihoods in Section~\ref{sec:whittle_llh}. Section~\ref{sec:rvga_whittle} introduces R-VGA-Whittle, while Section~\ref{sec:RVGA-BLOCK} proposes a blocking approach to improve computational efficiency. Section~\ref{sec:HMC_Whittle} gives details of HMC-Whittle.


\subsection{State space models \label{sec:SSM}}
An SSM consists of two sub-models: a data model that describes the relationship between the observations and the latent state variables, and a state evolution model that describes the (temporal) evolution of the latent states. 


In SSMs, the $d$-dimensional latent state variables $\{\x_t\}_{t = 1}^T$ are characterised by an initial density ${\x_1 \sim p(\x_1 \mid {\btheta})}$ and state transition densities 
\begin{equation}
\label{eq:ssm_state}
    \x_t \mid \x_{t-1}, \btheta \sim p(\x_t \mid \x_{t-1}, \btheta), \quad t = 2, \dots, T,
\end{equation}
where $\btheta$ is a vector of model parameters. 
The observations $\{\y_t\}_{t = 1}^T$ are related to the latent states via the data model
\begin{equation}
\label{eq:ssm_obs}
    \y_t \mid \x_t, \btheta \sim p(\y_t \mid \x_t, \btheta),
    \quad t = 1, \dots, T,
\end{equation}
where each $\y_t$ is assumed to be conditionally independent of the other observations, when conditioned on the latent state $\x_t$. 
\Copy{class_of_models}{In this article, we focus on linear SSMs with non-Gaussian error distributions in which the data model can be expressed in the linear additive form
\begin{equation}
\label{eq:linear_ssm1}
    \y_t = \x_t + \bepsilon_t, \quad t = 1, \dots, T,
\end{equation}
where $\mathbf{x}_t$ is the $d$-dimensional latent state variables and $\boldsymbol{\varepsilon}_t$ is a (possibly non-Gaussian) vector-valued noise process, both of which have closed-form spectral densities defined in Section \ref{sec:whittle_llh}.}

Let ${\y}_{1:T} \equiv ({\y}_1^\top, \dots, {\y}_T^\top)^{\top}$ and $\x_{1:T} \equiv (\x_1^\top, \dots, \x_T^\top)^{\top}$ be vectors of observations and latent states, respectively. The posterior density of the model parameters is $p(\btheta \mid {\y}_{1:T}) \propto p({\y}_{1:T} \mid \btheta) p(\btheta)$, where the likelihood function is
\begin{equation*}
    p({\y}_{1:T} \mid \btheta) = \int p({\y}_1 \mid \x_1, \btheta) p(\x_1 \mid \btheta) \prod_{t = 2}^T p({\y}_t \mid \x_t, \btheta) p(\x_t \mid \x_{t-1}, \btheta) \d \x_{1:T}.
\end{equation*}
This likelihood function is intractable, except for special cases such as the linear Gaussian SSM \citep[e.g.,][]{gibson2005robust}. In non-Gaussian SSMs, the likelihood is typically estimated using a particle filter \citep{gordon1993novel}, such as in the pseudo-marginal Metropolis-Hastings (PMMH) method of~\cite{andrieu2009pseudo}, which can be computationally expensive for large $T$.


The Whittle likelihood~\citep{whittle1953estimation} is a frequency-domain approximation of the exact likelihood, and offers a computationally efficient alternative for parameter inference in SSMs. Transforming the data to the frequency domain ``bypasses" the need for integrating out the states when computing the likelihood, as the Whittle likelihood only depends on the model parameters and not the latent variables. \Copy{linear_add}{The linear and additive structure in \eqref{eq:linear_ssm1} is important, as it enables a closed-form representation of the spectral density of ${\y}_t$ as the sum of the spectral densities of ${\x}_t$ and $\bepsilon_t$. In the following sections, we review the Whittle likelihood and its properties, and then propose a sequential VB approach that uses the Whittle likelihood for parameter inference.}

\subsection{The Whittle likelihood}
\label{sec:whittle_llh}



The Whittle likelihood has been widely used to estimate the parameters of {complex} time series or spatial models. It is more computationally efficient to evaluate than the exact likelihood, although it is approximate \citep[e.g.,][]{sykulski2019debiased}. Some applications of the Whittle likelihood include parameter estimation with long memory processes~\citep[e.g.,][]{robinson1995gaussian, hurvich2000efficient, giraitis2001whittle, abadir2007nonstationarity, shao2007local, giraitis2012large}, and with spatial models with irregularly spaced data~\citep[e.g.,][]{fuentes2007approximate, matsuda2009fourier}. The Whittle likelihood is also used for estimating parameters of vector autoregressive moving average (VARMA) models with heavy-tailed errors~\citep{she2022whittle}, and of vector autoregressive tempered fractionally integrated moving average (VARTFIMA) models~\citep{villani2022}.


    


\subsubsection{Univariate Whittle likelihood\label{sec:univariatewhittle}}
Consider a real-valued, zero-mean, second-order stationary, univariate time series $\{y_t\}_{t=1}^T$ with autocovariance function $C(h;\btheta) \equiv \Exp(y_t y_{t-h})$, where $0 \le h < t$ and $\Exp(\cdot)$ is the expectation operator. The power spectral density of the time series is the Fourier transform of the autocovariance function~\citep{brillinger2001time}:
\begin{equation}
    f(\omega; \btheta) = \sum_{h = -\infty}^{\infty} C(h;\btheta) \exp(-i \omega h),
\end{equation}
where $\omega \in (-\pi, \pi]$ is the angular frequency and $\btheta$ is a vector of model parameters. The periodogram, an asymptotically unbiased estimate of the spectral density, is given by
\begin{equation}
    I(\omega_k) = \frac{1}{T} \abs{J(\omega_k)}^2, \quad k =  -\ceil*{\frac{T}{2}} + 1, \dots, \floor*{\frac{T}{2}},
\end{equation}
where $\omega_k = \frac{2\pi k}{T}$, and $J(\omega_k)$ is the discrete Fourier transform (DFT) of $\{y_t\}_{t=1}^T$:
\begin{equation}
    J(\omega_k) = \sum_{t=1}^T y_t \exp(-i \omega_k t).
\end{equation}


For an observed time series, the Whittle likelihood, denoted by $l_{W}(\btheta)$, is~\citep[e.g.,][]{salomone2020spectral}
\begin{equation}
    l_{W} (\btheta) = -\sum_{k=1}^{\floor*{\frac{T-1}{2}}} \left( \log f(\omega_k; \btheta) + \frac{I(\omega_k)}{f(\omega_k; \btheta)} \right). \label{eq:whittle_llh_uni}
\end{equation}
Here, the Whittle likelihood is evaluated by summing over the non-negative frequencies only, since $f(\omega_k; \btheta)$ and $I(\omega_k)$ are both symmetric about the origin for real-valued time series.
The term $\omega_k = 0$ is not included since $I(0)=0$  when the time series is de-meaned. The term $\omega_k = \pi$ is also omitted since it has a different distribution from the other frequencies and its influence is asymptotically negligible~\citep{salomone2020spectral, villani2022}.

As $T$ tends to infinity, it can be shown that the parameter estimates obtained by maximising the exact Gaussian likelihood and those obtained by maximising the Whittle likelihood are  equivalent~\citep{dzhaparidze1983spectrum}. 
The calculation of the Whittle likelihood involves evaluation of the periodogram, which does not depend on the parameters $\btheta$ and can, therefore, be pre-computed at a cost of $O(T \log T)$ via the Fast Fourier Transform~\citep[FFT,][]{cooley1965algorithm}. Since the periodogram ordinates are asymptotically independent \citep{shao2007asymptotic}, the logarithm of the Whittle likelihood is a sum of the individual log-likelihood terms computed over different frequencies. 


\subsubsection{Multivariate Whittle likelihood\label{sec:multivariatewhittle}}
Consider a $d$-variate, real valued, second-order stationary, zero mean time series $\{\y_t\}_{t=1}^T$ with autocovariance function $\C(h;\btheta) \equiv \Exp(\y_t \y_{t-h}^\top)$, $0 \le h < t$. 
At a given angular frequency $\omega$, the power spectral density matrix of this series is
\begin{equation}
    \f(\omega;\btheta) = \sum_{h = -\infty}^\infty \C(h;\btheta) \exp(-i \omega h), \quad \omega \in (-\pi, \pi].
\end{equation}
The matrix-valued periodogram is 
\begin{equation}
        \I(\omega_k) = \frac{1}{T} \J(\omega_k) \J(\omega_k)^H, \quad k =  -\ceil*{\frac{T}{2}} + 1, \dots, \floor*{\frac{T}{2}},
\end{equation}
where $\omega_k = \frac{2\pi k}{T}$, $\A^H$ denotes the conjugate transpose of a complex matrix $\A$, and $\J(\omega_k)$, the DFT of the series, is given by
\begin{equation}
    \J(\omega_k) = \sum_{t=1}^{T} \y_t \exp(-i \omega_k t). 
\end{equation}


The Whittle likelihood for the multivariate series $\{\y_t\}_{t=1}^T$ is \citep{villani2022}:
\begin{equation}
    \label{eq:whittle_llh_multi}
    l_W(\btheta) = - \sum_{k=1}^{\floor{\frac{T-1}{2}}} \left( \log \lvert \f(\omega_k; \btheta) \rvert + \Tr(\f(\omega_k; \btheta)^{-1} \I(\omega_k)) \right).
\end{equation}



\subsection{R-VGA-Whittle}
\label{sec:rvga_whittle}

The original R-VGA algorithm of~\cite{lambert2022recursive} is a sequential VB algorithm for parameter inference in statistical models where the observations are independent, such as linear or logistic regression models. Given independent and exchangeable observations $\{\y_i: i = 1, \dots, N\}$, this algorithm sequentially approximates the ``pseudo-posterior'' distribution of the parameters given data up to the $i$th observation, $p(\btheta \mid \y_{1:i})$, with a Gaussian distribution $q_i(\btheta) = \Gau (\bmu_i, \bSigma_i)$. The updates for $\bmu_i$ and $\bSigma_i$ are available in closed form, as seen in Algorithm~\ref{algo:rvgal}. We describe this algorithm in more detail in Section~\ref{sec:rvga} of the online supplement.


At each iteration ${i = 1, \dots, N}$, the R-VGA algorithm requires the gradient and Hessian of the log-likelihood contribution from each observation, $\log p(\y_i \mid \btheta)$. 
In a setting with independent observations, these log-likelihood contributions are easy to obtain; however, in SSMs where the latent states follow a Markov process and are hence temporally dependent, the log-likelihood function is intractable, as shown in Section~\ref{sec:SSM}. The gradient and Hessian of $\log p(\y_t \mid \btheta)$, ${t=1, \dots, T}$, are hence also intractable. While these quantities can be estimated via particle filtering~\citep[e.g.,][]{doucet2012robust}, one would need to run a particle filter from the first time period to the current time period for every step in the sequential algorithm. This quickly becomes infeasible for long time series. 


\begin{algorithm} [t!]
\caption{R-VGA~\citep{lambert2022recursive}} 
\label{algo:rvgal}
\begin{algorithmic}
    \State Input: independent observations $\y_1, \dots, \y_N$, initial values $\bmu_0$ and $\bSigma_0$.
    \State Set $q_0(\btheta) = \Gau (\bmu_0, \bSigma_0)$.
    \For {$i = 1, \dots, N$}
    \State $\bmu_i = \bmu_{i-1} + \bSigma_i \Exp_{q_{i-1}} (\nabla_{\btheta} \log p(\y_i \mid \btheta))$ 
    \State $\bSigma_i^{-1} = \bSigma_{i-1}^{-1} - \Exp_{q_{i-1}} (\nabla_{\btheta}^2 \log p(\y_i \mid \btheta))$ 
    \State Set $q_{i}(\btheta) = \Gau (\bmu_i, \bSigma_i)$.
    \EndFor
    \State Output: variational parameters $\bmu_i$ and $\bSigma_i$, for $i=1,...,N$.
\end{algorithmic}
\end{algorithm}


We now extend the R-VGA algorithm of~\cite{lambert2022recursive} to state space models by replacing the time domain likelihood with a frequency domain alternative: the Whittle likelihood. 
From~\eqref{eq:whittle_llh_uni} and~\eqref{eq:whittle_llh_multi} we see that the log of the Whittle likelihood is a sum over individual log-likelihood components, a consequence of the asymptotic independence of the periodogram frequency components. We replace the quantities ${\nabla_{\btheta} \log p(\y_i \mid \btheta)}$ and ${\nabla_{\btheta}^2 \log p(\y_i \mid \btheta)}$ in Algorithm~\ref{algo:rvgal} with the gradient and Hessian of the Whittle log-likelihood evaluated at each frequency. In the univariate case, the individual log-likelihood contribution at frequency $\omega_k$ is
\begin{equation}
    l_{W_k} (\btheta) =  - \log f(\omega_k; \btheta) - \frac{I(\omega_k)}{f(\omega_k; \btheta)}, \quad k = 1, \dots, \floor*{\frac{T-1}{2}},
\end{equation}
while in the multivariate case,
\begin{equation}
    l_{W_k} (\btheta) = - \log \lvert \f (\omega_k; \btheta) \rvert - \Tr(\f(\omega_k; \btheta)^{-1} \I(\omega_k)), \quad k = 1, \dots, \floor*{\frac{T-1}{2}},
\end{equation} 
for $\omega_k = \frac{2\pi k}{T}$. The log-likelihood contributions $l_{W_k} (\btheta)$ are available for models where the power spectral density is available analytically, such as models where the latent states follow a VARMA~\citep{she2022whittle} or a VARTFIMA~\citep{villani2022} process. The gradient ${\nabla_{\btheta} l_{W_k} (\btheta)}$ and Hessian ${\nabla_{\btheta}^2 l_{W_k} (\btheta)}$ are therefore also available in closed form for these models. At each VB update, ${\nabla_{\btheta} l_{W_k} (\btheta)}$ and ${\nabla_{\btheta}^2 l_{W_k} (\btheta)}$ are used to update the variational parameters for $k = 1, \dots, \floor*{\frac{T-1}{2}}$. Their expectations with respect to $q_{k-1}(\btheta)$ are approximated using Monte Carlo samples ${\btheta^{(s)} \sim q_{k-1}(\btheta)}$, ${s = 1, \dots, S}$: 
\begin{equation*}
    \Exp_{q_{k-1}} (\nabla_{\btheta} l_{W_k} (\btheta)) \approx \frac{1}{S} \sum_{s=1}^S \nabla_{\btheta} l_{W_k} (\btheta^{(s)}), \text{ and }
    \Exp_{q_{k-1}} (\nabla_{\btheta}^2 l_{W_k} (\btheta)) \approx \frac{1}{S} \sum_{s=1}^S \nabla_{\btheta}^2 l_{W_k} (\btheta^{(s)}),
\end{equation*}
for $k = 1, \dots, \floor*{\frac{T-1}{2}}$. We set $S$ following an experiment on the sensitivity of the algorithm to the choice of $S$; see Section S4 of the online supplement for more details. 




We also implement the damping approach of \cite{vu2024r} to mitigate instability that may arise when traversing the first few frequencies. Damping is done by splitting an update into $D$ steps, where $D$ is user-specific. In each step, the gradient and the Hessian of the log of the Whittle likelihood are multiplied by a ``step size" $a= \frac{1}{D}$, and the variational parameters are updated $D$ times during one VB iteration. This has the effect of splitting a log-likelihood contribution into $D$ ``parts", and using them to update variational parameters one part at a time. In practice, we set $D$ through experimentation, and choose it so that initial algorithmic instability is reduced with only a small increase in computational cost. We summarise R-VGA-Whittle in Algorithm~\ref{algo:damped_rvgaw}.

\begin{algorithm} [t]
\caption{Damped R-VGA-Whittle} 
\label{algo:damped_rvgaw}
\begin{algorithmic}
\setstretch{1.5}
    \State Input: time series observations $\y_1, \dots, \y_T$, initial values $\bmu_0$ and $\bSigma_0$, number of frequencies to damp $n_{damp}$, number of damping steps $D$.
    \State Compute the periodogram $\I(\omega_k)$, $k =  -\ceil*{\frac{T}{2}} + 1, \dots, \floor*{\frac{T}{2}}$, via the FFT algorithm.
    \State Set $q_0(\btheta) = \Gau (\bmu_0, \bSigma_0)$.
    \For {$k = 1, \dots, \floor{\frac{T-1}{2}}$}
        \If {$k \leq n_{damp}$} 
        \State Set $a = 1/D$, $\bmu_{k, 0} = \bmu_{k-1}, \bSigma_{k, 0} = \bSigma_{k-1}$. 
            \For {$l = 1, \dots, D$}
            \State Set $q_{k, l-1}(\btheta) = \Gau (\bmu_{k, l-1}, \bSigma_{k, l-1})$,
            \State $\bmu_{k, l} = \bmu_{k, l-1} + a \bSigma_{k, l} \Exp_{q_{k, l-1}} (\nabla_{\btheta} l_{W_k} (\btheta))$, 
            \State $\bSigma_{k, l}^{-1} = \bSigma_{k, l-1}^{-1} - a \Exp_{q_{k, l-1}} (\nabla_{\btheta}^2 l_{W_k} (\btheta))$. 
            \EndFor
            \State Set $\bmu_{k} = \bmu_{k, D}, \bSigma_{k} = \bSigma_{k, D}$, $q_{k}(\btheta) = \Gau (\bmu_k, \bSigma_k)$.
        \Else 
            \State $\bmu_k = \bmu_{k-1} + \bSigma_k \Exp_{q_{k-1}} ({\nabla_{\btheta} l_{W_k} (\btheta)})$, 
            \State $\bSigma_k^{-1} = \bSigma_{k-1}^{-1} - \Exp_{q_{k-1}} ({\nabla_{\btheta}^2 l_{W_k} (\btheta)})$. 
            \State Set $q_{k}(\btheta) = \Gau (\bmu_k, \bSigma_k)$.
        \EndIf
    \EndFor
    \State Output: variational parameters $\bmu_k$ and $\bSigma_k$, for $k=1, \dots, \floor{\frac{T-1}{2}}$.
\end{algorithmic}
\end{algorithm}

\subsection{R-VGA-Whittle with block updates\label{sec:RVGA-BLOCK}}

\begin{figure}[t]
     \centering
     \begin{subfigure}[b]{0.49\textwidth}
         \centering
         \includegraphics[width=\textwidth]{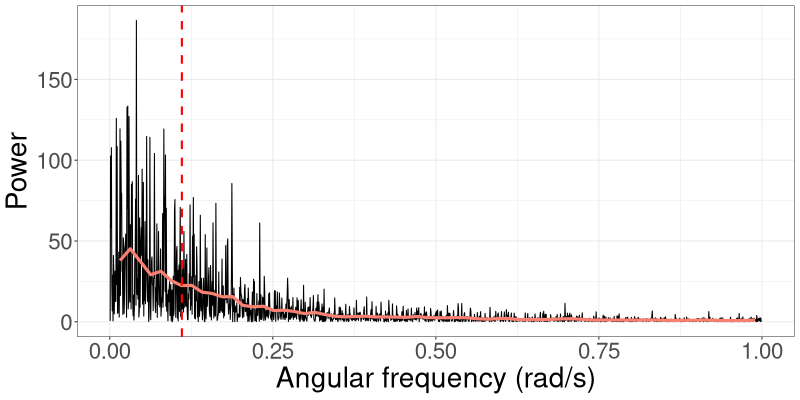}
         \caption{Linear Gaussian model.}
         \label{fig:periodogram_lgss_power2_20230525}
     \end{subfigure}
     \hfill
     \begin{subfigure}[b]{0.49\textwidth}
         \centering
         \includegraphics[width=\textwidth]{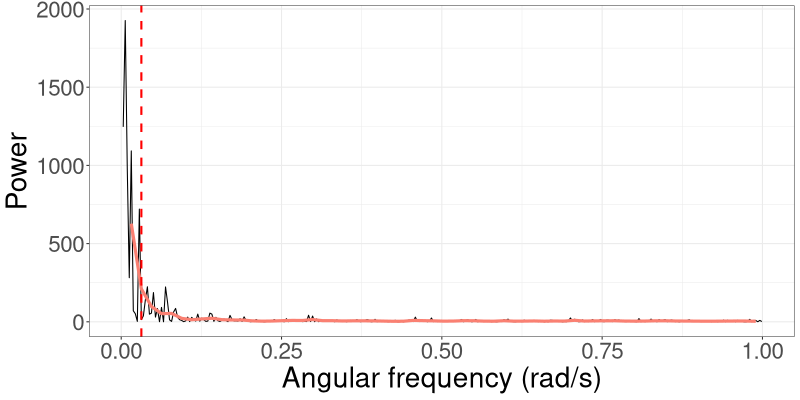}
         \caption{Univariate SV model.}
         \label{fig:periodogram_sv_sim_power2_20240214}
     \end{subfigure}
     \caption{Plot of raw periodograms of the data (black) and smoothed periodograms (light red) for the univariate linear Gaussian state space model of Section~\ref {sec:linear_gaussian} and the univariate stochastic volatility of Section~\ref{sec:sv_sim}. Vertical red dashed lines show the frequency cutoffs at half-power (3dB). Frequencies after this cutoff are processed in blocks.}
     \label{fig:cutoff_freqs_main}
\end{figure}

The spectral density of a process captures the distribution of signal power (variance) across frequency~\citep{percival1993spectral}. The periodogram is an estimate of the spectral density. In Figure~\ref{fig:periodogram_lgss_power2_20230525}, we show the periodogram for data generated from the linear Gaussian SSM that we consider in Section~\ref{sec:linear_gaussian}. The variance (power) of the signal gradually decreases as the frequency increases, which means that lower frequencies contribute more to the total power of the process. A similar pattern is seen in Figure~\ref{fig:periodogram_sv_sim_power2_20240214}, which shows the periodogram for data generated from a univariate stochastic volatility model that we consider in Section~\ref{sec:sv_sim}. In this case, there is an even larger concentration of power at the lower frequencies. \Copy{highlowfreq}{
For models where the power is concentrated at low frequencies, such as the ones presented in this paper, we can speed up the algorithm by processing frequency contributions with low power, specifically those above the 3dB cutoff point, together in ``blocks''. Here, we define the 3dB cutoff point as the angular frequency at which the power drops to half the maximum power in the (smoothed) periodogram, which we obtain using Welch's method \citep{welch1967use}. Section~\ref{sec:blocksize_test} studies several cutoff points and empirically justifies our choice of a 3dB cutoff. We note that for other models not examined in this paper, it is possible for higher frequencies to also exhibit substantial power. In such cases, we recommend evaluating all high-power frequencies individually.}

\begin{figure}[t]
     \centering
     \begin{subfigure}[b]{0.49\textwidth}
         \centering
         \includegraphics[width=\textwidth]{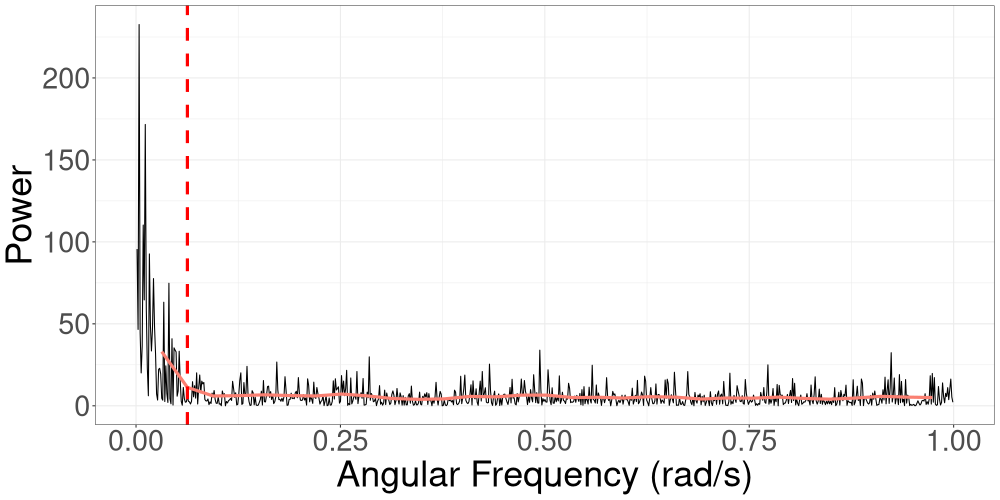}
         \caption{Variable 1.}
         \label{fig:bivariate_sv_pdg_plot_1}
     \end{subfigure}
     \hfill
     \begin{subfigure}[b]{0.49\textwidth}
         \centering
         \includegraphics[width=\textwidth]{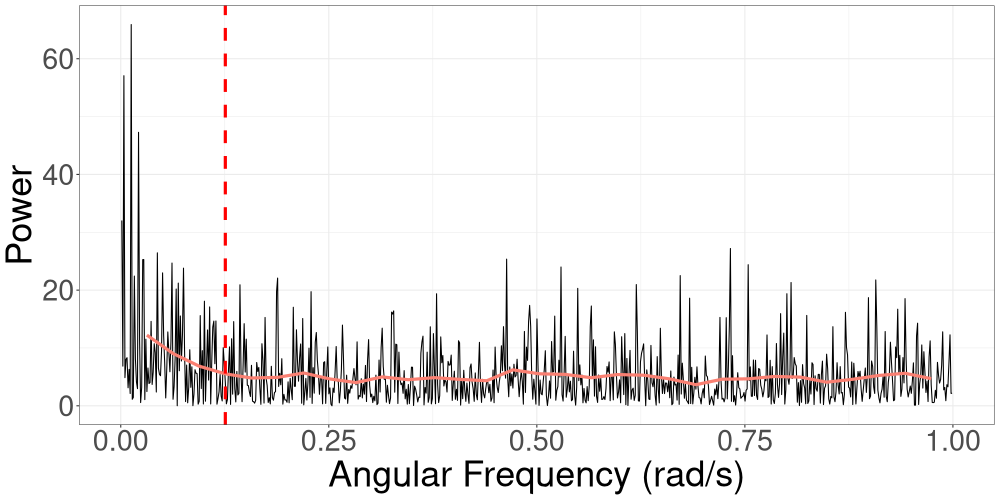}
         \caption{Variable 2.}
         \label{fig:bivariate_sv_pdg_plot_2}
     \end{subfigure}
     \caption{Plot of raw periodograms (black) and smoothed periodograms (light red) for the bivariate stochastic volatility model of Section~\ref{sec:bivariateSVmodel}. Vertical dashed lines show the 3dB frequency cutoff points for each series. In this bivariate case, the cutoff point we use for R-VGA-Whittle is the higher of the two.}
     \label{fig:bivariate_sv_pdg_plot}
\end{figure}

Assume the 3dB cutoff point occurs at $\omega_{\tilde{n}}$; we split the remaining $\floor{\frac{T-1}{2}} - \tilde{n}$ angular frequencies into blocks of approximately equal length $B$. Suppose that there are $n_b$ blocks, with $\Omega_b, b = 1, \dots, n_b$, denoting the set of angular frequencies in the $b$th block. The variational mean and precision matrix are then updated based on each block's Whittle log-likelihood contribution. For $b = 1, \dots, n_b$, this is given by
\begin{equation}
\label{eq:block_likelihood}
    l_{W}^b (\btheta) = \sum_{k \in \Omega_b} l_{W_k} (\btheta), 
\end{equation}
that is, by the sum of the Whittle log-likelihood contributions of the frequencies in the block.
For a multivariate time series, we compute the periodogram for each individual time series along with its corresponding 3dB cutoff point, and then use the highest cutoff point in R-VGA-Whittle. An example of bivariate time series periodograms along with their 3dB cutoffs points is shown in Figure~\ref{fig:bivariate_sv_pdg_plot}. 

We summarise the R-VGA-Whittle algorithm with block updates in Algorithm~\ref{algo:block_rvgaw}, under the assumption that there are no high-frequency components with substantial power beyond $\tilde{n}$.


\begin{algorithm} [t]
\caption{Block updates for R-VGA with the Whittle likelihood} 
\label{algo:block_rvgaw}
\begin{algorithmic}
\setstretch{1.5}
    \State Input: observations $\y_1, \dots, \y_T$, initial values $\bmu_0$ and $\bSigma_0$, number of angular frequencies to damp $n_{damp}$, number of damping steps $D$, number of angular frequencies to update individually $\tilde{n} > n_{damp}$, block length $B$. Assume that there are no high-frequency components with substantial power beyond $\tilde{n}$.
    \State Step 1: Compute the periodogram $\I(\omega_k)$, $k =  -\ceil*{\frac{T}{2}} + 1, \dots, \floor*{\frac{T}{2}}$, via the FFT algorithm.
    \State Step 2: Divide the last $\floor{\frac{T-1}{2}} - \tilde{n}$ frequencies into $n_b$ blocks of length (approximately) $B$.
    \State Step 3: Set $q_0(\btheta) = \Gau (\bmu_0, \bSigma_0)$.
    \For {$\tilde{k} = 1, \dots, \tilde{n}$}
        \State Update $\bmu_{\tilde{k}}, \bSigma_{\tilde{k}}$ as in Algorithm~\ref{algo:damped_rvgaw}.
    \EndFor
    \For {$b = 1, \dots, n_b$} 
        \State Set $\tilde{k} = \tilde{n} + b$,
        \State $\bmu_{\tilde{k}} = \bmu_{{\tilde{k}}-1} + \bSigma_{\tilde{k}} \Exp_{q_{{\tilde{k}}-1}} (\nabla_{\btheta} l_{W}^b (\btheta))$, where $l_{W}^b (\btheta)$ is given in~\eqref{eq:block_likelihood}, 
        \State $\bSigma_{\tilde{k}}^{-1} = \bSigma_{{\tilde{k}}-1}^{-1} - \Exp_{q_{{\tilde{k}}-1}} (\nabla_{\btheta}^2 l_{W}^b (\btheta))$. 
    \EndFor
    \State Output: variational parameters $\bmu_{\tilde{k}}$ and $\bSigma_{\tilde{k}}$, for $\tilde{k}=1, \dots, \tilde{n} + n_b$.
\end{algorithmic}
\end{algorithm}

\subsection{HMC-Whittle \label{sec:HMC_Whittle}}
We now discuss the use of HMC with the Whittle likelihood for estimating parameters in SSMs. Suppose we want to sample from the posterior distribution $p(\btheta \mid \y_{1:T})$. We introduce an auxiliary momentum variable $\r$ of the same length as $\btheta$ and assume $\r$ comes from a density $\Gau (\0, \M)$, where $\M$ is a \textit{mass matrix} that is often a multiple of the identity matrix. The joint density of $\btheta$ and $\r$ is $p(\btheta, \r \mid \y) \propto \exp(-H(\btheta, \r))$, where
\begin{equation}
    \label{eq:hamiltonian}
    H(\btheta, \r) = -\mathcal{L}(\btheta) + \frac{1}{2} \r^\top \M^{-1} \r
\end{equation}
is called the \textit{Hamiltonian}, and $\mathcal{L}(\btheta) = \log p(\btheta \mid \y_{1:T})$. HMC proceeds by moving $\btheta$ and $\r$ according to the Hamiltonian equations
\begin{equation}
\label{eq:hamiltonian_eq}
    \frac{\d \btheta}{\d t} = \frac{\partial H}{\partial \r} = \M^{-1} \r, \quad \frac{\d \r}{\d t} = - \frac{\partial H}{\partial \btheta} = \nabla_{\btheta} \mathcal{L}(\btheta),
\end{equation}
to obtain new proposed values $\btheta^*$ and $\r^*$. By Bayes' theorem, $p(\btheta \mid \y_{1:T}) \propto p(\y_{1:T} \mid \btheta) p(\btheta)$, so the gradient $\nabla_{\btheta} \mathcal{L}(\btheta)$ can be written as
\begin{equation}
\label{eq:hmc_gradient}
    \nabla_{\btheta} \mathcal{L}(\btheta) = \nabla_{\btheta} \log p(\y_{1:T} \mid \btheta) + \nabla_{\btheta} \log p(\btheta).
\end{equation}
The gradient $\nabla_{\btheta} \log p(\y_{1:T} \mid \btheta)$ on the right hand side of~\eqref{eq:hmc_gradient} is typically intractable in SSMs and needs to be estimated using a particle filter~\cite[e.g.,][]{nemeth2016particle}. However, if the exact likelihood is replaced with the Whittle likelihood, an approximation to ${\nabla_{\btheta} \mathcal{L}(\btheta)}$ can be obtained without the need for running an expensive particle filter repeatedly. Rather than targeting the true posterior distribution $p(\btheta \mid \y_{1:T})$, HMC-Whittle proceeds by targeting an approximation $\tilde{p}(\btheta \mid \y_{1:T}) \propto p_W(\y_{1:T} \mid \btheta) p(\btheta)$, where $p_W(\y_{1:T} \mid \btheta) \equiv \exp(l_W(\btheta))$ is the Whittle likelihood as defined in~\eqref{eq:whittle_llh_uni} for a univariate time series, or~\eqref{eq:whittle_llh_multi} for multivariate time series. Then $\mathcal{L}(\btheta)$ in~\eqref{eq:hamiltonian} is replaced by the quantity $\mathcal{L}_W(\btheta) = \log \tilde{p}(\btheta \mid \y_{1:T})$, and the term $\nabla_{\btheta} \mathcal{L}(\btheta)$ in~\eqref{eq:hamiltonian_eq} is replaced with 
\begin{equation}
    \nabla_{\btheta} \mathcal{L}_W(\btheta) = \nabla_{\btheta} l_W (\btheta) + \nabla_{\btheta} \log p(\btheta). 
\end{equation}
For models where the spectral density is available in closed form, the Whittle likelihood is also available in closed form, and the gradient $\nabla_{\btheta} \mathcal{L}_W(\btheta)$ can be obtained analytically. 

In practice, the continuous-time Hamiltonian dynamics in~\eqref{eq:hamiltonian_eq} are simulated over $L$ discrete time steps of size $\epsilon$ using a leapfrog integrator (see Algorithm~\ref{algo:leapfrog}). The resulting $\btheta^*$ and $\r^*$ from this integrator are then used as proposals and are accepted with probability 
\begin{equation*}
    \min \left(1, \frac{\exp(-H(\btheta^*, \r^*)}{\exp(-H(\btheta, \r)} \right).
\end{equation*}
This process of proposing and accepting/rejecting is repeated until the desired number of posterior samples has been obtained. We summarise HMC-Whittle in Algorithm~\ref{algo:hmc-whittle}.


\begin{algorithm} [t!]
\caption{One step of the Leapfrog algorithm, \text{Leapfrog}($\btheta, \r, \epsilon$)} 
\label{algo:leapfrog}
\begin{algorithmic}
    \State Input: parameters $\btheta$, $\r$, leapfrog step size $\epsilon$.
    \State Set $\r^{**} = \r +\epsilon\nabla_{\btheta} \mathcal{L}\left({\btheta}\right)/2$.
    \State Set $\btheta^{*} ={\btheta}+\epsilon\boldsymbol{{M}}^{-1}\r^{**}$.
    \State Set $\r^{*} = \r^{**} +\epsilon\nabla_{\btheta}\mathcal{L}\left(\btheta^{*}\right)/2$.
    \State Output: $\btheta^{*}$, $\r^{*}$.
    \end{algorithmic}
\end{algorithm}

\begin{algorithm} [t!]
\caption{HMC-Whittle} 
\label{algo:hmc-whittle}
\begin{algorithmic}
    \State Input: observations $\y_1, \dots, \y_T$, prior distribution $p(\btheta)$, mass matrix $\M$, number of HMC iterations $J$, leapfrog step size $\epsilon$, number of leapfrog steps $L$.
    \State Sample $\btheta^{(0)} \sim p(\btheta)$. 
    \For {$j = 1, \dots, J$} 
        \State Sample $\r^{(0)} \sim \Gau (\0, \M)$. 
        \State Set $\btheta^{*} = \btheta^{(j-1)}$, $\r^{*} = \r^{(0)}$.
        \For {$l = 1,...,L$} 
            \State $(\btheta^{*}, \r^{*}) \leftarrow \text{Leapfrog}(\btheta^{*}, \r^{*}, \epsilon)$. (see Algorithm \ref{algo:leapfrog})
        \EndFor
        \State With probability 
        \begin{equation*}
            \alpha = \text{min} \left( 1, \frac{\exp \left(\mathcal{L}_W(\btheta^{*}) - \frac{1}{2} \r^{*\top} \M^{-1} \r^*\right)}{\exp \left(\mathcal{L}_W(\btheta^{(j-1)}) - \frac{1}{2} \r^{(0)\top} \M^{-1} \r^{(0)} \right)} \right), 
        \end{equation*}
        \State set $\btheta^{(j)} = \btheta^*$, else $\btheta^{(j)} = \btheta^{(j-1)}$. 
    \EndFor
    \State Output: posterior samples $\btheta^{(1)}, \dots, \btheta^{(J)}$.
\end{algorithmic}
\end{algorithm}

\section{Applications}


\label{sec:applications}
This section demonstrates the use of R-VGA-Whittle to estimate parameters of various models: a univariate linear Gaussian state space model; univariate and bivariate stochastic volatility (SV) models; and a state space model with Student’s $t$ errors, where the latent states follow an autoregressive fractionally integrated moving average (ARFIMA) process. We evaluate R-VGA-Whittle on both simulated and real data, and compare the posterior densities obtained from R-VGA-Whittle to those from HMC-exact and HMC-Whittle. HMC-exact and HMC-Whittle are implemented using the CmdStanR interface to the Stan programming language~\citep{stan}. For the univariate SV model in Sections~\ref{sec:sv_sim} and~\ref{sec:real_data}, we also compare the posterior distributions from our algorithm to those obtained via the popular $\texttt{stochvol}$ R package~\citep{kastner2014ancillarity}. Reproducible R code for all examples is available from \url{https://github.com/bao-anh-vu/R-VGA-Whittle}. 




For all examples, we implement R-VGA-Whittle with damping and blocking as described in Algorithm~\ref{algo:block_rvgaw}. Damping is applied to the first $n_{damp} = 5$ angular frequencies with $D = 100$ damping steps each. These values were chosen to reduce the initial instability at the expense of a small additional computational cost. We select $\tilde{n}$ for each model by finding the 3dB frequency cutoff and process the remaining frequencies in blocks of length $B = 100$, following a study in Section~\ref{sec:blocksize_test} of the online supplement. At each iteration of R-VGA-Whittle, we use $S = 1000$ Monte Carlo samples to approximate the expectations of the gradient and Hessian of the Whittle log-likelihood. This value of $S$ was chosen because it results in very little variation between repeated runs of R-VGA-Whittle; see Section~\ref{sec:var_test} of the online supplement for an experiment exploring the algorithm's sensitivity to $S$. 

We run both the HMC-Whittle and HMC-exact algorithms using two parallel chains, each with a burn-in period of 1000 samples. The total number of iterations per chain is model-dependent, up to a maximum of 15000 samples (including burn-in).
We summarise the number of posterior samples used in both HMC-exact and HMC-Whittle in Table~\ref{tab:hmc_iters} in Section \ref{sec:traceplots} of the online supplement. These settings are chosen to target an effective sample size (ESS) of approximately $1000$ for all parameters. Detailed ESS results for each method are reported in Table~\ref{tab:hmc_convergence} in Section \ref{sec:traceplots} of the online supplement.
Posterior distributions estimated using HMC-exact are treated as ground truth for comparing the accuracy of the posteriors obtained using R-VGA-Whittle and HMC-Whittle. All experiments were carried out on a high-end server, with an NVIDIA Tesla A100 80GB graphics processing unit (GPU) used to parallelise over the $S = 1000$ Monte Carlo samples needed for estimating the expectations of the gradient and Hessian of the Whittle log-likelihood at each iteration of Algorithm~\ref{algo:damped_rvgaw}. 

Sections~\ref{sec:linear_gaussian} to~\ref{sec:arfima_ss} show results for simulated data examples involving a univariate linear Gaussian SSM, a univariate stochastic volatility (SV) model, a bivariate SV model, and an SSM with Student's $t$ error, where the latent states follow an ARFIMA process. Section~\ref{sec:real_data} applies R-VGA-Whittle to fit univariate and bivariate SV models to real data.

\subsection{Univariate linear Gaussian SSM}
\label{sec:linear_gaussian}
We generate $T = 10000$ observations from the linear Gaussian SSM, 
\begin{align}
    y_t &= x_t + \epsilon_t, \quad \epsilon_t \sim \Gau (0, \sigma_\epsilon^2), \quad t = 1, \dots, T, \\
    x_t &= \phi x_{t-1} + \eta_t, \quad \eta_t \sim \Gau (0, \sigma_\eta^2), \quad t = 2, \dots, T, \\
    x_1 &\sim \Gau \left(0, \frac{\sigma_\eta^2}{1-\phi^2}\right).
\end{align}
We set the true auto-correlation parameter to $\phi = 0.9$, the true state disturbance standard deviation to $\sigma_\eta = 0.7$ and the true measurement error standard deviation to $\sigma_\epsilon = 0.5$.

The parameters that need to be estimated, $\phi, \sigma_\eta$ and $\sigma_\epsilon$, take values in subsets of $\mathbb{R}$. We therefore instead estimate the transformed parameters ${\btheta = (\theta_\phi, \theta_\eta, \theta_\epsilon)^\top}$, where 
$\theta_\phi = \tanh^{-1}(\phi)$,
$\theta_\eta = \log(\sigma_\eta^{2})$, and $\theta_\epsilon = \log(\sigma_\epsilon^{2})$, which are unconstrained. 


The spectral density of the observed time series $\{y_t\}_{t=1}^{T}$ is 
\begin{align}
    f_y(\omega_k; \btheta) &= f_x(\omega_k; \btheta) + f_\epsilon(\omega_k; \btheta), \quad \omega_k = \frac{2\pi k}{T}, \quad k =  -\ceil*{\frac{T}{2}} + 1, \dots, \floor*{\frac{T}{2}}, 
\end{align}
where
\begin{equation*}
    f_x(\omega_k; \btheta) = \frac{\sigma_\eta^2}{1 + \phi^2 - 2 \phi \cos(\omega_k)}
\end{equation*}
is the spectral density of the AR(1) process $\{x_t\}_{t=1}^{T}$, and 
\begin{equation*}
    f_\epsilon(\omega_k; \btheta) = \sigma_\epsilon^2
\end{equation*}
is the spectral density of the white noise process $\{\epsilon_t\}_{t=1}^{T}$. The Whittle log-likelihood for this model is given by~\eqref{eq:whittle_llh_uni},
where the periodogram $\{I(\omega_k): k=-\ceil*{\frac{T}{2}} + 1, \dots, \floor*{\frac{T}{2}}\}$ is the discrete Fourier transform (DFT) of the observations $\{y_t\}_{t=1}^T$, which we compute using the $\texttt{fft}$ function in R. 
We evaluate the gradient and Hessian of the Whittle log-likelihood at each frequency, $\nabla_{\theta} l_{W_k} (\btheta)$ and $\nabla_{\theta}^2 l_{W_k} (\btheta)$, using automatic differentiation via the $\texttt{tensorflow}$ library~\citep{tensorflow2015-whitepaper} in R. The initial/prior distribution we use is 
\begin{equation}
    p(\btheta) = q_0(\btheta) = \Gau \left(
    \begin{bmatrix} 0 \\ -1 \\ -1 \end{bmatrix}, 
    \begin{bmatrix} 1 & 0 & 0 \\
    0 & 1 & 0 \\
    0 & 0 & 1 \\
    \end{bmatrix}
    \right).
\end{equation}
This leads to 95\% probability intervals of $(-0.96,  0.96)$ for $\phi$, and $(0.23, 1.59)$ for $\sigma_\eta$ and $\sigma_\epsilon$.


\begin{figure}[t]
    \centering
    \includegraphics[width = \linewidth]{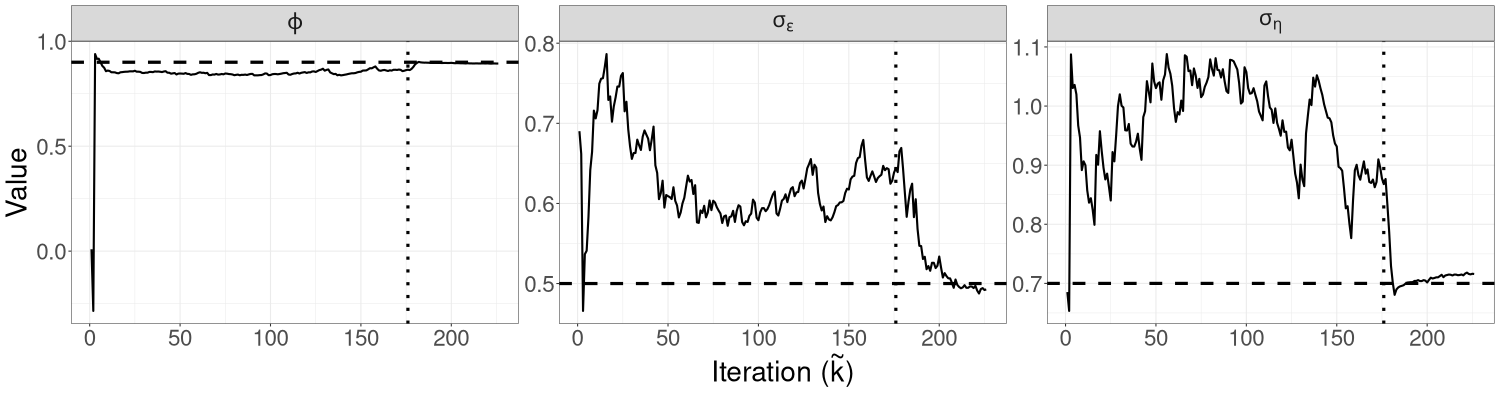}
    \caption{Trajectories of the variational means for parameters in the linear Gaussian SSM with simulated data. True parameter values are marked with dashed lines. The vertical line is at $\tilde{n}$, the point beyond which updates are done in blocks.}
    \label{fig:trajectories_lgss_blocksize100_176indiv}
\end{figure}

Figure~\ref{fig:trajectories_lgss_blocksize100_176indiv} shows the trajectory of the R-VGA-Whittle variational mean for each parameter. There are only 226 R-VGA-Whittle updates, as the 3dB cutoff point is the 176th angular frequency, and there are 50 blocks thereafter. The figure shows that the trajectory of $\phi$ moves towards the true value within the first 10 updates. This is expected, since most of the information on $\phi$, which is reasonably large in our example, is in the low-frequency components of the signal. The trajectories of $\sigma_\eta$ and $\sigma_\epsilon$ are more sporadic before settling close to the true value around the last 25 (block) updates. This later convergence is not unexpected since high-frequency log-likelihood components contain non-negligible information on these parameters that describe fine components of variation (the spectral density of white noise is a constant function of frequency). This behaviour is especially apparent for the measurement error standard deviation $\sigma_\epsilon$.



Figures~\ref{fig:lgss_hmc_traceplot} and \ref{fig:lgss_hmcw_traceplot} in Section~\ref{sec:traceplots} of the online supplement show the trace plots for each parameter obtained from HMC-exact and HMC-Whittle. Trace plots from both methods show good mixing for all three parameters but, for $\sigma_\eta$ and $\sigma_\epsilon$ in particular, HMC-Whittle has slightly better mixing than HMC-exact. 

\begin{figure}[t]
    \centering
    \includegraphics[width = 0.7\textwidth]{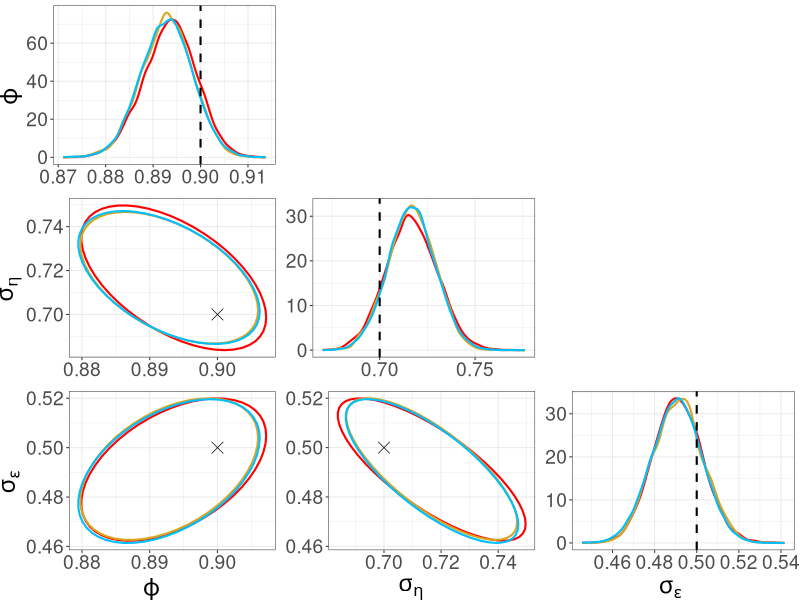}
    \caption{Posterior distributions from HMC-exact (blue), and approximate posterior distributions from R-VGA-Whittle (red) and HMC-Whittle (yellow) for data simulated from the univariate linear Gaussian SSM with $T = 10000$. Diagonal panels: Marginal posterior distributions with true parameters denoted using dotted lines. Off-diagonal panels: Bivariate posterior distributions with true parameters denoted using the symbol $\times$.}
    \label{fig:lgss_posterior_10000_temperfirst5_blocksize100_176indiv_arctanh_thinned_20230525}
\end{figure}

Figure~\ref{fig:lgss_posterior_10000_temperfirst5_blocksize100_176indiv_arctanh_thinned_20230525} shows the marginal posterior distributions of the parameters, along with bivariate posterior distributions. The three methods yield very similar posterior distributions for all parameters. We also show in Section~\ref{sec:blocksize_test} of the online supplement that the posterior densities obtained from R-VGA-Whittle using block size $B = 100$ are highly similar to those obtained without block updates. This suggests that block updates are useful for speeding up R-VGA-Whittle without compromising the accuracy of the approximate posterior densities.



\begin{table}[t]
\centering
\caption{Computing time (in seconds) for R-VGA-Whittle, HMC-Whittle and HMC-exact for each model we consider in this study. The number of observations $T$ for each example is also shown.}
\begin{tabular}{lrrrrr}
\hline
      & $T$ & R-VGA-Whittle & HMC-Whittle & HMC-exact & MCMC-stochvol \\ 
      \hline
Linear Gaussian  & 10000  &    8   &   10   &   348 & -  \\ 
SV (sim. data)  &  2000 &    8   &   4   &  383 & 11  \\ 
2D SV (sim. data) &  5000 &   53    &   1464   &  35421 & - \\ 
ARFIMA (sim. data) &  50000 &  823    &   14220   &  - & -   \\
SV (real data)   &  2649 &    8   &   3   &  1508 & 14  \\ 
2D SV (real data) &  2649 &  49    &   1136   &  37227 & -   \\ 
\hline
\end{tabular}
\label{tab:comp_time}
\end{table}

A comparison of the computing time between R-VGA-Whittle, HMC-Whittle and HMC-exact for this model can be found in Table~\ref{tab:comp_time}. The table shows that R-VGA-Whittle is slightly faster than HMC-Whittle and more than 40 times faster than HMC-exact. 

\subsection{Univariate stochastic volatility model \label{sec:sv_sim}}

\Copy{Kim}{SV models are a popular class of non-Gaussian financial time series models used for modelling the volatility of stock returns. Under a log-squared transformation of the data, a univariate SV model can be written in a linear additive form, where the error follows a log $\chi^{2}_{1}$ distribution~\citep{ruiz1994quasi}. A popular approach for estimating this type of model was proposed by \cite{kim1998stochastic}, in which the distribution of the error terms in the log-transformed data is approximated using a Gaussian mixture. This approach can be implemented via the R package \texttt{stochvol} \citep{kastner2014ancillarity}; we use this package for comparing to our proposed approach, and refer to this method as {MCMC-stochvol}.}  




In this section, we follow~\cite{fearnhead2011mcmc} and consider the following univariate SV model:
\begin{align}
    y_t &= \sigma_t \epsilon_t, \quad \epsilon_t \sim \Gau (0, 1), \quad t = 1, \dots, T, \label{eq:sv_y} \\
    \sigma_t &= \kappa \exp \left(\frac{x_t}{2}\right), \quad t = 1, \dots, T, \\
    x_t &= \phi x_{t-1} + \eta_t, \quad \eta_t \sim \Gau (0, \sigma_\eta^2), \quad t = 2, \dots, T,  \\
    x_1 &\sim \Gau \left(0, \frac{\sigma_\eta^2}{1 - \phi^2} \right) \label{eq:sv_x1}.
\end{align}


We apply a log-squared transformation~\cite{ruiz1994quasi},
\begin{align}
    \log (y_t^2) &= \log(\kappa^2) + x_t + \log(\epsilon_t^2) \nonumber \\ 
 &=\log(\kappa^2) +  \Exp(\log(\epsilon_t^2)) + x_t + \log(\epsilon_t^2) - \Exp(\log(\epsilon_t^2)) \nonumber \\
    &= \mu + x_t + \xi_t, \label{eq:log_squared}
\end{align}
where $\mu \equiv \log(\kappa^2) + \Exp(\log(\epsilon_t^2))$ and $\xi_t \equiv \log(\epsilon_t^2) - \Exp(\log(\epsilon_t^2))$, for $t = 1, \dots, T$. The terms $\{\xi_t\}_{t=1}^T$ are independent and identically distributed (iid) non-Gaussian terms with mean zero and variance $\frac{\pi^2}{2}$ (see~\cite{ruiz1994quasi} for a derivation, which is reproduced in Section~\ref{sec:add_maths} of the online supplement for completeness).

Rearranging~\eqref{eq:log_squared} yields the de-meaned series
\begin{equation}
\label{eq:log_squared_demeaned}
    z_t \equiv \log(y_t^2) - \mu = x_t + \xi_t, \quad t = 1, \dots, T.
\end{equation}
Since $\Exp(\log(y_t^2)) = \mu$, the parameter $\mu$ can be estimated using the sample mean~\citep{ruiz1994quasi}:
\begin{equation*}
    \hat{\mu} = \frac{1}{T} \sum_{t = 1}^T \log(y_t^2).
\end{equation*}
As $\mu = \log(\kappa^2) + \Exp(\log(\epsilon_t^2))$, once an estimate $\hat{\mu}$ is obtained, $\kappa$ can be estimated as
\begin{equation}\label{eq:kappaestimate}
    \hat{\kappa} = \sqrt{\exp(\hat{\mu} - \Exp(\log(\epsilon_t^2)))},
\end{equation}
where $\Exp(\log(\epsilon_t^2))) = \psi(1/2) + \log(2)$, and $\psi(\cdot)$ is the digamma function~\citep{abramowitz1968handbook}. 
\Copy{kappa}{A limitation of using the Whittle likelihood in place of the time-domain likelihood is that the parameter \( \kappa \) must be specified using the plug-in estimate given in~\eqref{eq:kappaestimate}, as the Whittle likelihood does not depend on \( \kappa \). In contrast, \( \kappa \) appears in the likelihood used by HMC-exact. Since the Whittle likelihood-based methods do not estimate \( \kappa \), we fix it to its plug-in estimate \( \hat{\kappa} \) when using HMC-exact in order to ensure comparability between methods.}





The remaining parameters to estimate are $\phi$ and $\sigma_\eta$. As with the linear Gaussian state space model, we work with the transformed parameters $\btheta = (\theta_\phi, \theta_\eta)^\top$, where 
$\theta_\phi = \tanh^{-1}(\phi)$
and $\theta_\eta = \log(\sigma_\eta^2)$, which are unconstrained. We fit the spectral density
\begin{align}
    f_{z}(\omega_k; \btheta) &= f_x(\omega_k; \btheta) + f_\xi(\omega_k; \btheta),
\end{align}
to the log-squared, demeaned series $\{z_t\}_{t=1}^T$, where 
\begin{equation*}
    f_x(\omega_k; \btheta) = \frac{\sigma_\eta^2}{1 + \phi^2 - 2 \phi \cos(\omega_k)},
\end{equation*}
and $f_\xi(\omega_k; \btheta)$ is the Fourier transform of the covariance function $C_{\xi}(h; \btheta)$ of $\{\xi_t\}_{t=1}^T$:
\begin{equation*}
    C_{\xi}(h; \btheta) = \Exp(\xi_t \xi_{t-h}) = 
    \begin{cases}
        \frac{\pi^2}{2}, & h = 0, \\
        0, & h \neq 0,
    \end{cases}
\end{equation*}
and is thus given by
\begin{equation} 
\label{eq:variancepsi}
    f_\xi(\omega_k; \btheta) = \sum_{h = -\infty}^\infty C_{\xi}(h; \btheta) \exp(-i \omega_k h) = C_{\xi}(0) = \frac{\pi^2}{2}, 
    \quad k =  -\ceil*{\frac{T}{2}} + 1, \dots, \floor*{\frac{T}{2}}. 
\end{equation}




The prior/initial variational distribution we use is 
\begin{equation}
    p(\btheta) = q_0(\btheta) = \Gau \left(
    \begin{bmatrix} 2 \\ -3 \end{bmatrix}, 
    \begin{bmatrix} 0.5 & 0 \\
    0 & 0.5 \\
    \end{bmatrix}
    \right), \label{eq:univariate_sv_prior}
\end{equation}
which leads to 95\% probability intervals of $(0.57, 0.99)$ for $\phi$, and $(0.11, 0.44)$ for $\sigma_\eta$.  The interval for 
$\phi$ aligns closely with those used in similar studies of the stochastic volatility (SV) model. For example, \citet{kim1998stochastic} employed the transformation $\phi = 2 \phi^* - 1$ with $\phi^* \sim \textrm{Beta}(20, 1.5)$, resulting in a 95\% probability interval for $\phi$ of $(0.59, 0.99)$.
Additionally, \citet{kim1998stochastic} utilised an inverse gamma prior for $\sigma_\eta^2$ with shape and rate parameters of $2.5$ and $0.025$, respectively, corresponding to a 95\% probability interval $\sigma_\eta$ for $(0.06, 0.24)$, which is comparable to the interval implied by our chosen prior. 

The \texttt{stochvol} package uses a Beta$(a_0, b_0)$ prior for $\phi$ and a Gamma$(1/2, \lambda_0/2)$ prior for $\sigma_\eta^2$, where $\lambda_0$ is the rate parameter. To make the priors comparable between \texttt{stochvol} and R-VGA-Whittle, we tune the hyperparameters $a_0, b_0$, and $\lambda_0$ by minimising the sum of squared differences between the 2.5th, 50th and 97.5th quantiles of the prior distributions used in $\texttt{stochvol}$ and those of the prior distribution in R-VGA-Whittle. The resulting prior hyperparameters used with $\texttt{stochvol}$ are $a_0 = 10.777, b_0 = 0.459$, and $\lambda_0 = 19.445$.  


We first generate $T = 2000$ observations from the univariate SV model given in \eqref{eq:sv_y}--\eqref{eq:sv_x1} with the true parameter values: $\phi = 0.99$, $\sigma_\eta = 0.4$, and $\kappa = 2$. We compare the performance of R-VGA-Whittle to the HMC-exact, HMC-Whittle, and MCMC-stochvol methods.

\begin{figure}[t]
    \centering
    \includegraphics[width = 0.7\linewidth]{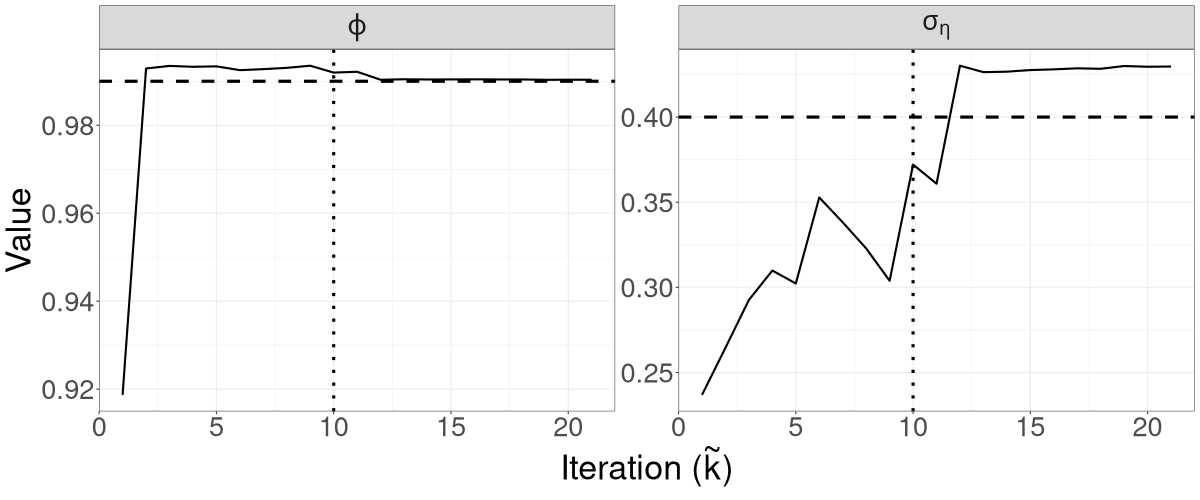}
    \caption{Trajectories of the variational means for the parameters in the univariate SV model fitted to simulated data with \( T = 2000 \), where \( \phi = 0.99 \) and \( \sigma_\eta = 0.4 \). The true parameter values are indicated by dashed lines, and the vertical line marks \( \tilde{n} \), the point beyond which updates are performed in blocks.}
    \label{fig:trajectories_sv_sim_blocksize100_10indiv}
\end{figure}

Figure~\ref{fig:trajectories_sv_sim_blocksize100_10indiv} shows the trajectories of the R-VGA-Whittle variational means for all parameters. The 3dB cutoff point for this example is the 10th frequency, after which there are 11 blocks, for a total of 21 updates. Similar to the case of the LGSS model in Section \ref{sec:linear_gaussian}, the variational mean of $\phi$ is close to the true value from the early stages of the algorithm, while that of $\sigma_\eta$ requires more iterations to achieve convergence.

Figures~\ref{fig:sv_sim_hmc_traceplot} -- \ref{fig:sv_sim_hmcw_traceplot} in Section~\ref{sec:traceplots} of the online supplement show the HMC-exact, HMC-Whittle, and MCMC-stochvol trace plots for all three methods. In this example, all MCMC methods yield well-mixed chains. However, both HMC-exact and MCMC-stochvol yield posterior samples with higher inefficiency factors, measured as the total number of MCMC iterations after burn-in divided by the effective sample size (ESS), compared to HMC-Whittle, particularly for \(\sigma_\eta\). This difference arises because HMC-Whittle targets the marginal posterior distribution of the parameters only, whereas HMC-exact samples from the full joint posterior of both the latent states and parameters, which is high-dimensional and more challenging to explore efficiently. Similarly, MCMC-stochvol employs a Gibbs sampling scheme that alternates between sampling the latent states \(\{x_t\}_{t=1}^T\) given the parameters \(\boldsymbol{\theta}\), and sampling \(\boldsymbol{\theta}\) given the latent states, which also introduces dependence between successive draws.



\begin{figure}[t]
    \centering
    \includegraphics[width = 0.65\textwidth]{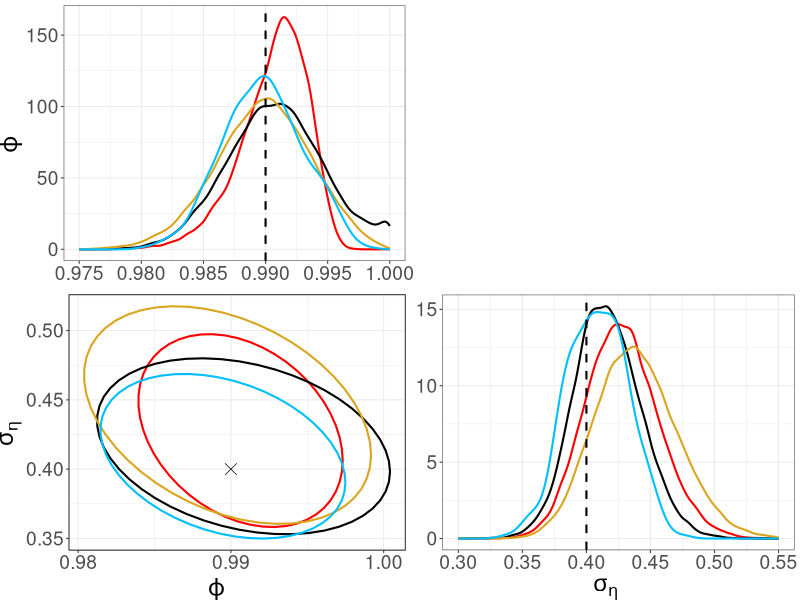}
    \caption{Posterior distributions from HMC-exact (blue), MCMC-stochvol (black), and approximate posterior distributions from R-VGA-Whittle (red) and HMC-Whittle (yellow) for data simulated from the univariate SV model with $T = 2000$, $\phi = 0.99$, and $\sigma_\eta = 0.4$. Diagonal panels: Marginal posterior distributions with true parameters denoted using dotted lines. Off-diagonal panels: Bivariate posterior distributions with true parameters denoted using the symbol $\times$.}
    \label{fig:sv_sim_posterior}
\end{figure}


\Copy{stochvol}{We compare the posterior densities from R-VGA-Whittle, HMC-Whittle, HMC-exact, and MCMC-stochvol in Figure~\ref{fig:sv_sim_posterior}. For the parameter \( \phi \), the posterior distributions obtained from HMC-Whittle closely resemble those from HMC-exact and MCMC-stochvol, whereas that from R-VGA-Whittle is slightly narrower. The posterior distributions for the parameter \( \sigma_\eta \) are similar across all four methods.}

Table~\ref{tab:comp_time} shows that R-VGA-Whittle is a few seconds slower than HMC-Whittle in this example, but faster than MCMC-stochvol, and nearly 50 times faster than HMC-exact. 

We now examine the empirical coverage of the 95\% credible intervals of R-VGA-Whittle and compare them to those obtained from HMC-exact and HMC-Whittle.
We consider different values of $\phi$ in the set $\{0.7, 0.8, 0.9, 0.99\}$, with $\sigma_\eta$ now fixed at 0.2. For each setting, we simulate 100 time series, each of length $T = 2000$. 
For each value of $\phi$, the $\hat{R}$ statistic and effective sample sizes averaged across 100 datasets are reported in Table~\ref{tab:rhatrepeated} in the online supplement. 

\begin{figure}[t]
     \centering
     \begin{subfigure}[b]{0.49\linewidth}
         \centering
         \centering
        \includegraphics[width=\linewidth]{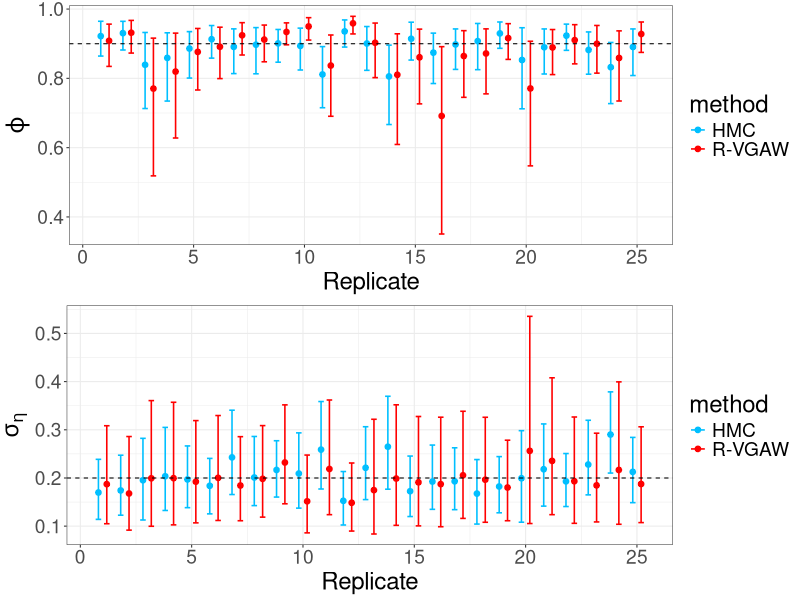}
         \caption{HMC-exact and R-VGA-Whittle.}
         \label{fig:ci_hmc_rvgaw_phi09}
     \end{subfigure}
     \hfill
     \begin{subfigure}[b]{0.49\linewidth}
         \centering
         \includegraphics[width=\linewidth]{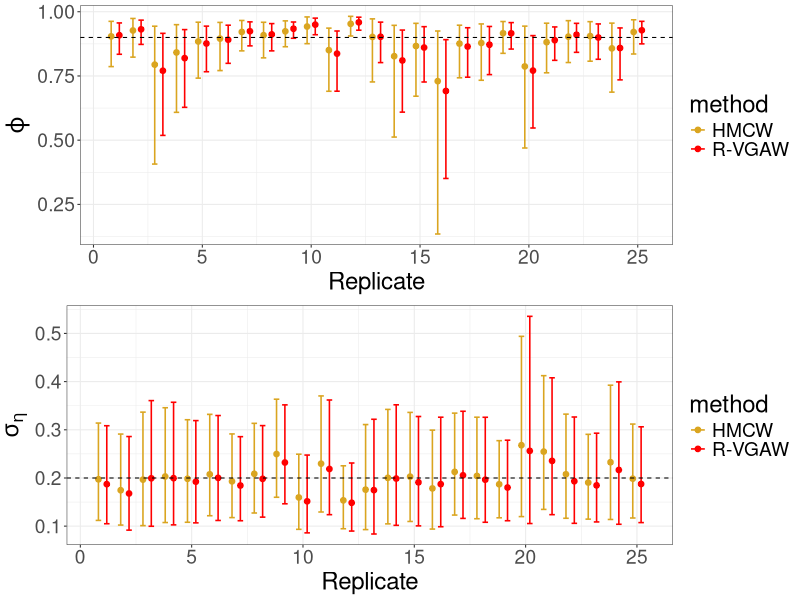}
         \caption{HMC-Whittle and R-VGA-Whittle.}
         \label{fig:ci_hmcw_rvgaw_phi09}
     \end{subfigure}
     \caption{Comparison of the 95\% credible intervals from R-VGA-Whittle and HMC-exact (left) and R-VGA-Whittle and HMC-Whittle (right) for the first 25 simulated datasets from the univariate SV model with $T=2000$, $\phi = 0.9$ and $\sigma_\eta = 0.2$. Posterior means are marked with points.}
     \label{fig:cis_phi09}
\end{figure}

Table \ref{tab:coverageSVmodel} gives the empirical coverage of the 95\% credible intervals for $\phi$ and $\sigma_\eta$, over 100 simulated datasets, for different values of $\phi$ and $\sigma_\eta = 0.2$.
The empirical coverages for $\phi$ are generally close to the nominal 95\% level across all methods, indicating good calibration. R-VGA-Whittle performs comparably to the HMC-Whittle and HMC-exact methods, although it slightly undercovers $\phi$ when $\phi = 0.9$, while HMC-exact and HMC-Whittle maintain more stable coverage across all values of $\phi$. The empirical coverage for $\sigma_\eta$ is consistently above 95\% across all three methods when $\phi = 0.7$, $0.8$, and $0.9$. At $\phi = 0.99$, both R-VGA-Whittle and HMC-Whittle exhibit undercoverage, while HMC-exact shows overcoverage. Figure \ref{fig:cis_phi09} show the comparisons of the 95\% credible intervals from R-VGA-Whittle and HMC-exact (left) and R-VGA-Whittle and HMC-Whittle (right) in the case of $\phi = 0.9$ for the first 25 simulated datasets. Figures \ref{fig:cis_phi07}--\ref{fig:cis_phi099} in Section \ref{sec:other_param_vals} of the online supplement show similar plots for $\phi \in \{0.7, 0.8, 0.99\}$ and $\sigma_\eta = 0.2$. The figures show that the widths of the 95\% credible intervals obtained from R-VGA-Whittle and HMC-Whittle are similar, but slightly different from those obtained with HMC-exact.

\begin{table}[t]
\caption{Empirical coverage of the 95\% credible intervals for \( \phi \) and \( \sigma_\eta \), computed over 100 simulated datasets of length \( T = 2000 \), with \( \phi \in \{0.7, 0.8, 0.9, 0.99\} \) and \( \sigma_\eta = 0.2 \). Results are reported for R-VGA-Whittle, HMC-Whittle, and HMC-exact.}
\centering
\begin{tabular}{|l|rr|rr|rr|}
\hline
$\phi$  & \multicolumn{2}{c|}{R-VGA-Whittle}                                            & \multicolumn{2}{c|}{HMC-Whittle}                                              & \multicolumn{2}{c|}{HMC}                                                      \\ \hline
     & \multicolumn{1}{c|}{Coverage of $\phi$} & \multicolumn{1}{c|}{Coverage of $\sigma_\eta$} & \multicolumn{1}{c|}{Coverage of $\phi$} & \multicolumn{1}{c|}{Coverage of $\sigma_\eta$} & \multicolumn{1}{c|}{Coverage of $\phi$} & \multicolumn{1}{c|}{Coverage of $\sigma_\eta$} \\  \hline
0.7  & 0.96                                 & 0.99                                   & 0.97                                 & 0.99                                   & 0.99                                 & 0.99                                   \\ 
0.8  & 0.97                                 & 0.99                                   & 0.99                                 & 0.99                                   & 0.98                                 & 0.99                                   \\ 
0.9  & 0.91                                  & 0.99                                   & 0.97                                 & 0.98                                   & 0.96                                 & 0.97                                   \\ 
0.99 & 0.94                                 & 0.92                                   & 0.97                                 & 0.89                                    & 0.96                                 & 0.97                                   \\ 
\hline
\end{tabular}
 \label{tab:coverageSVmodel}
\end{table}




\subsection{Bivariate SV model \label{sec:bivariateSVmodel}}
We now consider the following bivariate extension of the SV model in Section~\ref{sec:sv_sim} and simulate $T=5000$ observations as follows:
\begin{align}
    \y_t &= \V_t \bepsilon_t, \quad \bepsilon_t \sim \Gau (\0, \I_2), \quad t = 1, \dots, T, \label{eq:bivariate_sv_Y} \\
    \x_t &= \bPhi \x_{t-1} + \boldeta_t, \quad \boldeta_t \sim \Gau (\0, \bSigma_\eta), \quad t = 2, \dots, T, \\
    \x_1 &\sim \Gau (\0, \bSigma_{\eta_1}), \label{eq:bivariate_sv_X}
\end{align}
where $\x_t \equiv (x_{1,t}, x_{2,t})^\top$, $\y_t \equiv (y_{1,t}, y_{2,t})^\top$, ${\V_t \equiv \diag(\exp(\tfrac{x_{1,t}}{2}), \exp(\tfrac{x_{2,t}}{2}))}$, and $\I_m$ is the $m \times m$ identity matrix. The covariance matrix $\bSigma_{\eta_1}$ is the solution to the equation $\vec(\bSigma_{\eta_1}) = (\I_4 - \bPhi \otimes \bPhi)^{-1} \vec(\bSigma_{\eta})$, where $\textrm{vec}(\cdot)$ is the vectorisation operator  
and $\otimes$ denotes the Kronecker product of two matrices. 
We let $\bPhi$ be a diagonal matrix but let $\bSigma_\eta$ have off-diagonal entries so that there is dependence between the two series. We set these matrices to 
\begin{equation}
    \bPhi = \begin{bmatrix}
        0.99 & 0 \\
        0 & 0.98
    \end{bmatrix}, \quad 
    \bSigma_\eta = \begin{bmatrix}
        0.02  & 0.005 \\
        0.005 & 0.01
    \end{bmatrix}.
\end{equation}

As with the univariate model, we take a log-squared transformation of~\eqref{eq:bivariate_sv_Y} to find
\begin{equation}
\label{eq:y_tilde}
    \tilde{\y}_t = \bmu + \x_t + \bxi_t, 
\end{equation}
where 
\begin{equation*}
    \tilde{\y}_t = \begin{bmatrix}
        \log(y_{1,t}^2) \\
        \log(y_{2,t}^2)
    \end{bmatrix}, \quad 
    \bmu = \begin{bmatrix}
        \mu_1 \\
        \mu_2
    \end{bmatrix} =
    \begin{bmatrix}
        \Exp(\log(\epsilon_{1,t}^2)) \\ 
        \Exp(\log(\epsilon_{2,t}^2))
    \end{bmatrix}, \quad 
    \bxi_t = \begin{bmatrix}
        \xi_{1,t} \\
        \xi_{2,t}
    \end{bmatrix} =
    \begin{bmatrix}
        \log(\epsilon_{1,t}^2) - \Exp(\log(\epsilon_{1,t}^2)) \\
        \log(\epsilon_{2,t}^2) - \Exp(\log(\epsilon_{2,t}^2)) \\
    \end{bmatrix}.
\end{equation*}
The terms $\xi_{1,t}$ and $\xi_{2,t}$ are iid non-Gaussian uncorrelated terms, each with mean zero and variance $\sigma_\xi^2 = \frac{\pi^2}{2}$. We express \eqref{eq:y_tilde} as 
\begin{equation*}
    \z_t \equiv \tilde{\y}_t - \bmu = \x_t + \bxi_t, 
\end{equation*}
and use the sample means
\begin{equation*}
    \hat{\mu}_1 = \frac{1}{T} \sum_{t=1}^T \log(y_{1,t}^2), \quad \hat{\mu}_2 = \frac{1}{T} \sum_{t=1}^T \log(y_{2,t}^2)
\end{equation*}
as estimates of $\mu_1$ and $\mu_2$, respectively. The parameters in this model are $\bPhi$ and $\bSigma_\eta$. To estimate these parameters, we fit the following spectral density to $\{\z_t\}_{t=1}^T$:
\begin{equation*}
    \f_{\z} (\omega_k; \btheta) = \f_{\x} (\omega_k; \btheta) + \f_{\bxi} (\omega_k; \btheta),
\end{equation*}
where $\f_{\x} (\omega_k; \btheta)$ is the spectral density matrix of the VAR(1) process $\{\x_t\}_{t=1}^T$,
\begin{equation}
    \f_\x(\omega_k; \btheta) = (\I_2 - \bPhi e^{-i\omega_k})^{-1} \bSigma_\eta (\I_2 - \bPhi e^{-i\omega_k})^{-H}, 
    \quad k =  -\ceil*{\frac{T}{2}} + 1, \dots, \floor*{\frac{T}{2}},
\end{equation}
with $\A^{-H}$ denoting the inverse of the conjugate transpose of the matrix $\A$, and $\f_{\bxi} (\omega_k; \btheta) = \tfrac{\pi^2}{2} \I_2$ is the spectral density of $\{\bxi_t\}_{t=1}^T$ (following a similar derivation as that for the univariate SV model in Equation \eqref{eq:variancepsi}).

We parameterise $\theta_{ii} = \tanh^{-1}(\Phi_{ii}), i = 1, 2$, ${\bSigma_\eta = \L\L^\top}$, where $\L$ is the lower Cholesky factor of $\bSigma_\eta$, and the diagonal elements of $\L$ as $\gamma_{ii} = \log(l_{ii}), i = 1, 2$. The vector of transformed parameters is ${\btheta = (\theta_{11}, \theta_{22}, \gamma_{11}, \gamma_{22}, l_{21})^\top}$.

The initial/prior distribution we use is 
\begin{equation}
    p(\btheta) = q_0(\btheta) = \Gau \left(
    \begin{bmatrix} 2 \\ 2 \\ -2 \\ -3 \\ 0 \end{bmatrix}, 
    \begin{bmatrix} 
    0.5 & 0 & 0 & 0 & 0 \\
    0 & 0.5 & 0 & 0 & 0 \\
    0 & 0 & 0.5 & 0 & 0 \\
    0 & 0 & 0 & 0.05 & 0 \\
    0 & 0 & 0 & 0 & 0.05 \\
    \end{bmatrix}
    \right). \label{eq:bivariate_sv_prior}
\end{equation}
For $\Phi_{11}$ and $\Phi_{22}$, this prior leads to 95\% probability intervals of $(0.52, 0.99)$. For $\Sigma_{\eta_{11}}$, $\Sigma_{\eta_{21}}$, and $\Sigma_{\eta_{22}}$, the 95\% probability intervals are $(0.001, 0.286)$, $(-0.101, 0.101)$, and $(0.002, 0.261)$, respectively. 

\begin{figure}[t]
    \centering
    \includegraphics[width = 0.7\linewidth]{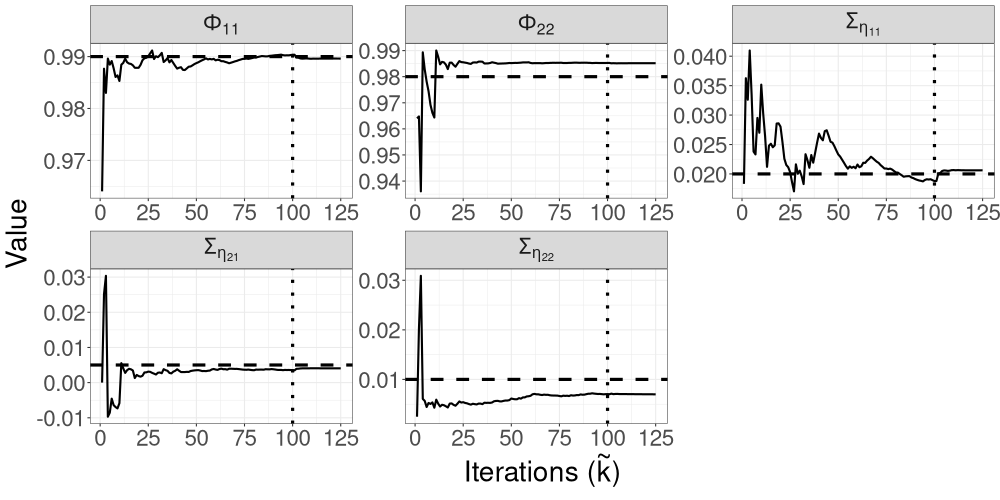}
    \caption{Trajectories of the variational means for the parameters in the bivariate SV model fitted to simulated data with \( T = 5000 \). The true parameter values are indicated by dashed lines, and the vertical line marks \( \tilde{n} \), the point beyond which updates are performed in blocks.}
    \label{fig:trajectories_multi_sv_sim_blocksize100_100indiv}
\end{figure}

Figure~\ref{fig:trajectories_multi_sv_sim_blocksize100_100indiv} shows the trajectories of the R-VGA-Whittle variational means for all parameters in this model. For most parameters, the trajectories quickly move towards the true value within the first 10 to 20 updates, with only small adjustments afterwards. The only exception is for parameter $\Sigma_{\eta_{11}}$, which exhibits more gradual changes before settling close to the true value in the last 25 iterations; this is somewhat expected since this parameter corresponds to a high-frequency signal component. R-VGA-Whittle slightly overestimated $\Phi_{22}$ and underestimated $\Sigma_{\eta_{22}}$.

Figures~\ref{fig:multi_sv_hmc_traceplot}--\ref{fig:multi_sv_hmcw_traceplot} in Section~\ref{sec:traceplots} of the online supplement show the HMC-exact and HMC-Whittle trace plots for this model. HMC-exact samples are highly correlated, especially those of $\Sigma_{\eta_{22}}$. We find that it is necessary to thin the HMC-exact samples by a factor of 50 for the samples to be considered approximately uncorrelated draws from the posterior distribution. In contrast, HMC-Whittle produces well-mixed posterior samples. Section~\ref{sec:traceplots} of the online supplement also shows that the inefficiency factors from HMC-Whittle are much lower than those from HMC-exact for all parameters.

\begin{figure}[t]
    \centering
    \includegraphics[width=\textwidth]{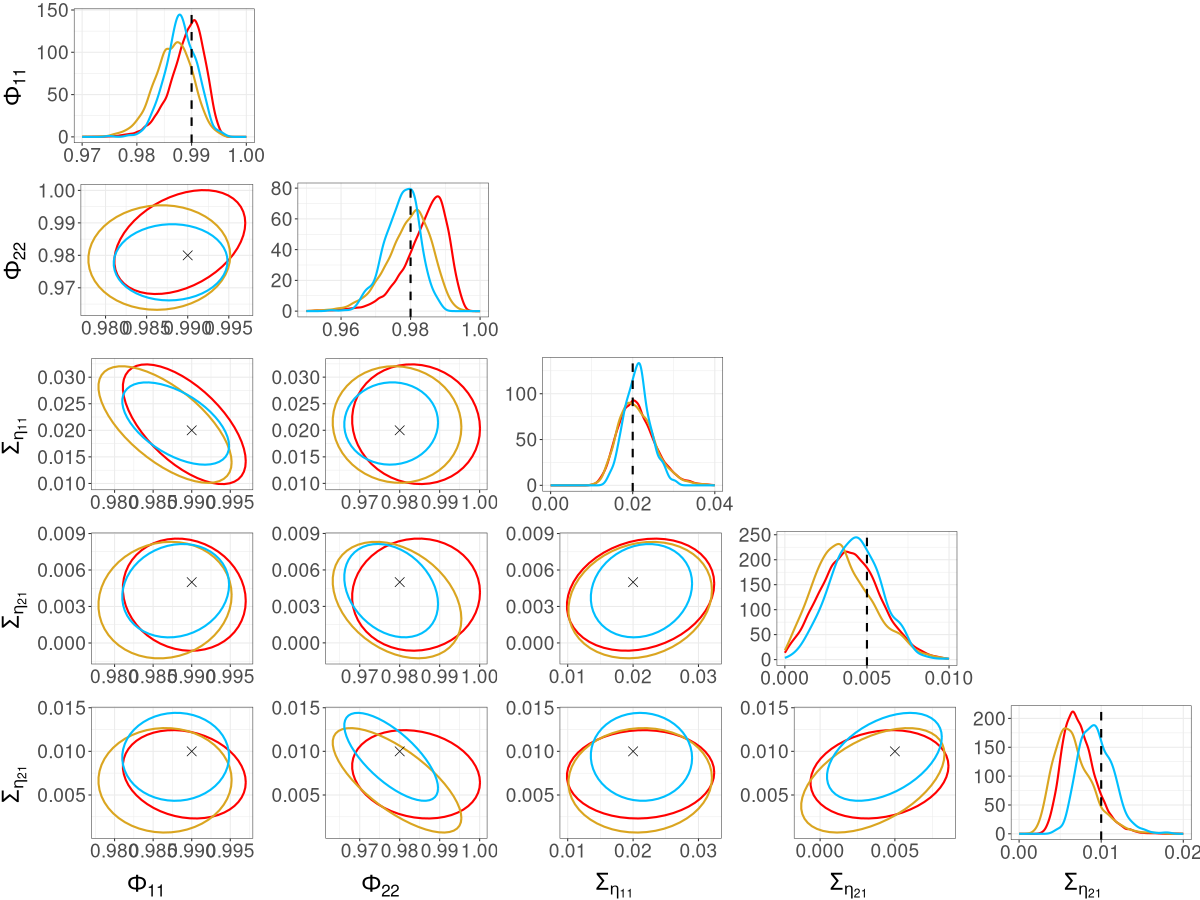}
    \caption{Posterior distributions from HMC-exact (blue), and approximate posterior distributions from R-VGA-Whittle (red) and HMC-Whittle (yellow) for data simulated from the bivariate SV model with $T=5000$. Diagonal panels: Marginal posterior distributions with true parameters denoted using dotted lines. Off-diagonal panels: Bivariate posterior distributions with true parameters denoted using the symbol $\times$.}
    \label{fig:multi_sv_sim_posterior_5000_temper5_blocksize100_100indiv_arctanh_thinned_20240613}
\end{figure}

We compare the posterior densities from R-VGA-Whittle, HMC-Whittle, and HMC-exact in Figure \ref{fig:multi_sv_sim_posterior_5000_temper5_blocksize100_100indiv_arctanh_thinned_20240613}. 
In this example, we observe broad agreement between the three methods, though the posterior densities from R-VGA-Whittle and HMC-Whittle are wider than those from HMC-exact. The posterior density of $\Phi_{22}$ from R-VGA-Whittle puts more probability on values close to one than those from the other two methods, while the posterior density of $\Sigma_{\eta_{21}}$ from both R-VGA-Whittle and HMC-Whittle appear slightly shifted towards zero compared to that from HMC-exact. As shown in Table~\ref{tab:comp_time}, R-VGA-Whittle is more than 25 times faster than HMC-Whittle, and over 600 times faster than HMC-exact in this example.


\subsection{ARFIMA model with t-distributed measurement noise}
\label{sec:arfima_ss}
Autoregressive Fractionally Integrated Moving Average (ARFIMA) models~\citep{granger1980introduction} are a class of long-memory time series models that extend the well-known class of Autoregressive Integrated Moving Average (ARIMA) models~\citep{box2015time} to non-integer values of the differencing parameter $d$. Of particular interest is the case $-0.5 < d < 0.5$, where the ARFIMA process is stationary and has long-range dependence with an autocovariance function that decays very slowly. ARFIMA processes are particularly useful for modeling macroeconomic and financial time series; see, for example, ~\citet[][]{bhardwaj_empirical_2006} for a comparison of the predictive accuracy (evaluated based on real datasets of stock index returns) of ARFIMA models to other conventional time series models such as autoregressive (AR), moving average (MA), and autoregressive moving average (ARMA) models. 


We use the $\texttt{fracdiff}$ R package \citep{maechler_fracdiff_2024} to simulate a time series of length $T = 50000$ from a model in which an underlying ARFIMA$(1,d,1)$ process is contaminated by measurement noise following a Student's $t$-distribution with $\nu$ degrees of freedom:
\begin{align}
    y_t &= x_t + \epsilon_t, \quad \epsilon_t \sim t_{\nu}(0,1), \nonumber \\
    \bPsi(L) (1 - L)^d x_t &= \bTheta(L) \eta_t, \quad \eta_t \sim N(0, \sigma_\eta^2), \quad t = 1, \dots, T,
\end{align}
where $L$ is the lag operator such that $L x_t = x_{t-1}$, $\bPsi(L) = 1 - \phi L$, $\bTheta(L) = 1 + \theta L$, and the fractional differencing operator $(1-L)^d$ is defined through a binomial series expansion~\citep{hosking_fractional_1981}. We set the true parameters to $\phi = 0.3, \theta = 0.7, d = 0.25, \sigma_\eta = 1$, and $\nu = 4$. This model is computationally challenging to fit using standard likelihood-based methods \citep{martin1999indirect}; estimating parameters in ARFIMA-type models is more computationally feasible in the frequency domain using the Whittle likelihood \citep{salomone2020spectral}. As a result, we only compare R-VGA-Whittle and HMC-Whittle for this example.

The spectral density for $\{y_t\}_{t=1}^T$ is given by
\begin{equation}
    f_y(\omega_k) = f_x(\omega_k) + f_{\epsilon}(\omega_k),
\end{equation}
where the spectral density for the underlying ARFIMA$(1, d, 1)$ process is given by
\begin{equation}
    f_x(\omega_k) = \sigma^2_\eta \abs{1 - e^{-i \omega_k}}^{-2d} \abs{\frac{\bTheta (e^{-i \omega_k})}{\bPsi (e^{-i \omega_k})}}^2 , \quad k =  -\ceil*{\frac{T}{2}} + 1, \dots, \floor*{\frac{T}{2}},
\end{equation}
and the spectral density for the Student's $t$ noise process with degree of freedom $\nu$ is
\begin{equation}
    f_{\epsilon}(\omega_k) = \frac{\nu}{\nu - 2}.
\end{equation}
As in previous examples, we use the following transformations to make the parameters unconstrained: ${\tilde{\phi} = \tanh^{-1}(\phi)}$, $\tilde{\theta} = \tanh^{-1}(\theta), \tilde{d} = \tanh^{-1}(2d), \tilde{\sigma}_\eta = \log(\sigma_\eta^2)$, and $\tilde{\nu} = \log(\nu - 2)$. 

To improve the convergence of HMC-Whittle and R-VGA-Whittle, we construct prior distributions for the parameters based on their maximum-Whittle-likelihood estimates. 
The prior distribution we use is 
\begin{equation}
    p(\btheta) = q_0(\btheta) = \Gau \left(
    \begin{bmatrix} \tanh^{-1} (\hat{\phi}) \\ \tanh^{-1} (\hat{\theta}) \\ \tanh^{-1} (2\hat{d}) \\ \log(\hat{\sigma}_\eta) \\ \log(\hat{\nu} - 2) \end{bmatrix}, 
    \begin{bmatrix} 
    0.25 & 0 & 0 & 0 & 0 \\
    0 & 0.25 & 0 & 0 & 0 \\
    0 & 0 & 0.25 & 0 & 0 \\
    0 & 0 & 0 & 1 & 0 \\
    0 & 0 & 0 & 0 & 1 \\
    \end{bmatrix}
    \right), \label{eq:arfima_prior}
\end{equation}
where $\hat{\phi}, \hat{\theta}, \hat{d}, \hat{\sigma}_\eta$, and $\hat{\nu}$ are the maximum-Whittle-likelihood estimates of $\phi, \theta, d, \sigma_\eta$, and $\nu$. This prior distribution results in $95\%$ probability intervals of $(-0.56, 0.86)$ for $\phi$, $(-0.23, 0.94)$ for $\theta$, $(-0.21, 0.45)$ for $d$, $(0.39, 2.76)$ for $\sigma_\eta$, and $(2.29, 16.98)$ for $\nu$.

\begin{figure}[ht]
    \centering
    \includegraphics[width=0.8\linewidth]{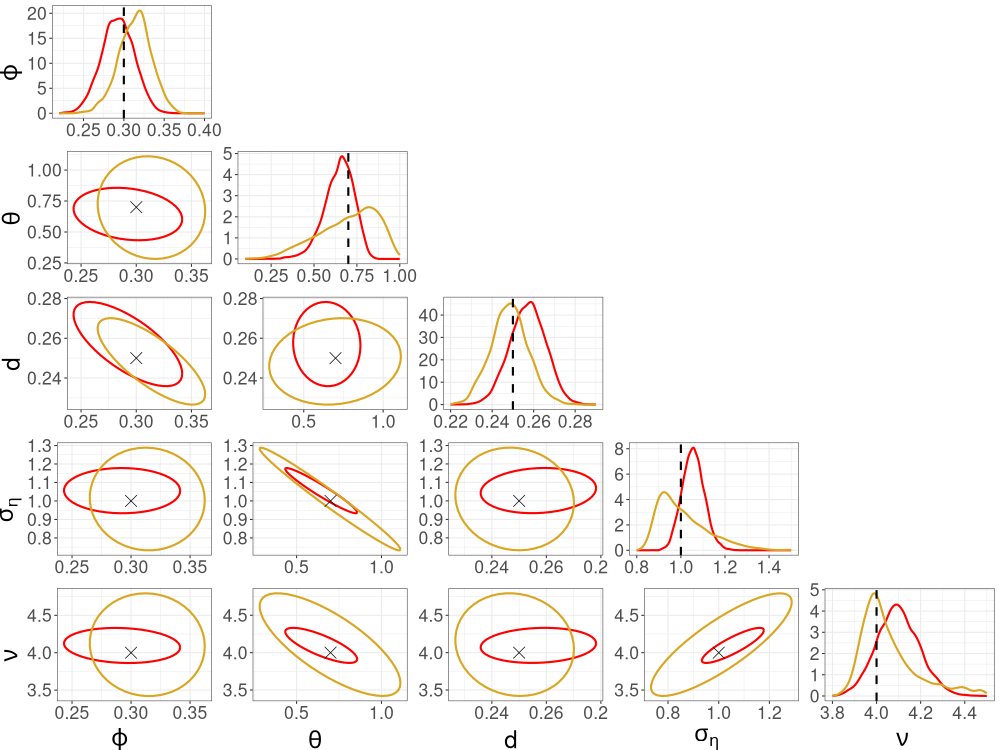}
    \caption{Approximate posterior distributions from R-VGA-Whittle (red) and HMC-Whittle (yellow) for data simulated from the SSM with Student's $t$ error, where the latent states follow an ARFIMA$(1,d,1)$ process, with $T=50000$ observations. Diagonal panels: Marginal posterior distributions with true parameters denoted using dotted lines. Off-diagonal panels: Bivariate posterior distributions with true parameters denoted using the symbol $\times$.}
    \label{fig:arfima_ss_posterior_n50000}
\end{figure}

For this model, we find it necessary to increase the number of damping observations in R-VGA-Whittle to ${n_{damp} = 100}$. All other tuning parameters for R-VGA-Whittle are kept the same as in other examples. We thin the HMC-Whittle posterior samples by a factor of 10 to reduce autocorrelation in the Markov chain. A comparison of the posterior distributions from R-VGA-Whittle and HMC-Whittle is shown in Figure~\ref{fig:arfima_ss_posterior_n50000}. We find that the posterior distributions from both R-VGA-Whittle and HMC-Whittle contain the true parameters, but that the posterior distributions from R-VGA-Whittle are slightly narrower than those from HMC-Whittle.

R-VGA-Whittle takes less than 15 minutes to fit this model, whereas HMC-Whittle takes nearly 4 hours (see Table~\ref{tab:comp_time}), making R-VGA-Whittle more than 17 times faster in this case. This further highlights the computational efficiency of R-VGA-Whittle for larger datasets. For ARFIMA models, larger datasets are not uncommon. \cite{salomone2020spectral}, for instance, uses an SV model with an underlying ARFIMA process to analyse a time series of one-minute Bitcoin returns with an order of magnitude more observations than the time series in this simulation example.



\subsection{Univariate and bivariate SV models with real data}
\label{sec:real_data}
We apply R-VGA-Whittle to estimate parameters in univariate and bivariate SV models from real data. We first apply it to daily JPY-EUR exchange rate data~\citep{kastner2017efficient} from April 1st, 2005 to August 6th, 2015, which consists of $2650$ observations. Typically, the data used in SV models are de-meaned log-returns, which are computed from the raw exchange rate values as:
\begin{equation*}
    y_t = y_t^{*} - \frac{1}{T} \sum_{t'=1}^T y_{t'}^{*}, \quad 
    y_t^{*} = \log \left(\frac{r_{t+1}}{r_t} \right), \quad t = 1, \dots, 2649,
\end{equation*}
where $r_t$ is the exchange rate on day $t$, and $y_t^{*}$ is the log-return on day $t$. We fit the model \eqref{eq:sv_y}--\eqref{eq:sv_x1} to this time series of de-meaned log-returns with $T = 2649$, and use the initial variational distribution specified in \eqref{eq:univariate_sv_prior}. For the univariate SV model, we also run MCMC-stochvol with the same settings as in Section~\ref{sec:sv_sim}.  



The HMC-exact trace plots in Figures~\ref{fig:sv_real_hmc_traceplot} of the online supplement show that posterior samples of $\phi$ are relatively well-mixed, but that the posterior samples of $\sigma_\eta$ are highly correlated. We find it necessary to thin these samples by a factor of 50 to reduce autocorrelation and obtain approximately uncorrelated draws. Similarly, MCMC-stochvol also generates highly autocorrelated samples and requires thinning by a factor of 50. In contrast, the HMC-Whittle approach produces well-mixed posterior samples with low autocorrelation. 

\begin{figure}[t]
    \centering
    \includegraphics[width = 0.65\textwidth]{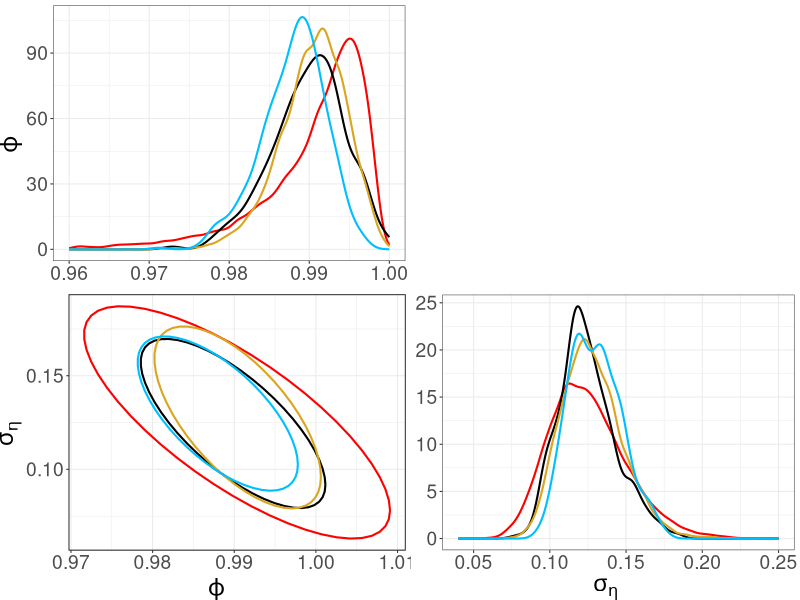}
    \caption{Posterior distributions from HMC-exact (blue), MCMC-stochvol (black), and approximate posterior distributions from R-VGA-Whittle (red) and HMC-Whittle (yellow) for the univariate SV model fitted using JPY-EUR exchange rate data with $T=2649$. Diagonal panels: Marginal posterior distributions of the parameters. Off-diagonal panels: Bivariate posterior distributions of the parameters.}
    \label{fig:sv_real_posterior_2649_temperfirst5_blocksize100_20indiv_arctanh_thinned_20230918}
\end{figure}

We compare the posterior densities from R-VGA-Whittle, HMC-Whittle and HMC-exact in Figure~\ref{fig:sv_real_posterior_2649_temperfirst5_blocksize100_20indiv_arctanh_thinned_20230918}. The posterior densities obtained from R-VGA-Whittle are similar, but slightly wider than those obtained from the other three methods.


We also apply R-VGA-Whittle to a bivariate time series of daily GBP-EUR and USD-EUR exchange rates~\citep{kastner2017efficient} from April 1st, 2005, to August 6th, 2015, which consists of $2650$ observations per exchange rate. We compute the de-meaned log-returns for both series and then fit the model \eqref{eq:bivariate_sv_Y}--\eqref{eq:bivariate_sv_X} to these log-returns. The prior distribution we use is specified in~\eqref{eq:bivariate_sv_prior}.



Figures~\ref{fig:multi_sv_real_hmc_traceplot}--\ref{fig:multi_sv_real_hmcw_traceplot} in Section~\ref{sec:traceplots} of the online supplement show the HMC-exact and HMC-Whittle trace plots for the bivariate SV model with exchange rate data. Similar to the simulation study, we find that HMC-exact samples are highly correlated, with samples of parameters in $\bPhi$ less correlated than samples of parameters in $\bSigma_\eta$. We thin the HMC-exact samples by a factor of 50 so that the remaining samples can be treated as approximately uncorrelated. Posterior samples from HMC-Whittle have low correlation and are mixed well for all parameters, so no thinning is necessary. 

\begin{figure}[t]
    \centering
    \includegraphics[width = 0.8\textwidth]{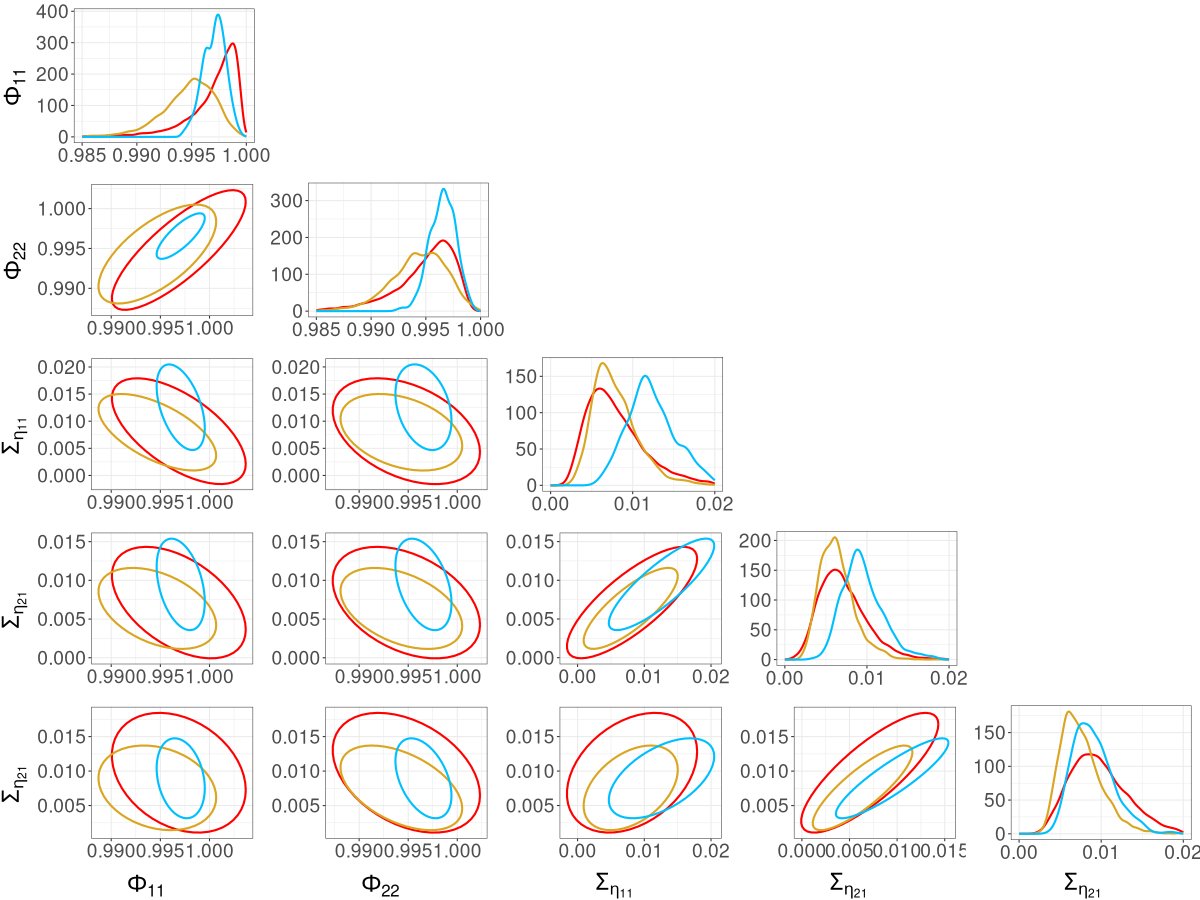}
    
    \caption{Posterior distributions from HMC-exact (blue), and approximate posterior distributions from R-VGA-Whittle (red) and HMC-Whittle (yellow) for the bivariate SV model fitted using GBP-EUR and USD-EUR exchange rate data with $T=2649$. Diagonal panels: Marginal posterior distributions of the parameters. Off-diagonal panels: Bivariate posterior distributions of the parameters.}
    \label{fig:multi_sv_real_posterior_GBP_USD_temper5_blocksize100_50indiv_arctanh_thinned_20240613}
\end{figure}

We compare the posterior densities from R-VGA-Whittle, HMC-Whittle, and HMC-exact in Figure~\ref{fig:multi_sv_real_posterior_GBP_USD_temper5_blocksize100_50indiv_arctanh_thinned_20240613}. The marginal posterior densities from R-VGA-Whittle and HMC-Whittle are similar, but noticeably wider than the posterior densities from HMC-exact. 
Nevertheless, there is a substantial overlap between the posterior densities from all three methods. 


Table~\ref{tab:comp_time} shows that in the univariate case, R-VGA-Whittle is slightly slower than HMC-Whittle, but about twice as fast as MCMC-stochvol, and more than 180 times faster than HMC-exact. In the bivariate case, R-VGA-Whittle is more than 23 times faster than HMC-Whittle and more than 750 times faster than HMC-exact.


\section{Conclusion}
\label{sec:conclusion}

We develop a novel sequential variational Bayes algorithm that uses the Whittle likelihood, called {R-VGA-Whittle}, for parameter inference with linear non-Gaussian state space models. The Whittle likelihood is a frequency-domain approximation of the exact likelihood. It is available in closed form 
and its log can be expressed as a sum of individual log-likelihood frequency components. At each iteration, R-VGA-Whittle sequentially updates the variational parameters by processing one frequency or a block of frequencies at a time. We also demonstrate the use of the Whittle likelihood in HMC (HMC-Whittle), and compare the parameter estimates from R-VGA-Whittle and HMC-Whittle to those of HMC with the exact likelihood (HMC-exact).
We show that the posterior densities from the three methods are generally similar, though posterior densities from R-VGA-Whittle and HMC-Whittle tend to be wider compared to those from HMC-exact. As HMC-Whittle only targets the parameters and not the states, it tends to produce posterior samples of the parameters that are less correlated than those from HMC-exact. We also show that R-VGA-Whittle and HMC-Whittle are much more computationally efficient than HMC-exact. R-VGA-Whittle is particularly efficient when dealing with bivariate time series, where in our study it is more than 25 times faster than HMC-Whittle and 750 times faster than HMC-exact. R-VGA-Whittle is also shown to be highly computationally efficient in an example involving a long memory (ARFIMA) latent process with a large sample size, which is challenging to fit using classic likelihood-based approaches.


\Copy{theoretical}{There are several avenues for future research. First, the work in this paper is largely computational and empirical, and an important next step is to derive theoretical properties of the R-VGA-Whittle algorithm.} Second, R-VGA-Whittle assumes a Gaussian variational distribution; future work could attempt to extend this framework to more flexible variational distributions for approximating skewed or multimodal posterior distributions. Third, some tuning parameters of R-VGA-Whittle (such as the number of damping observations and number of damping steps) are currently user-defined, so an adaptive tuning scheme could be developed to select these parameters. Fourth, as the Whittle likelihood is known to produce biased parameter estimates for finite sample sizes, using R-VGA-Whittle with the debiased Whittle likelihood of~\cite{sykulski2019debiased} may improve parameter estimates. 





\section*{Acknowledgements}

This work was supported by ARC SRIEAS Grant SR200100005 Securing Antarctica’s Environmental Future. 

\clearpage
\bibliographystyle{apalike}
\bibliography{bibliography}

\ifarxiv
    \newpage
    \renewcommand{\thealgorithm}{S\arabic{algorithm}}
    \renewcommand{\theequation}{S\arabic{equation}}
    \renewcommand{\thesection}{S\arabic{section}}
    \renewcommand{\thepage}{S\arabic{page}}
    \renewcommand{\thetable}{S\arabic{table}}
    \renewcommand{\thefigure}{S\arabic{figure}}
    \setcounter{page}{1}
    \setcounter{section}{0}
    \setcounter{equation}{0}
    \setcounter{algorithm}{0}
    \setcounter{table}{0}
    \setcounter{figure}{0}
    \section{The R-VGA algorithm}
\label{sec:rvga}

In this section, we provide a sketch of the derivations for the R-VGA algorithm of~\cite{lambert2022recursive}, which serves as the basis for our R-VGA-Whittle algorithm. 

The R-VGA algorithm of~\cite{lambert2022recursive} requires iid observations, denoted by ${\y_{1:N} \equiv (\y_1^\top, \dots, \y_N^\top)^\top}$, and approximates the posterior distribution $p(\btheta \mid \y_{1:N})$ with a Gaussian distribution $N(\bmu, \bSigma)$. By assumption of conditional independence between observations $\y_{1:N}$ given the parameters $\btheta$, the KL divergence between the variational distribution $q_i(\btheta)$ and the posterior distribution $p(\btheta \mid \y_{1:i})$ can be expressed as
\begin{align*}
    \text{KL} \divergence{q_i(\btheta)}{p(\btheta \mid \y_{1:i})} 
    &\equiv \int q_i(\btheta) \log \frac{q_i(\btheta)}{p(\btheta \mid \y_{1:i})} \d \btheta \\
    &= \Exp_{q_i} \left(\log q_i(\btheta) - \log p(\btheta \mid \y_{1:i-1}) - \log p(\y_i \mid \btheta) \right) + \log p(\y_{1:i}) - \log p(\y_{1:i-1}).
\end{align*}
In a sequential VB framework, the posterior distribution after incorporating the first $i-1$ observations, $p(\btheta \mid \y_{1:i-1})$, is approximated by the variational distribution $q_{i-1} (\btheta)$ to give
\begin{equation}
    \label{eq:KL_approx}
    \text{KL} \divergence{q_i(\btheta)}{p(\btheta \mid \y_{1:i})} \approx \Exp_{q_i}(\log q_i(\btheta) - \log q_{i-1} (\btheta) - \log p(\y_i \mid \btheta)) + \log p(\y_{1:i}) - \log p(\y_{1:i-1}).
\end{equation}
As the last two terms on the right hand side of~\eqref{eq:KL_approx} do not depend on $\btheta$, the KL-minimisation problem is equivalent to solving 
\begin{equation}
    \label{eq:exp_to_minimise}
    \argmin_{\bmu_i, \bSigma_i} \, \Exp_{q_i}(\log q_i(\btheta) - \log q_{i-1} (\btheta) - \log p(\y_i \mid \btheta)).
\end{equation}
Differentiating the expectation in~\eqref{eq:exp_to_minimise} with respect to $\bmu_i$ and $\bSigma_i$, setting the derivatives to zero, and rearranging the resulting equations, yields the following recursive updates for the variational mean~$\bmu_i$ and precision matrix~$\bSigma_i^{-1}$:
\begin{align}
    \bmu_i &= \bmu_{i-1} + \bSigma_{i-1} \nabla_{\bmu_i} \Exp_{q_i}(\log p(\y_i \mid \btheta)), \label{eq:bmu_1}\\
    \bSigma_i^{-1} &= \bSigma_{i-1}^{-1} - 2\nabla_{\bSigma_i} \Exp_{q_i}(\log p(\y_i \mid \btheta)). \label{eq:bSigma_1}
\end{align}
Then, using Bonnet's Theorem~\citep{bonnet1964transformations} on~\eqref{eq:bmu_1} and Price's Theorem~\citep{price1958useful} on~\eqref{eq:bSigma_1}, we rewrite the gradient terms as
\begin{align}
    \nabla_{\bmu_i} \Exp_{q_i}(\log p(\y_i \mid \btheta)) &= \Exp_{q_i}(\nabla_{\btheta} \log p(\y_i \mid \btheta)), \\
    \nabla_{\bSigma_i} \Exp_{q_i}(\log p(\y_i \mid \btheta)) &= \frac{1}{2} \Exp_{q_i}(\nabla_{\btheta}^2 \log p(\y_i \mid \btheta)).
\end{align}
Thus the updates~\eqref{eq:bmu_1} and~\eqref{eq:bSigma_1} become
\begin{align}
    \bmu_i &= \bmu_{i-1} + \bSigma_{i-1} \Exp_{q_i}(\nabla_{\btheta} \log p(\y_i \mid \btheta)), \label{eq:bmu_2}\\
    \bSigma_i^{-1} &= \bSigma_{i-1}^{-1} - \Exp_{q_i}(\nabla_{\btheta}^2 \log p(\y_i \mid \btheta)). \label{eq:bSigma_2}
\end{align}
These updates are implicit as they require the evaluation of expectations with respect to $q_i(\btheta)$. Under the assumption that $q_i(\btheta)$ is close to $q_{i-1}(\btheta)$, \cite{lambert2022recursive} propose replacing $q_i(\btheta)$ with $q_{i-1}(\btheta)$ in~\eqref{eq:bmu_2} and~\eqref{eq:bSigma_2}, and replacing $\bSigma_{i-1}$ with $\bSigma_i$ on the right hand side of~\eqref{eq:bmu_2}, to yield an explicit scheme
\begin{align}
    \bmu_i &= \bmu_{i-1} + \bSigma_i \Exp_{q_{i-1}}(\nabla_{\btheta} \log p(\y_i \mid \btheta)), \label{eq:bmu_3}\\
    \bSigma_i^{-1} &= \bSigma_{i-1}^{-1} - \Exp_{q_{i-1}}(\nabla_{\btheta}^2 \log p(\y_i \mid \btheta)). \label{eq:bSigma_3}
\end{align}
Equations~\eqref{eq:bmu_3} and~\eqref{eq:bSigma_3} form the so-called R-VGA algorithm of~\cite{lambert2022recursive}.


    \section{Additional details on the SV model}
\label{sec:add_maths}
In this section, we follow~\cite{ruiz1994quasi} and show that the terms $\{\xi_t\}_{t=1}^T$ in the log-squared transformation \eqref{eq:log_squared} in the main paper have mean zero and variance $\pi^2/2$.

Since $\epsilon_t \sim N(0, 1)$, we have $\epsilon_t^2 \sim \chi^2_1$, that is, $\epsilon^2_t$ is gamma distributed with shape and scale parameters equal to $1/2$ and $2$, respectively.
Therefore, $\log(\epsilon_t^2)$ has mean $\psi(1/2) + \log(2)$, where $\psi(\cdot)$ denotes the \textit{digamma} function (the first derivative of the logarithm of the gamma function), and variance $\psi^{(1)} \left( \frac{1}{2} \right) = \frac{\pi^2}{2}$,
where $\psi^{(1)}(\cdot)$ is the \textit{trigamma} function (the second derivative of the logarithm of the gamma function; see, e.g.,~\cite{abramowitz1968handbook}). Thus
\begin{equation*}
    \Exp(\xi_t) = \Exp\left( \log(\epsilon_t^2) - \Exp(\log(\epsilon_t^2)) \right) = 0,
\end{equation*}
and
\begin{equation*}
    \Var(\xi_t) = \Var\left( \log(\epsilon_t^2) - \Exp(\log(\epsilon_t^2)) \right) = \Var\left( \log(\epsilon_t^2) \right) = \frac{\pi^2}{2}.
\end{equation*}


    \section{Experiments involving different blocking strategies}
\label{sec:blocksize_test}
In this section, we discuss the tuning parameters in Algorithm~\ref{algo:block_rvgaw}: the choice of block length $B$ and the number of frequencies to be processed individually before blocking, $\tilde{n}$. We base our discussion on inspection of the periodogram.

Recall that the periodogram is an estimate of the spectral density and can be viewed as a measure of the relative contributions of the frequencies to the total variance of the process~\citep{percival1993spectral}. For example,  Figure~\ref{fig:cutoff_freqs} shows the periodograms for the linear Gaussian SSM in Section~\ref{sec:linear_gaussian} and the univariate SV model in Section~\ref{sec:sv_sim}. This figure shows that for the examples considered, the power generally decreases as frequency increases.  
In Figure ~\ref{fig:trajectory_lgss}, the trajectory of $\phi$ moves towards the true parameter value within the first few frequencies, with only minor adjustments afterwards. The trajectories for $\sigma_\eta$ and $\sigma_\epsilon$ behave in a similar manner, but with more gradual changes. This behaviour can also be observed in Figure~\ref{fig:trajectories_sv_sim_blocksize0}, in which the trajectories of both parameters in the univariate SV model exhibit large changes in the first few iterations, then stabilise afterwards. Higher frequencies that have little effect on the variational mean can thus be processed together in ``blocks". The R-VGA-Whittle updates can then be made based on the Whittle log-likelihood for a ``block" of frequencies, which we compute as the sum of the individual log-likelihood contribution evaluated at each frequency within that block.




\begin{figure}
     \centering
     \begin{subfigure}[b]{0.49\textwidth}
         \centering
         \includegraphics[width=\textwidth]{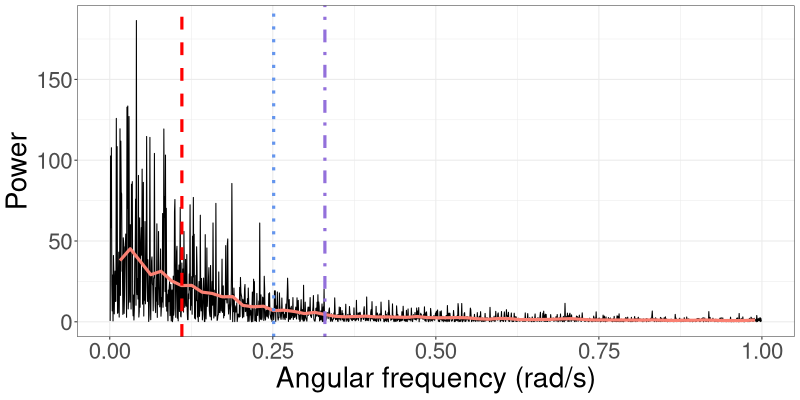}
         \caption{Linear Gaussian model.}
         \label{fig:cutoff_freqs_lgss}
     \end{subfigure}
     \hfill
     \begin{subfigure}[b]{0.49\textwidth}
         \centering
         \includegraphics[width=\textwidth]{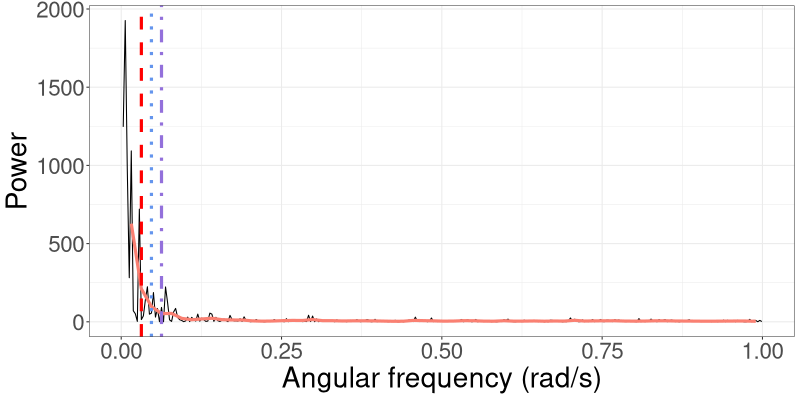}
         \caption{Univariate SV model.}
         \label{fig:cutoff_freqs_sv_sim}
     \end{subfigure}
     \caption{Plot of raw periodograms of the data (black) and smoothed periodograms (light red) for the univariate linear Gaussian state space model in Section~\ref {sec:linear_gaussian} and univariate SV model (with $T=2000$, $\phi = 0.99$, and $\sigma_\eta = 0.4$) in Section~\ref{sec:sv_sim}. Vertical lines show the frequency cutoffs at half-power (3dB, red dashed line), one-fifth-power (7dB, blue dotted line) and one-tenth-power (10dB, purple dash-dotted line).}
     \label{fig:cutoff_freqs}
\end{figure}

\begin{figure}
    \centering
    \includegraphics[width = \linewidth]{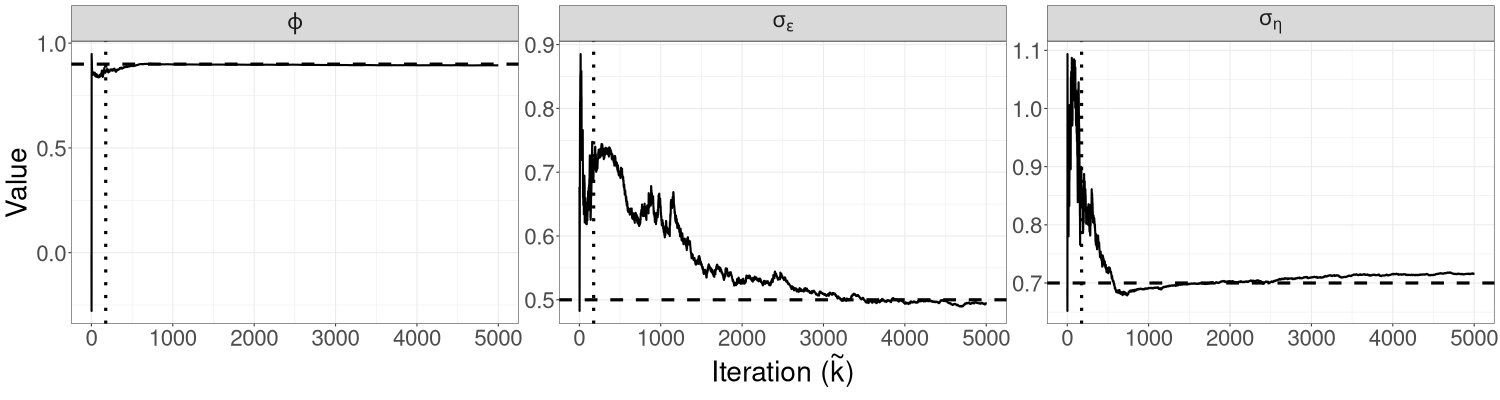}
    \caption{Trajectories of the variational mean for the parameters in the linear Gaussian SSM of Section~\ref{sec:linear_gaussian}, estimated using damped R-VGA-Whittle (Algorithm~\ref{algo:damped_rvgaw}) with no blocking. The true parameter values are marked with dashed lines.}
    \label{fig:trajectory_lgss}
\end{figure}

\begin{figure}
    \centering
    \includegraphics[width = 0.7\linewidth]{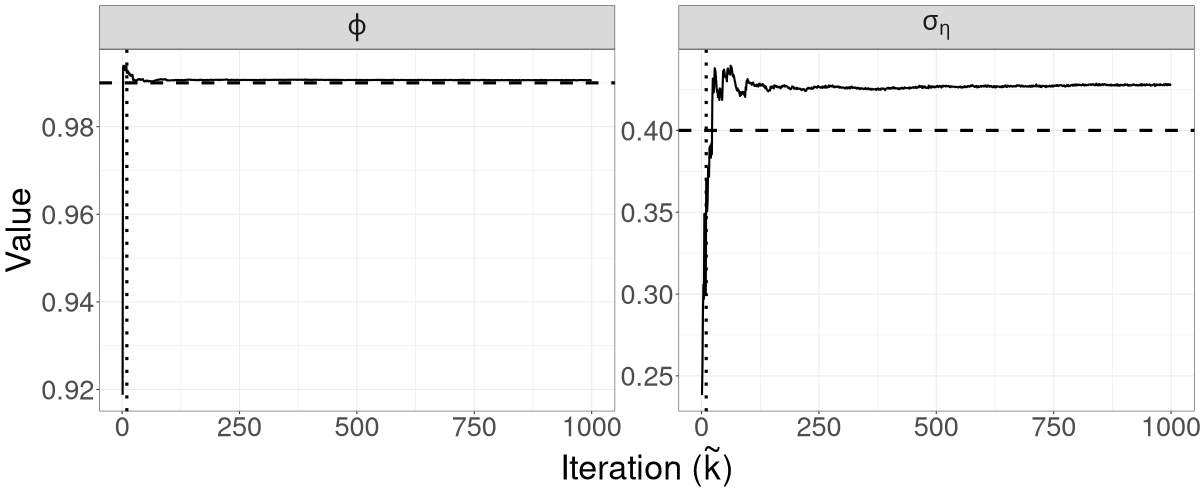}
    \caption{Trajectories of the variational mean for the parameters in the univariate SV model in Section~\ref{sec:sv_sim} (with $T=2000$, $\phi = 0.99$, and $\sigma_\eta = 0.4$), estimated using damped R-VGA-Whittle (Algorithm~\ref{algo:damped_rvgaw}) with no blocking. True parameter values are marked with dashed lines.}
    \label{fig:trajectories_sv_sim_blocksize0}
\end{figure}


It is important to choose an appropriate point to start blocking, as well as an appropriate length for each block of frequencies. To determine the frequency at which to begin blocking, 
we use the \textit{3dB frequency cutoff}, which is the frequency on the power spectrum at which the power drops to approximately half the maximum power of the spectrum. The value 3 comes from the fact that, on the decibel scale, a drop from a power level $P_1$ to another level $P_2 = \frac{1}{2} P_1$ results in a drop of 3 decibels. The change in decibels is calculated as
\begin{equation*}
    \Delta \textrm{dB} = 10 \log_{10} \frac{P_2}{P_1} = 10 \log_{10}\left( \frac{1}{2} \right) = -3.0103.
\end{equation*}
We employ a similar idea here, where we define the ``cutoff" frequency as the frequency at which the power drops to half of the maximum power in a smoothed version of the periodogram. Specifically, we seek the frequency $\omega^*$ such that $\tilde{I}(\omega^*) = 1/2 \max(\tilde{I}(\omega))$, where $\tilde{I}(\omega)$ is the periodogram smoothed using Welch's method~\citep{welch1967use}. Briefly, Welch's method works by dividing the frequencies into equal segments, computing the periodogram for each segment, and then averaging over these segments to obtain a smoother periodogram with lower variance. 
We also experiment with other cutoffs, such as the frequencies at which the power is one-fifth and one-tenth of the maximum power in the smoothed periodogram, which correspond to a drop of approximately 7dB and 10dB, respectively. We mark these cutoff frequencies in Figure~\ref{fig:cutoff_freqs}. 

Using each of the 3dB, 7dB and 10dB cutoffs, we run R-VGA-Whittle with no blocking, and then with varying block sizes ($B = 10, 50, 100, 300, 500$, and $1000$ frequencies), and examine the posterior densities of the parameters under these different settings. For all runs, we keep the settings for other tuning parameters the same as in Section~\ref{sec:applications} and use $S = 1000$ Monte Carlo samples, $n_{damp} = 5$, and $D = 100$ damping steps. For the linear Gaussian SSM, we plot the R-VGA-Whittle posteriors for these different cutoffs and block sizes in Figures~\ref{fig:compare_blocksizes_lgss_10000_temperfirst5_power2_arctanh_20230525}, \ref{fig:compare_blocksizes_lgss_10000_temperfirst5_power5_arctanh_20230525} and~\ref{fig:compare_blocksizes_lgss_10000_temperfirst5_power10_arctanh_20230525}. The corresponding plots for the univariate SV model are shown in Figures~\ref{fig:compare_blockvsnot_sv_sim_2000_temperfirst5_power2_arctanh_20240214}, \ref{fig:compare_blockvsnot_sv_sim_2000_temperfirst5_power5_arctanh_20240214}, and~\ref{fig:compare_blockvsnot_sv_sim_2000_temperfirst5_power10_arctanh_20240214}. The plots for the bivariate SV model are similar to those from the univariate SV model and not shown. 

We find that for the linear Gaussian SSM, block sizes of $100$ and below result in R-VGA-Whittle posterior densities that are very similar to each other, similar to the posterior densities obtained without blocking, and also similar to the posterior densities from HMC-Whittle and HMC-exact. For block sizes of $300$ and $500$, the posterior densities of the parameters $\sigma_\eta$ and $\sigma_\epsilon$ begin to differ slightly compared to the posterior densities obtained without blocking. When the block size is $1000$, the posterior densities for all three parameters become noticeably biased, although this bias reduces when the frequency cutoff point increases, as expected. However, for the SV model, the R-VGA-Whittle posterior densities are highly similar for all block sizes.
To achieve the best reduction in computing time while retaining estimation accuracy, we choose to begin blocking at the earliest cutoff point we consider in this experiment (3dB), with the block size set to 100.


\begin{figure}
     \centering
     \begin{subfigure}[b]{\linewidth}
         \centering
         \includegraphics[width = \linewidth]{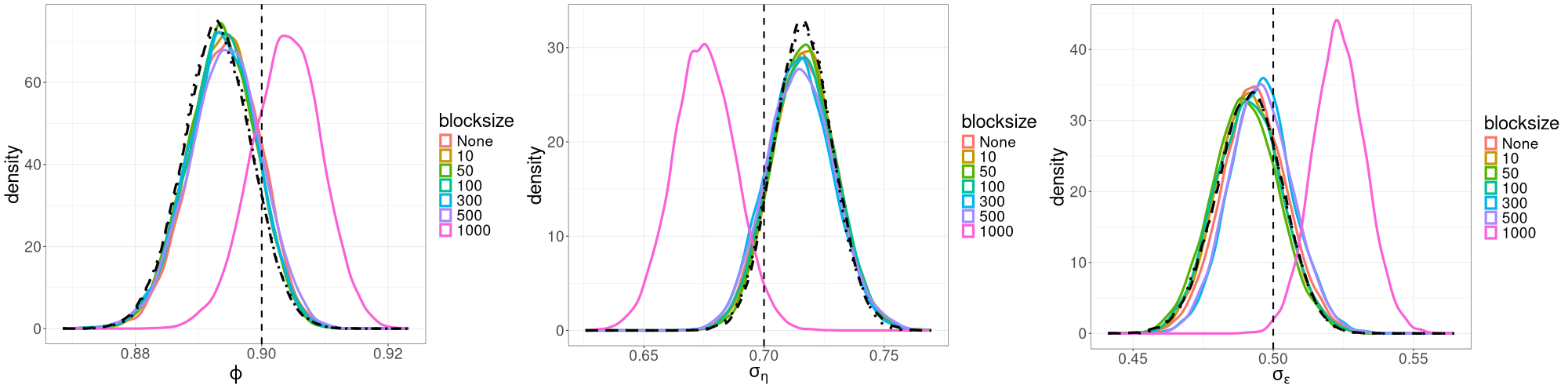}
        \caption{Blocking begins at the 3dB cutoff point.}
        \label{fig:compare_blocksizes_lgss_10000_temperfirst5_power2_arctanh_20230525} 
        \vspace{0.5cm}
     \end{subfigure}

     \begin{subfigure}[b]{\linewidth}
         \centering
         \includegraphics[width = \linewidth]{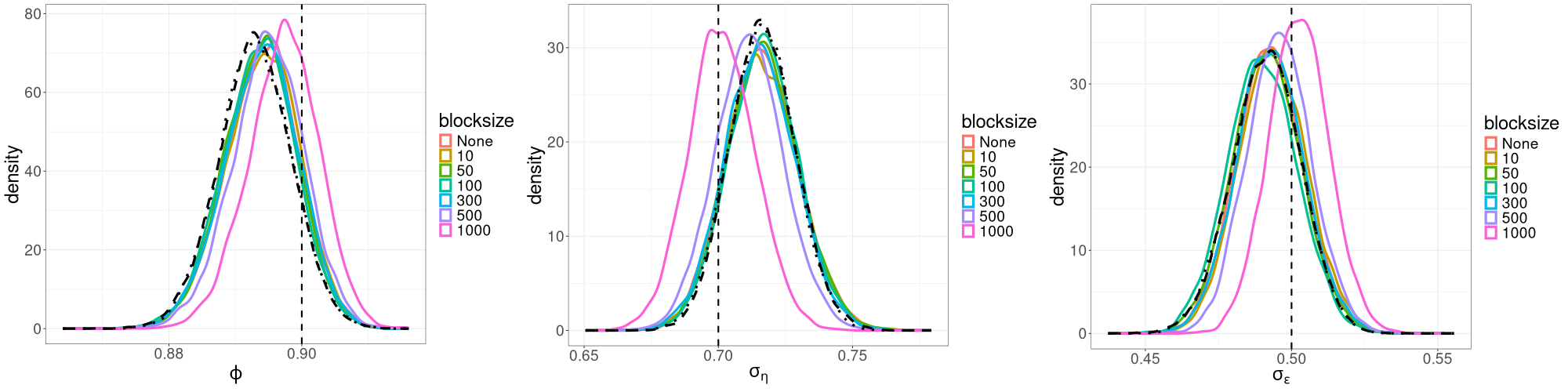}
        \caption{Blocking begins at the 7dB cutoff point.}
        \label{fig:compare_blocksizes_lgss_10000_temperfirst5_power5_arctanh_20230525} 
        \vspace{0.5cm}
     \end{subfigure}

     \begin{subfigure}[b]{\linewidth}
         \centering
         \includegraphics[width = \linewidth]{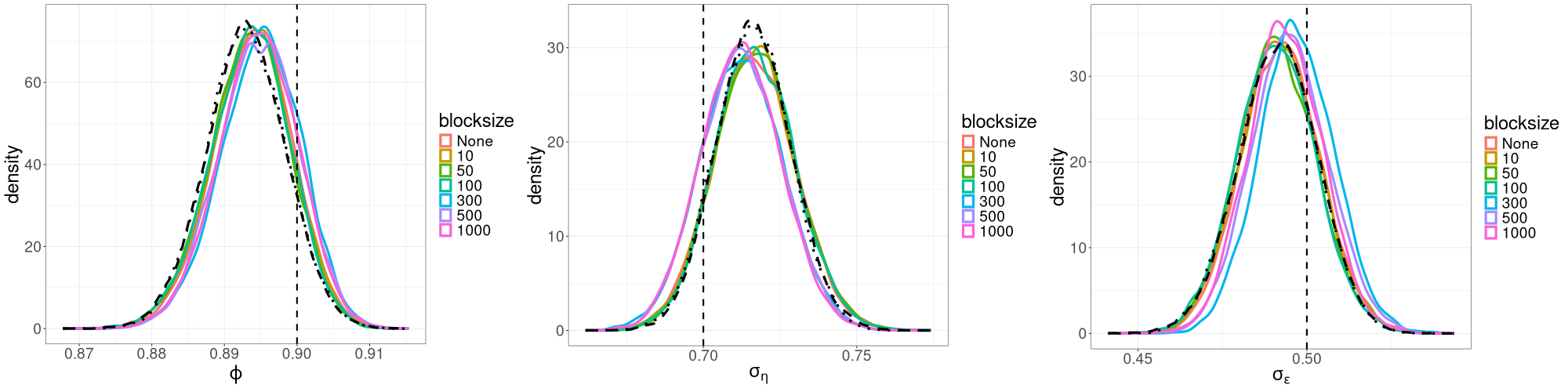}
        \caption{Blocking begins at the 10dB cutoff point.}
        \label{fig:compare_blocksizes_lgss_10000_temperfirst5_power10_arctanh_20230525} 
     \end{subfigure}
     
    \caption{Posterior densities of the parameters of the linear Gaussian SSM in Section \ref{sec:linear_gaussian} estimated using R-VGA-Whittle without blocking (denoted as ``None") and with blocking starting at different cutoff points for several block sizes (coloured lines), HMC-Whittle (dashed lines), and HMC-exact (dotted lines). True parameter values are denoted using vertical dashed lines.}
    \label{fig:compare_blocksizes_lgss}
\end{figure}

\begin{figure}
     \centering
     \begin{subfigure}[b]{\linewidth}
         \centering
         \includegraphics[width = 0.8\linewidth]{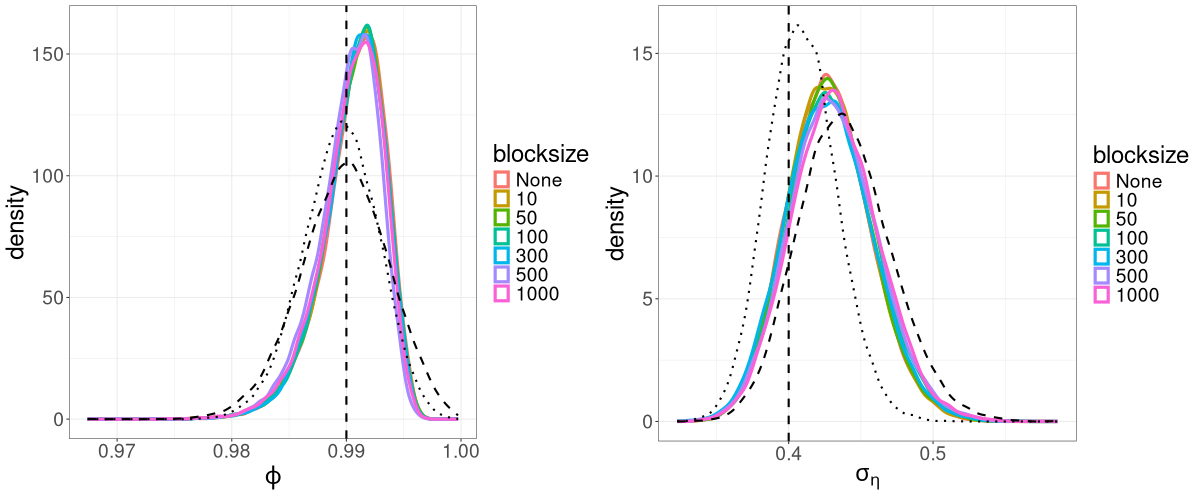}
        \caption{Blocking begins at the 3dB cutoff point.}
        \label{fig:compare_blockvsnot_sv_sim_2000_temperfirst5_power2_arctanh_20240214} 
        \vspace{0.5cm}
     \end{subfigure}

     \begin{subfigure}[b]{\linewidth}
         \centering
         \includegraphics[width = 0.8\linewidth]{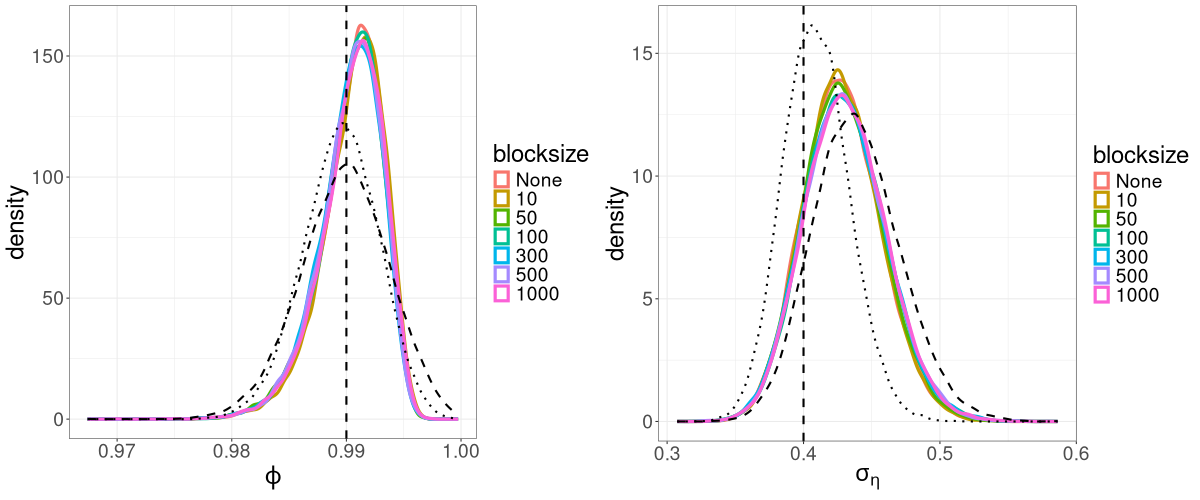}
        \caption{Blocking begins at the 7dB cutoff point.}
        \label{fig:compare_blockvsnot_sv_sim_2000_temperfirst5_power5_arctanh_20240214} 
        \vspace{0.5cm}
     \end{subfigure}

     \begin{subfigure}[b]{\linewidth}
         \centering
         \includegraphics[width = 0.8\linewidth]{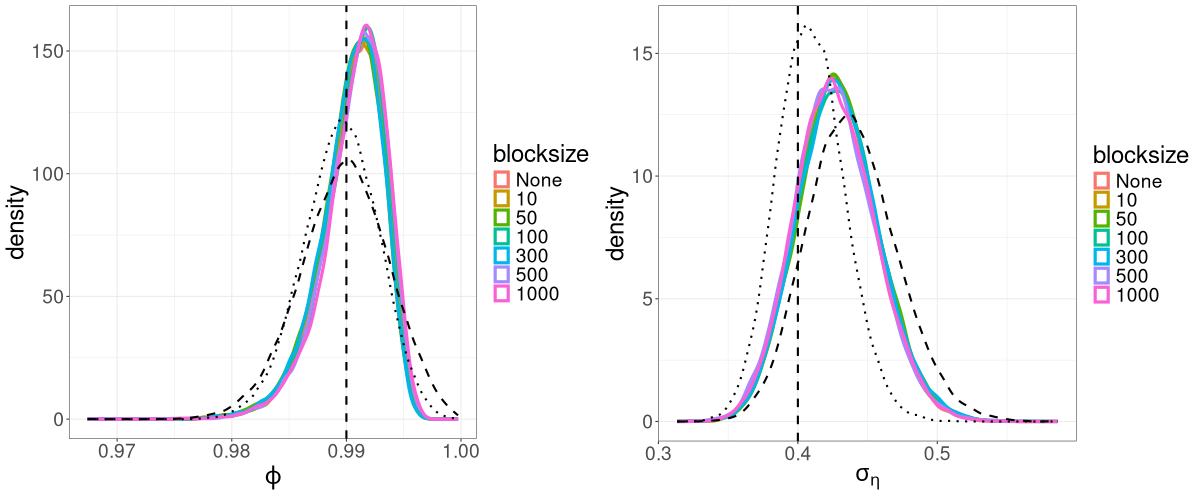}
        \caption{Blocking begins at the 10dB cutoff point.}
        \label{fig:compare_blockvsnot_sv_sim_2000_temperfirst5_power10_arctanh_20240214} 
     \end{subfigure}
     
    \caption{Posterior densities of the parameters of the univariate SV model with $T=2000$ in Section~\ref{sec:sv_sim} estimated using R-VGA-Whittle without blocking (denoted as ``None") and with blocking starting at different cutoff points for several block sizes (coloured lines), HMC-Whittle (dashed lines), and HMC-exact (dotted lines). True parameter values ($\phi = 0.99$ and $\sigma_\eta = 0.4$) are denoted using vertical dashed lines.}
    \label{fig:compare_blocksizes_sv_sim}
\end{figure}

    \clearpage
\section{Variance of the R-VGA-Whittle posterior densities for various Monte
Carlo sample sizes}
\label{sec:var_test}

In this section, we study the robustness of R-VGA-Whittle posterior distributions to different values of $S$, the Monte Carlo sample size used to approximate the expectations of the gradient and Hessian in Algorithm~\ref{algo:damped_rvgaw}. We consider sample sizes $S = 100, S = 500$ and $S = 1000$. We apply blocking as in Algorithm~\ref{algo:block_rvgaw}, and fix the block length to $B = 100$. We set $\tilde{n} = 176$ for the linear Gaussian SSM, and $\tilde{n} = 10$ for the univariate SV model, which corresponds to the 3dB cutoff point as discussed in Section~\ref{sec:blocksize_test}.

Figures~\ref{fig:var_test_lgss_S100_power2_blocksize100_176indiv_20230525}--\ref{fig:var_test_lgss_S1000_power2_blocksize100_176indiv_20230525} show the posterior densities of the parameters of the linear Gaussian SSM from 10 runs of R-VGA-Whittle with $S = 100$, $S = 500$, and $S = 1000$, respectively, while Figures~\ref{fig:var_test_sv_S100_power2_blocksize100_10indiv_20240214}--\ref{fig:var_test_sv_S1000_power2_blocksize100_10indiv_20240214} show the corresponding plots for the univariate SV model with $\phi = 0.99$ and $\sigma_\eta = 0.4$. As the Monte Carlo sample size increases, the variance in the posterior densities across different runs decreases. When $S = 1000$, the variance across different runs is significantly reduced, so we choose this as our Monte Carlo sample size for all examples in Section~\ref{sec:applications} of the main paper. 

\begin{figure}
     \centering
     \begin{subfigure}[b]{\linewidth}
         \centering
         \includegraphics[width = \linewidth]{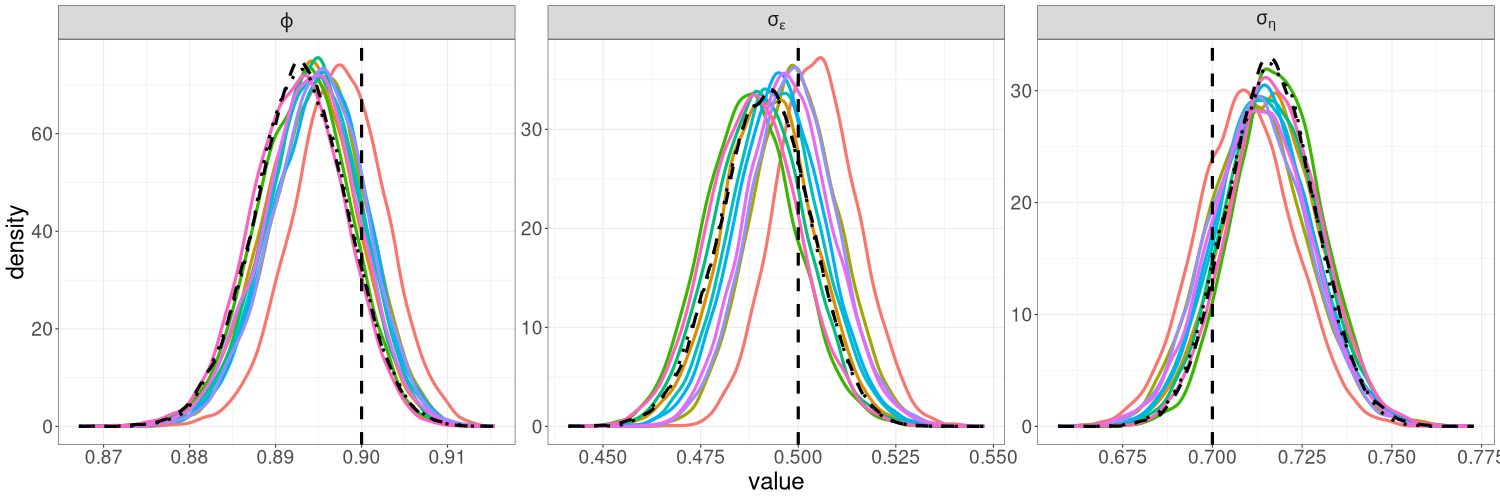}
        \caption{$S = 100$.}
        \label{fig:var_test_lgss_S100_power2_blocksize100_176indiv_20230525} 
        \vspace{0.5cm}
     \end{subfigure}

     \begin{subfigure}[b]{\linewidth}
         \centering
         \includegraphics[width = \linewidth]{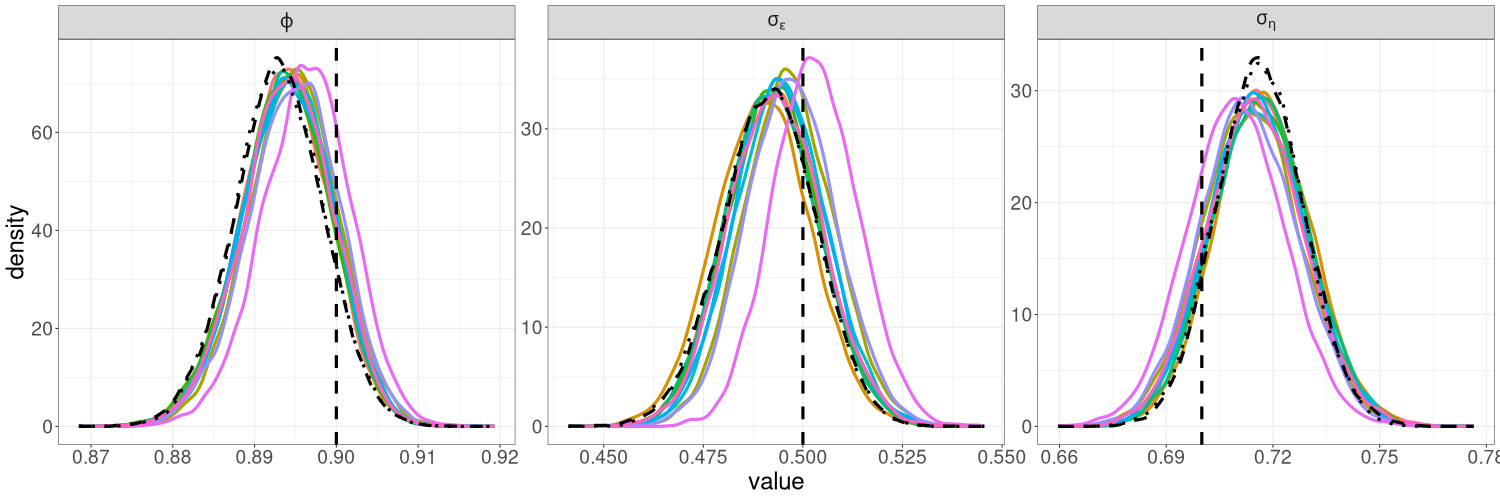}
        \caption{$S = 500$.}
        \label{fig:var_test_lgss_S500_power2_blocksize100_176indiv_20230525} 
        \vspace{0.5cm}
     \end{subfigure}

     \begin{subfigure}[b]{\linewidth}
         \centering
         \includegraphics[width = \linewidth]{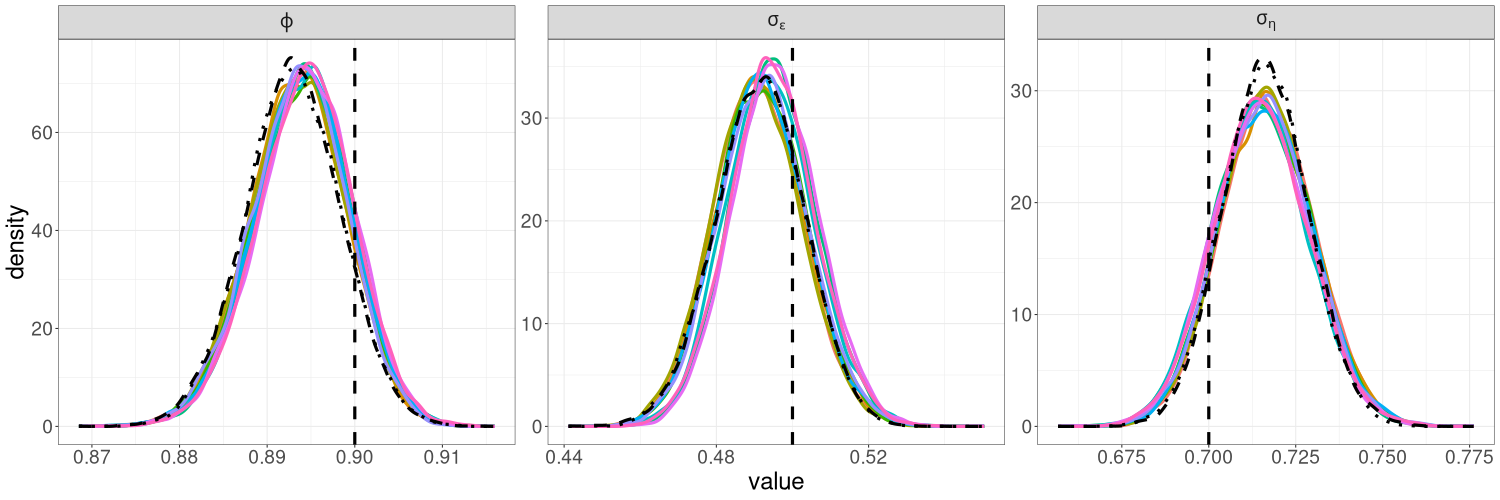}
        \caption{$S = 1000$.}
        \label{fig:var_test_lgss_S1000_power2_blocksize100_176indiv_20230525} 
        
     \end{subfigure}
    \caption{Posterior densities of the parameters of the linear Gaussian SSM in Section~\ref{sec:linear_gaussian} from 10 runs of R-VGA-Whittle (coloured solid lines) with different Monte Carlo sample sizes $S$, from a single run of HMC-Whittle (dashed lines), and from a single run of HMC-exact (dotted lines). True parameter values are marked with dashed lines.}
    \label{fig:var_test_lgss}
    \end{figure}

\begin{figure}
     \centering
     \begin{subfigure}[b]{\linewidth}
         \centering
         \includegraphics[width = 0.7\linewidth]{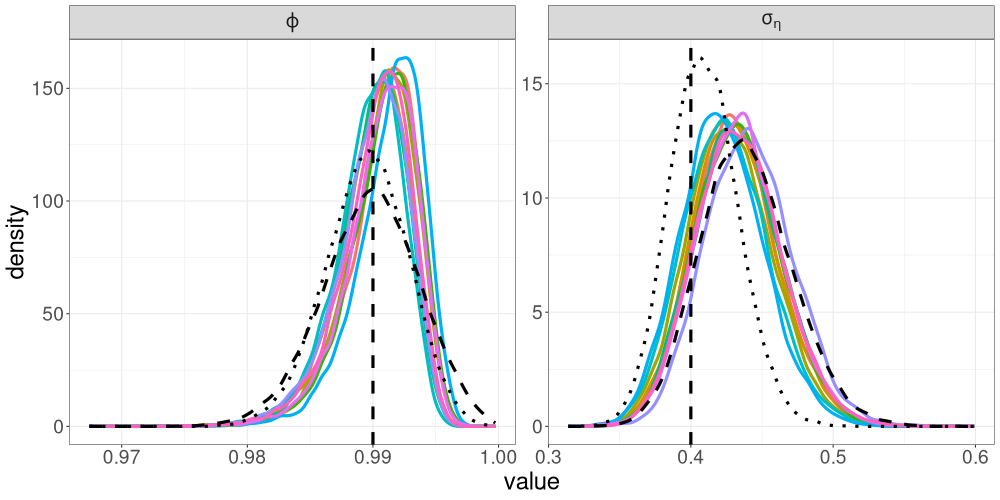}
        \caption{$S = 100$.}
        \label{fig:var_test_sv_S100_power2_blocksize100_10indiv_20240214} 
        \vspace{0.5cm}
     \end{subfigure}

     \begin{subfigure}[b]{0.7\linewidth}
         \centering
         \includegraphics[width = \linewidth]{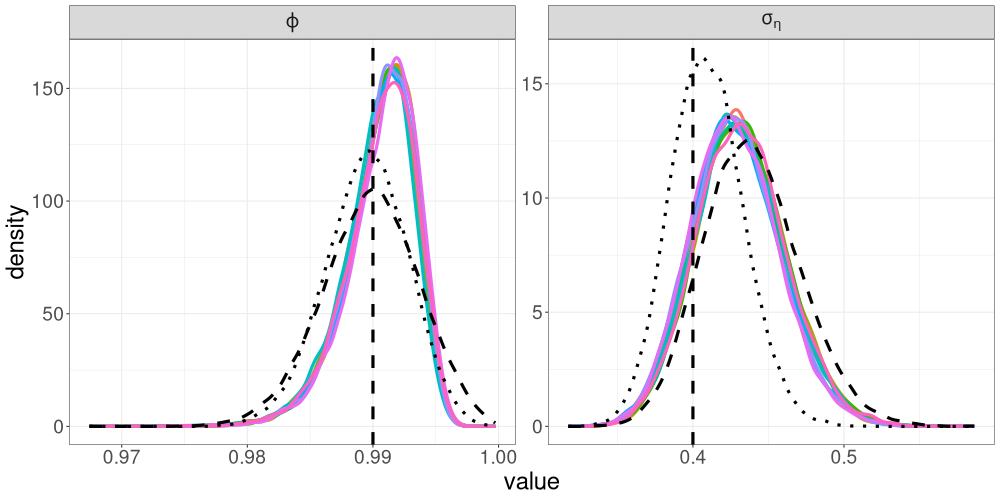
         }
        \caption{$S = 500$.}
        \label{fig:var_test_sv_S500_power2_blocksize100_10indiv_20240214} 
        \vspace{0.5cm}
     \end{subfigure}

     \begin{subfigure}[b]{0.7\linewidth}
         \centering
         \includegraphics[width = \linewidth]{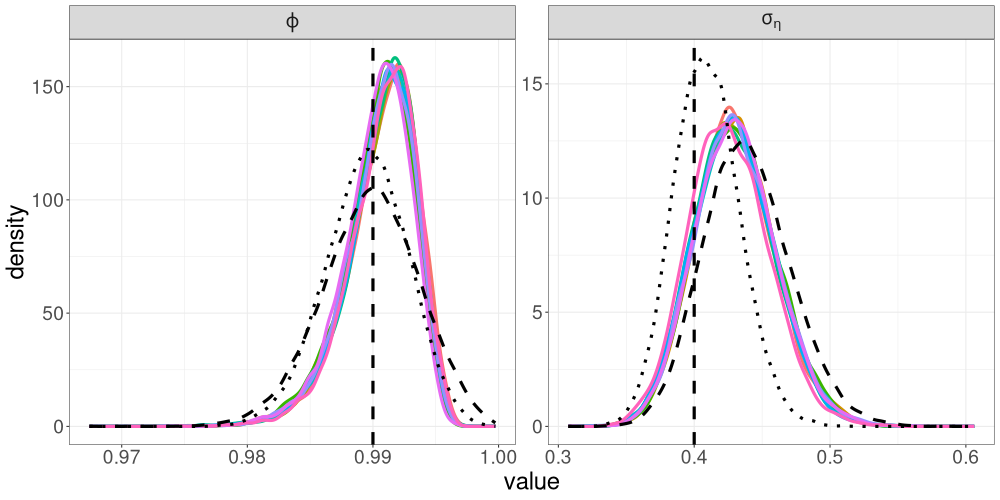}
        \caption{$S = 1000$.}
        \label{fig:var_test_sv_S1000_power2_blocksize100_10indiv_20240214} 
        
     \end{subfigure}
    \caption{Posterior densities of the parameters of the univariate SV model with $T=2000$ in Section \ref{sec:sv_sim} from 10 runs of R-VGA-Whittle (coloured solid lines) with different Monte Carlo sample sizes $S$, from a single run of HMC-Whittle (dashed lines), and from a single run of HMC-exact (dotted lines). True parameter values ($\phi = 0.99$, $\sigma_\eta = 0.4$) are marked with dashed lines.}
    \label{fig:var_test_sv_sim}
    \end{figure}

    \clearpage
\section{Trace plots \label{sec:traceplots}}
This section contains trace plots, effective samples sizes (ESS), and $\hat{R}$ values of HMC-exact and HMC-Whittle for all the models we consider (except the ARFIMA model with Student's $t$ measurement noise, for which we only run HMC-Whittle). Diagnostics for MCMC-stochvol for the univariate SV example in Section~\ref{sec:sv_sim} of the main paper are also included. The ESS is a measure of the number of uncorrelated posterior samples, and the higher the ESS the better;
see~\cite{gelman2003bayesian} for further details on how Stan calculates ESS. The $\hat{R}$ statistic proposed by~\cite{gelman1992inference} measures the ratio of the average variance of samples within each chain to the variance of the pooled samples across chains; an $\hat{R}$ close to 1 indicates that the chains have converged. 

Figure~\ref{fig:lgss_traceplots} shows trace plots for the linear Gaussian SSM in Section \ref{sec:linear_gaussian}; Figures~\ref{fig:sv_sim_traceplots} and~\ref{fig:multi_sv_sim_traceplots} show trace plots for the univariate and bivariate SV models with simulated data in Sections \ref{sec:sv_sim} and \ref{sec:bivariateSVmodel}; Figure~\ref{fig:hmcw_arfima_ss_trace} shows the trace plots from the HMC-Whittle sampler for the state space model with Student's $t$ errors and latent states following an ARFIMA process in Section \ref{sec:arfima_ss}; and Figures~\ref{fig:sv_real_traceplots} and~\ref{fig:multi_sv_real_traceplots} show trace plots for the univariate and bivariate SV models with real data in Section \ref{sec:real_data}. Table~\ref{tab:hmc_convergence} contains the ESS and $\hat{R}$ corresponding to each parameter in our examples. Table~\ref{tab:hmc_iters} shows the total number of posterior samples after burn-in, for HMC-exact, HMC-Whittle, and MCMC-stochvol.

For all examples, the \(\hat{R}\) values for all parameters are very close to 1, indicating good convergence of the MCMC chains. HMC-Whittle consistently produces posterior samples with effective sample sizes (ESS) exceeding 1000 across all examples. In contrast, HMC-exact yields ESS values below 1000 for some of the parameters of the univariate SV model with real data and for the bivariate SV models with both simulated and real data, even when we run with a maximum of $30000$ iterations (two chains). Similarly, MCMC-stochvol produces ESS values below 1000 for some parameters in the univariate SV model, for both simulated and real datasets. HMC-Whittle is more computationally efficient than both HMC-exact and MCMC-stochvol because it targets only the posterior distribution of the parameters, whereas HMC-exact targets the full joint posterior of both parameters and latent states, which is high-dimensional.

\begin{table}
\centering
\caption{The effective sample sizes (ESS) and \(\hat{R}\) statistics from HMC-exact and HMC-Whittle are reported for all parameters in the examples described in the main text. For the univariate SV examples, results from MCMC-stochvol are also included. Examples for which a method was not run (and thus ESS and \(\hat{R}\) are not available) are indicated with a dash (--).}
\begin{tabular}{|l|l|rr|rr|rr|}
\hline
\multicolumn{1}{|c|}{\multirow{2}{*}{Example}} & \multicolumn{1}{c|}{\multirow{2}{*}{Parameter}} & \multicolumn{2}{c|}{HMC-Whittle}                     & \multicolumn{2}{c|}{HMC-exact}             & \multicolumn{2}{c|}{MCMC-stochvol}                \\ \cline{3-8} 
\multicolumn{1}{|c|}{}                         & \multicolumn{1}{c|}{}                           & \multicolumn{1}{c|}{$\hat{R}$} & \multicolumn{1}{c|}{ESS} & \multicolumn{1}{c|}{$\hat{R}$} & \multicolumn{1}{c|}{ESS} & \multicolumn{1}{c|}{$\hat{R}$} & \multicolumn{1}{c|}{ESS} \\ \hline
Linear Gaussian                                & $\phi$                                             & 1.0000                      & 2172                   & 1.0001                  & 4160       & - & -           \\
                                               & $\sigma_\eta$                                      & 0.9999                  & 1655                   & 1.0000                  & 1801                    & - & - \\
                                               & $\sigma_\epsilon$                                  & 0.9998                  & 1659                  & 1.0004                 & 1350                   & - & - \\
\hline
Univariate SV (sim. data with               & $\phi$                                             & 1.0006                     & 1709                & 1.0000                     & 8137   & 1.0029    & 1413              \\
$\phi=0.99$ and $\sigma_\eta=0.4 $)         & $\sigma_\eta$                                      & 1.0002                    & 1710                & 1.0004                    & 1211 & 1.0001 & 945               \\
\hline                                               
Bivariate SV (sim. data)                       & $\Phi_{11}$                                         & 1.0006                         & 2490                      & 1.0009                         & 1528                       & - & - \\
                                               & $\Phi_{22}$                                         & 1.0007                     & 2267                  & 1.0001                     & 467                    & - & - \\
                                               & $\Sigma_{\eta_{11}}$                                  & 1.0013                     & 2871                  & 1.0019                     & 568                    & - & - \\
                                               & $\Sigma_{\eta_{21}}$                                  & 1.0007                     & 2886                 & 1.0005                     & 599                    & - & - \\
                                               & $\Sigma_{\eta_{22}}$                                  & 1.0002                    & 2048                 & 1.0001                    & 274                  & - & - \\
\hline
ARFIMA (sim. data)                       & $\phi$                                         & 1.0006                      & 6234                    & -                    & -                  & - & -  \\
& $\theta$                                         & 1.0014                      & 2299                    & -                    & -                  & - & - \\
& $d$                                  & 1.0008                      & 6053                   & -                   & -                  & - & - \\
& $\sigma_\eta$                                  & 1.0014                      & 2083                    & -                    & -                  & - & - \\
& $\nu$                                  & 1.0016                    & 1637                 & -                    & -            & - & - \\
\hline
Univariate SV (real data)                      & $\phi$                                             & 1.0049                      & 1166                     & 1.0023                     & 482                     & 1.0003 & 573 \\
                                               & $\sigma_\eta$                                      & 1.0019                    & 1150                   & 1.0059                    & 184                  & 1.0007 & 296 \\
\hline
Bivariate SV (real data)                       & $\Phi_{11}$                                         & 0.9999                        & 1665                       & 1.0007                      & 1018                         & - & - \\
                                               & $\Phi_{22}$                                         & 0.9999                     & 1558                   & 1.0004                      & 1233                    & - & - \\
                                               & $\Sigma_{\eta_{11}}$                                  & 1.0001                     & 1882                  & 1.0037                      & 277                    & - & - \\
                                               & $\Sigma_{\eta_{21}}$                                  & 1.0008                     & 1708                   & 1.0050    & 218                    & - & - \\
                                               & $\Sigma_{\eta_{22}}$                                  & 1.0013                    & 1622                 & 1.0024                    & 295             & - & - \\
\hline
\end{tabular}
    \label{tab:hmc_convergence}
\end{table}
    
\begin{table}
\centering
\caption{Total number of posterior samples used after burn-in (combined across two chains). For all examples and all methods, a burn-in period of 1000 samples is used.}
\begin{tabular}{|l|r|r|r|}
\hline
\multicolumn{1}{|c|}{Example} & \multicolumn{1}{c|}{HMC-Whittle} & \multicolumn{1}{c|}{HMC-exact} & \multicolumn{1}{c|}{MCMC-stochvol} \\ \hline
Linear Gaussian               & 4000                             & 20000                          & -                                  \\
Univariate SV (sim. data with $\phi=0.99$ and $\sigma_\eta=0.4 $)     & 4000                             & 28000                          & 28000                              \\
Bivariate SV (sim. data)      & 4000                             & 28000                            & -                                  \\
ARFIMA (sim. data)            & 20000                             & -                              & -                                  \\
Univariate SV (real data)     & 4000                             & 28000                            & 28000                                \\
Bivariate SV (real data)      & 4000                             & 28000                            & -                                  \\
\hline
\end{tabular}
\label{tab:hmc_iters}
\end{table}

\begin{figure}
     \centering
     \begin{subfigure}[b]{\linewidth}
         \centering
         \includegraphics[width=\textwidth]{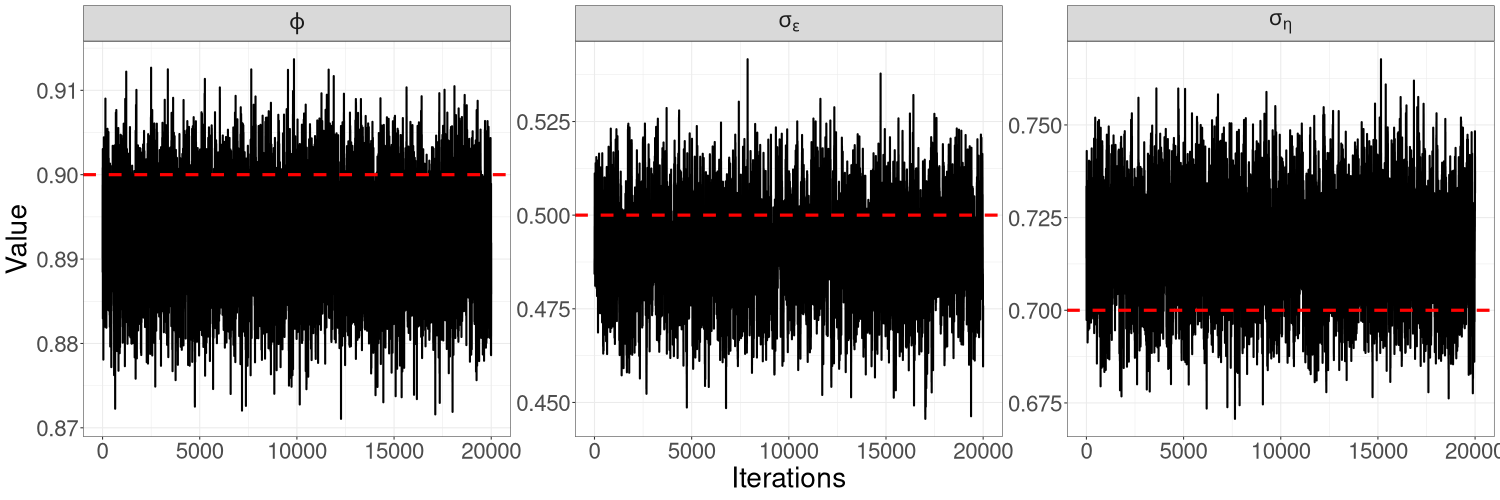}
         \caption{HMC-exact trace plots (samples not thinned).}
         \label{fig:lgss_hmc_traceplot}
         \vspace{0.5cm}
     \end{subfigure}
     
     
     \begin{subfigure}[b]{\linewidth}
         \centering
         \includegraphics[width=\textwidth]{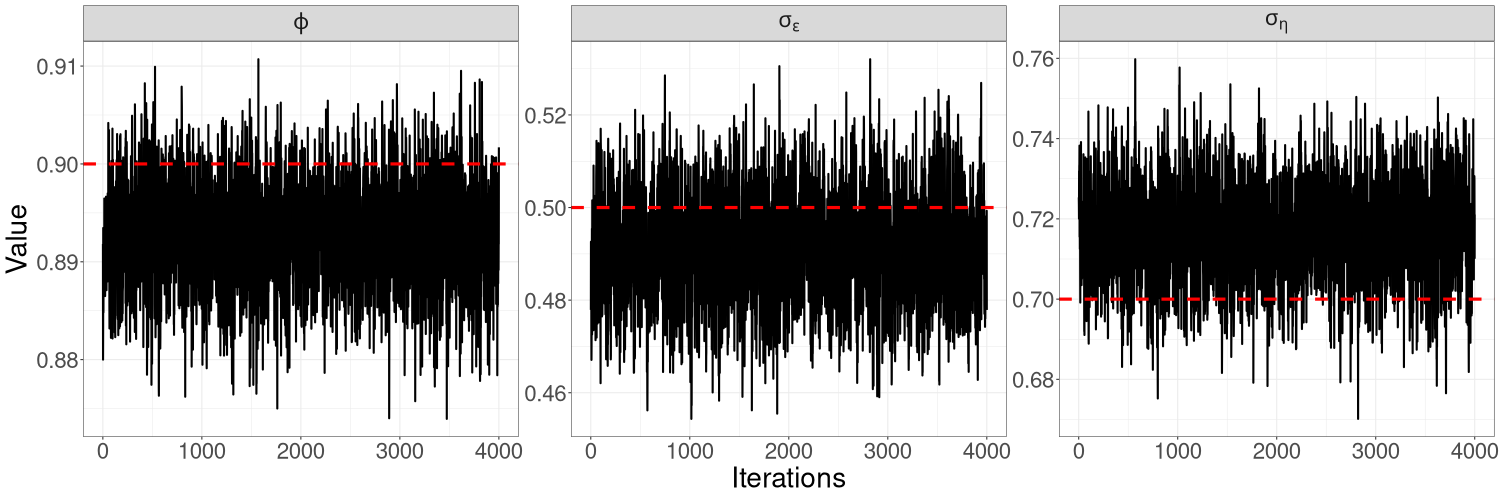}
         \caption{HMC-Whittle trace plots (samples not thinned).}
         \label{fig:lgss_hmcw_traceplot}
     \end{subfigure}
     \caption{Trace plots for the linear Gaussian SSM with simulated data in Section~\ref{sec:linear_gaussian} of the main paper. Red dashed lines denote the true parameter values.}
     \label{fig:lgss_traceplots}
\end{figure}


\begin{figure}
     \centering
     \begin{subfigure}[b]{\linewidth}
         \centering
         \includegraphics[width = \linewidth]{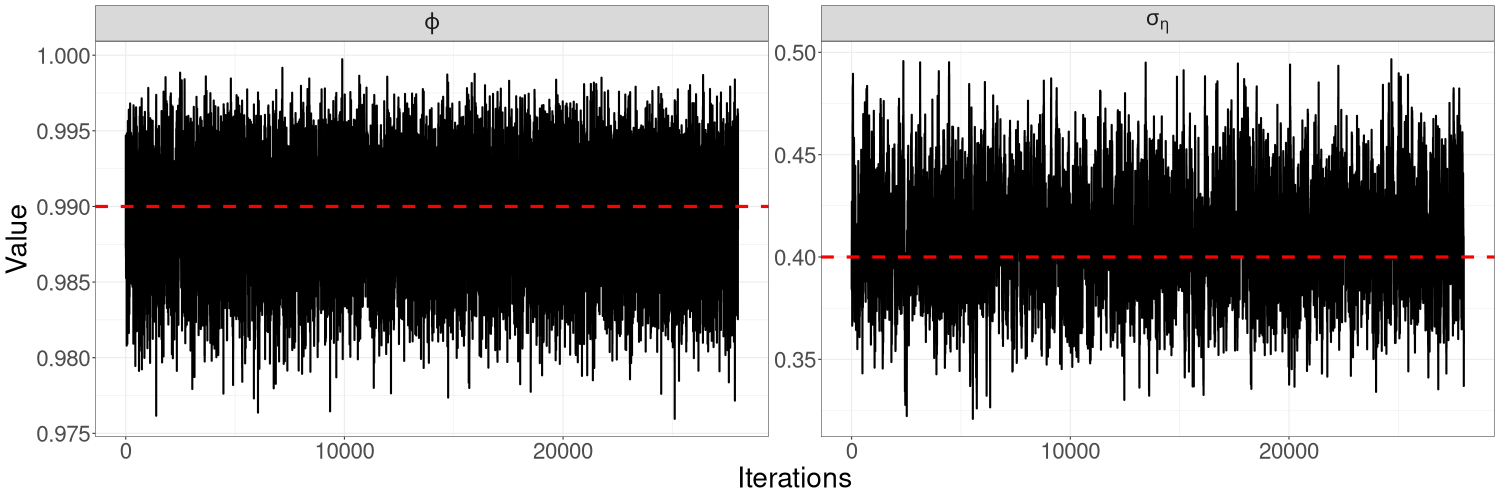}
        \caption{HMC-exact trace plots (samples not thinned).}
        \label{fig:sv_sim_hmc_traceplot} 
        \vspace{0.5cm}
     \end{subfigure}


     \begin{subfigure}[b]{\linewidth}
         \centering
         \includegraphics[width = \linewidth]{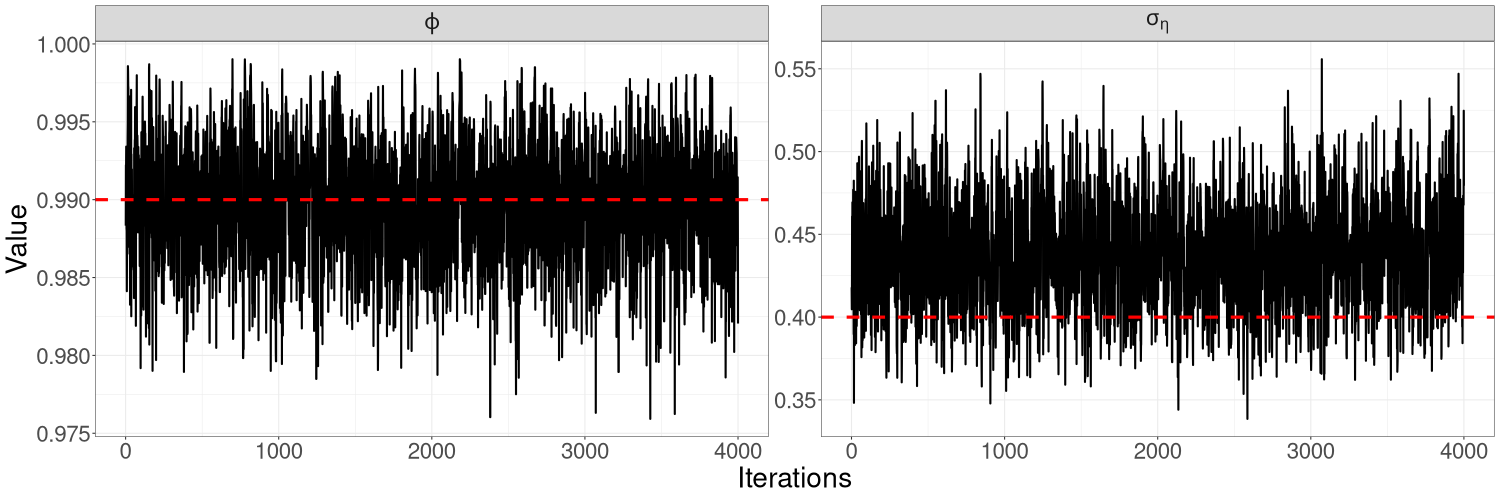}
        \caption{HMC-Whittle trace plots (samples not thinned).}
        \label{fig:sv_sim_hmcw_traceplot} 
     \end{subfigure}

      \begin{subfigure}[b]{\linewidth}
         \centering
         \includegraphics[width = \linewidth]{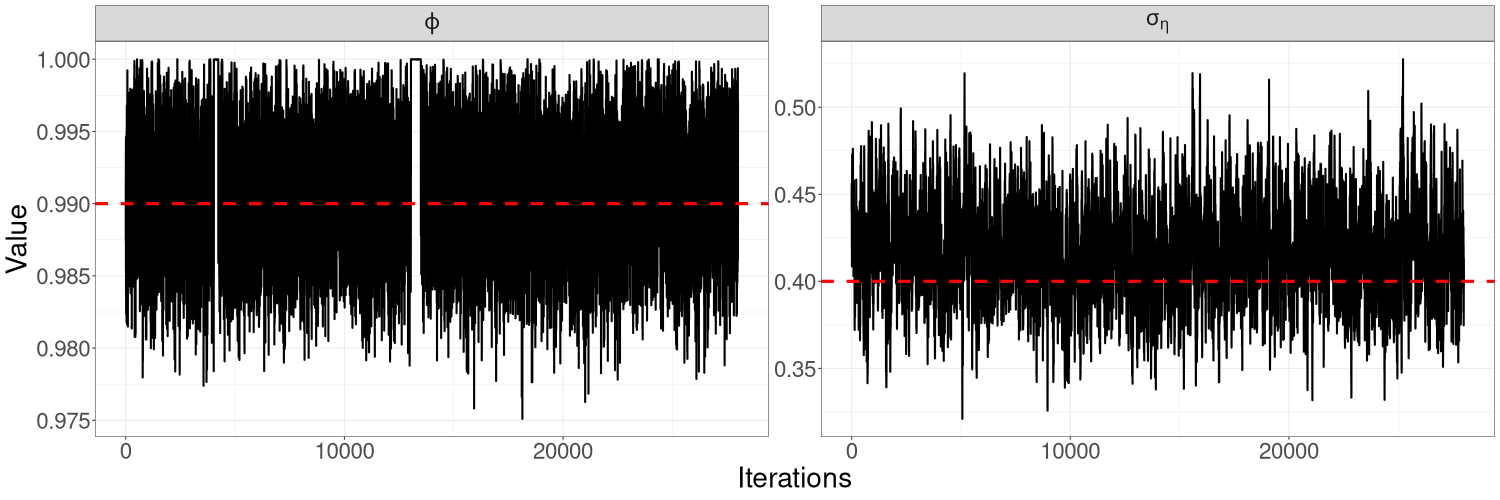}
        \caption{MCMC-stochvol trace plots (samples not thinned).}
        \label{fig:sv_sim_stv_traceplot} 
     \end{subfigure}
     \caption{Trace plots for the univariate SV model with simulated data in Section~\ref{sec:sv_sim} of the main paper. Red dashed lines denote the true parameter values ($\phi = 0.99$ and $\sigma_\eta = 0.4$).}
     \label{fig:sv_sim_traceplots}
 \end{figure}

\begin{figure}
     \centering
     \begin{subfigure}[b]{\linewidth}
         \centering
         \includegraphics[width = \linewidth]{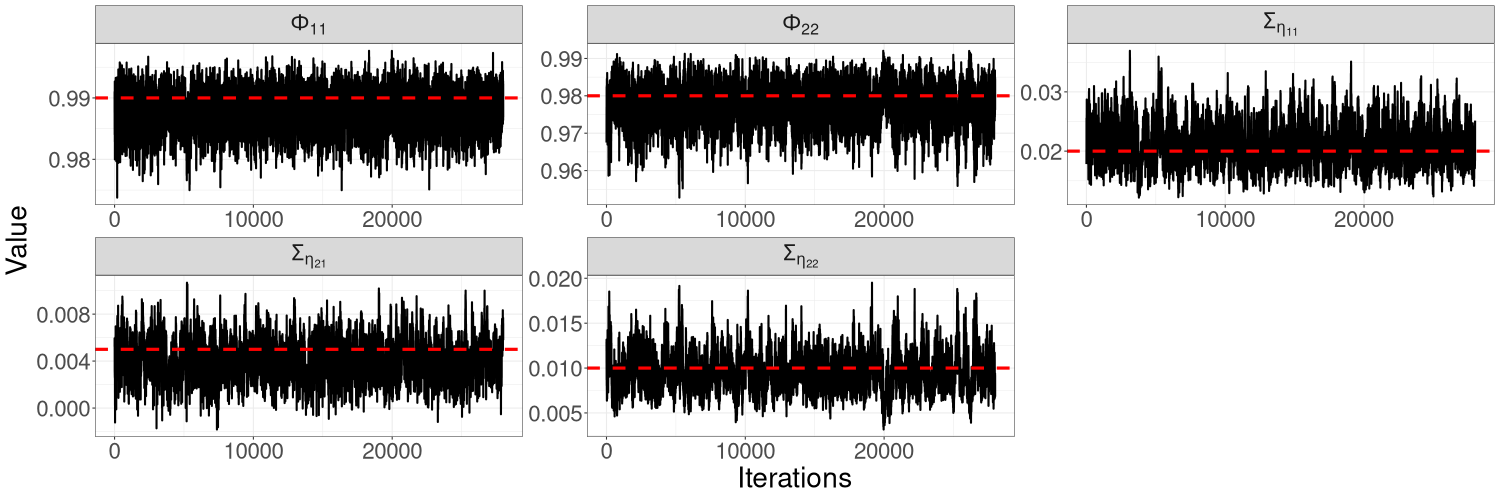}
        \caption{HMC-exact trace plots (samples not thinned).}
        \label{fig:multi_sv_hmc_traceplot} 
        \vspace{0.5cm}
     \end{subfigure}

     \begin{subfigure}[b]{\linewidth}
         \centering
         \includegraphics[width = \linewidth]{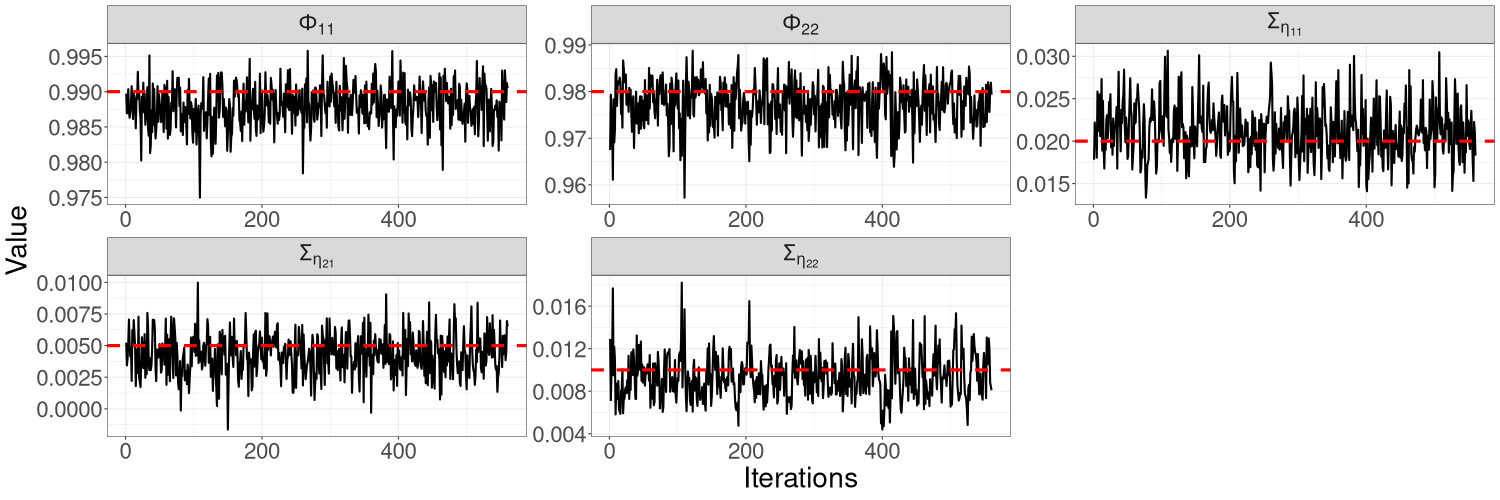}
        \caption{HMC-exact trace plots (samples thinned by a factor of 50).}
        \label{fig:multi_sv_hmc_traceplot_thin} 
        \vspace{0.5cm}
     \end{subfigure}

     \begin{subfigure}[b]{\linewidth}
         \centering
         \includegraphics[width = \linewidth]{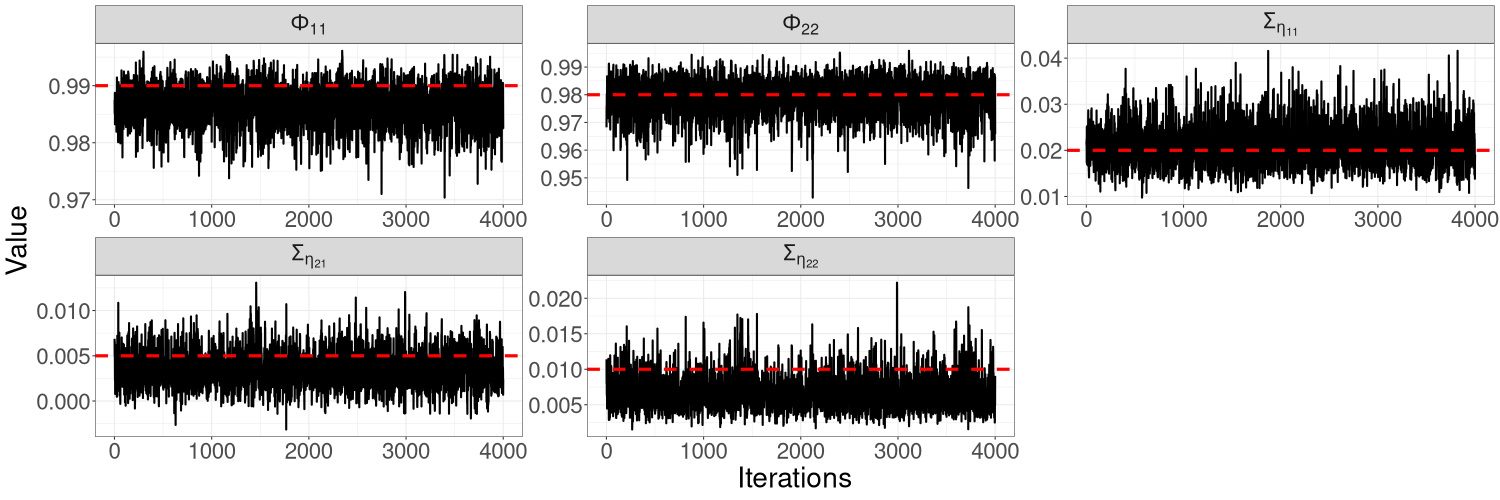}
        \caption{HMC-Whittle trace plots (samples not thinned).}
        \label{fig:multi_sv_hmcw_traceplot} 
     \end{subfigure}
     \caption{Trace plots for the bivariate SV model with simulated data in Section~\ref{sec:bivariateSVmodel} of the main paper. Red dashed lines denote the true parameter values.}
     \label{fig:multi_sv_sim_traceplots}
 \end{figure}

\begin{figure}
    \centering
     \begin{subfigure}[b]{\linewidth}
         \centering
         \includegraphics[width = \linewidth]{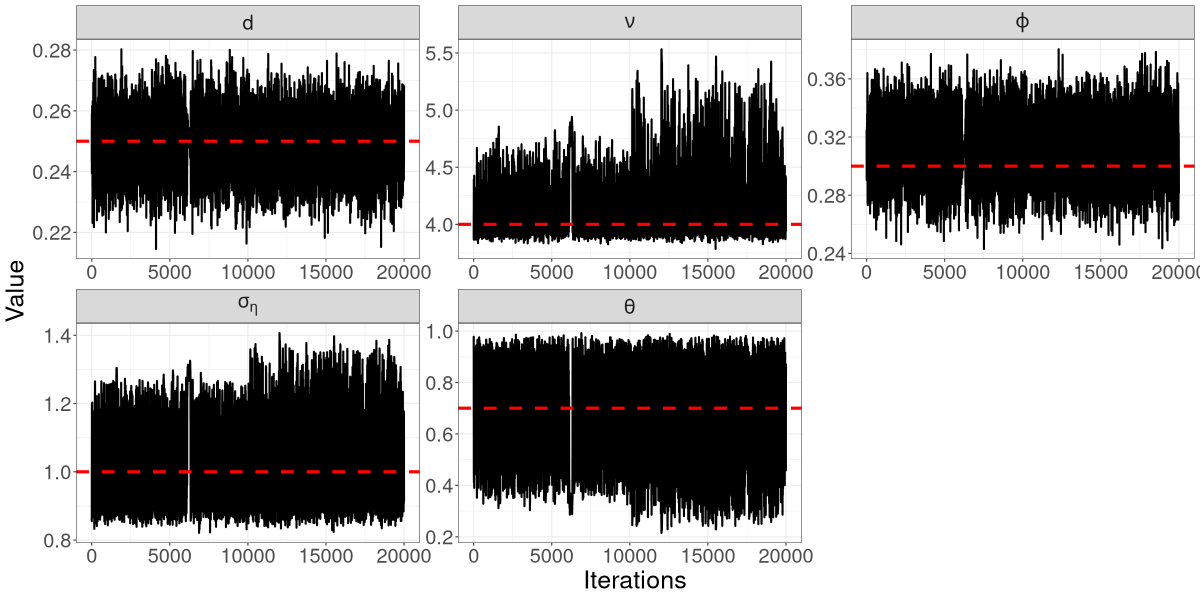}
        \caption{HMC-Whittle trace plots (samples not thinned).}
        \label{fig:hmcw_arfima_ss_trace_n50000} 
        \vspace{0.5cm}
     \end{subfigure}

     \begin{subfigure}[b]{\linewidth}
         \centering
         \includegraphics[width = \linewidth]{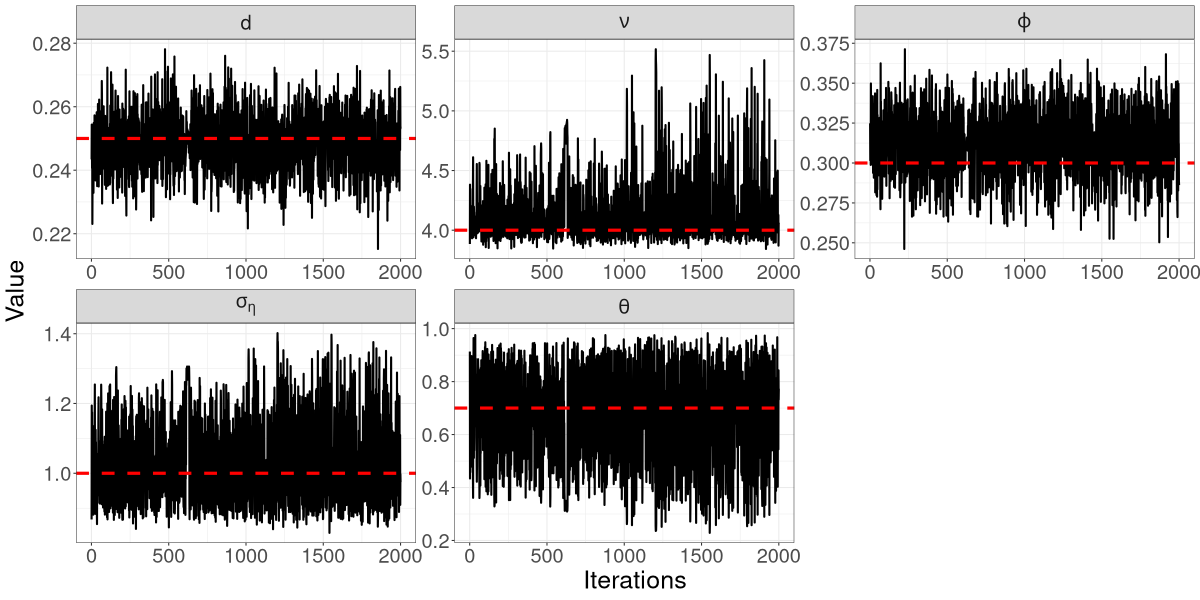}
        \caption{HMC-Whittle trace plots (samples thinned by a factor of 10).}
        \label{fig:hmcw_arfima_ss_trace_n50000_thinned} 
        \vspace{0.5cm}
     \end{subfigure}
     \caption{Trace plots for the state space model with Student's $t$ errors and latent states following an ARFIMA process in Section~\ref{sec:arfima_ss} of the main paper. Red dashed lines denote the true parameter values.}
     \label{fig:hmcw_arfima_ss_trace}
\end{figure}
 



\begin{figure}
     \centering
     \begin{subfigure}[b]{0.8\linewidth}
         \centering
         \includegraphics[width = \linewidth]{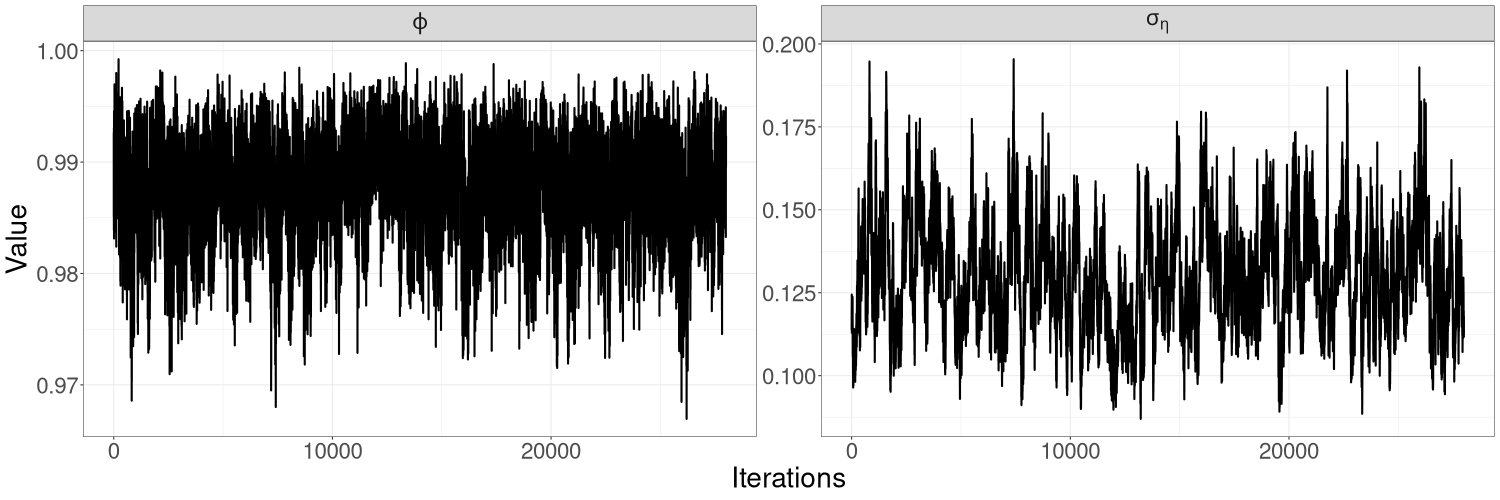}
        \caption{HMC-exact trace plots (samples not thinned).}
        \label{fig:sv_real_hmc_traceplot} 
        \vspace{0.5cm}
     \end{subfigure}

     \begin{subfigure}[b]{0.8\linewidth}
         \centering
         \includegraphics[width = \linewidth]{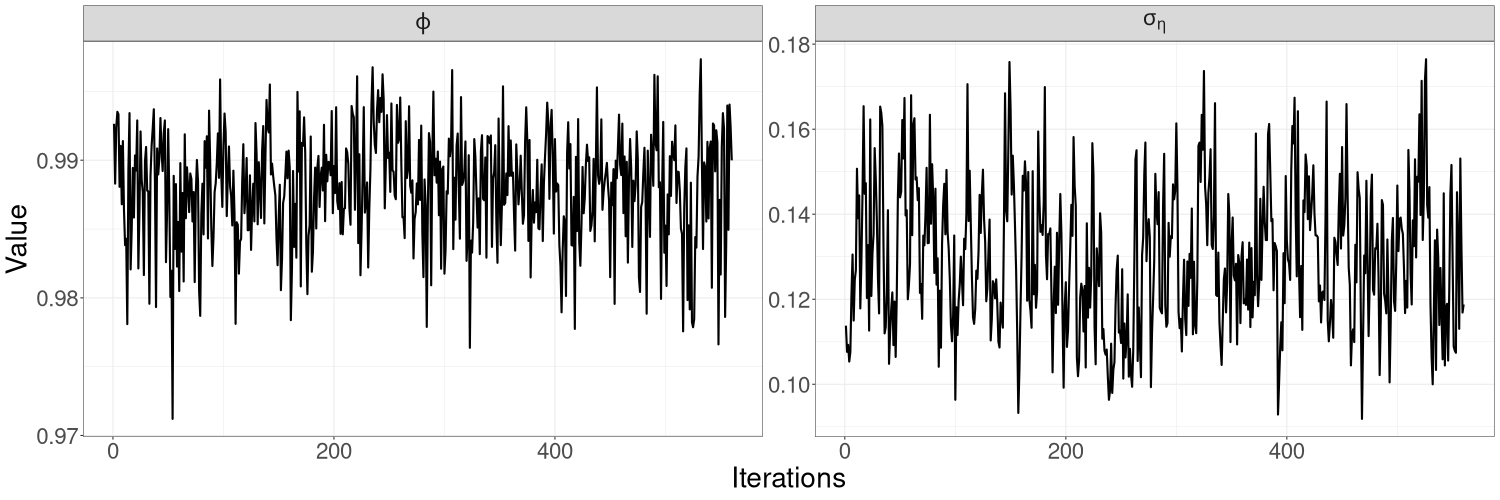}
        \caption{HMC-exact trace plots (samples thinned by a factor of 50).}
        \label{fig:sv_real_hmc_traceplot_thin} 
        \vspace{0.5cm}
     \end{subfigure}

     \begin{subfigure}[b]{0.8\linewidth}
         \centering
         \includegraphics[width = \linewidth]{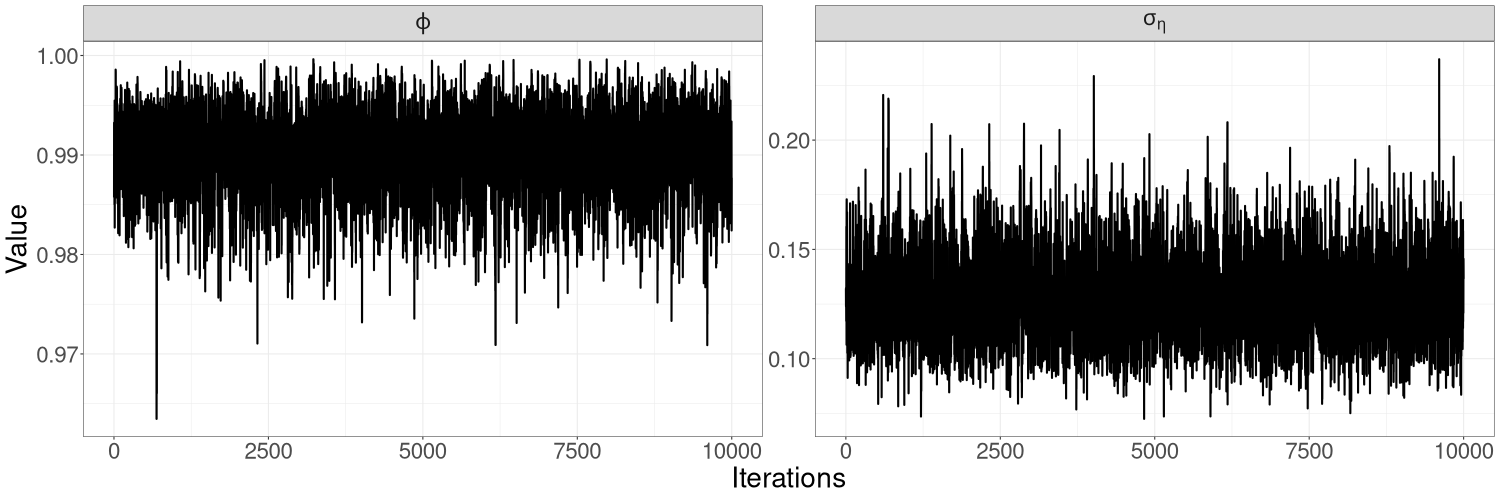}
        \caption{HMC-Whittle trace plots (samples not thinned).}
        \label{fig:sv_real_hmcw_traceplot} 
     \end{subfigure}

     \begin{subfigure}[b]{0.8\linewidth}
         \centering
         \includegraphics[width = \linewidth]{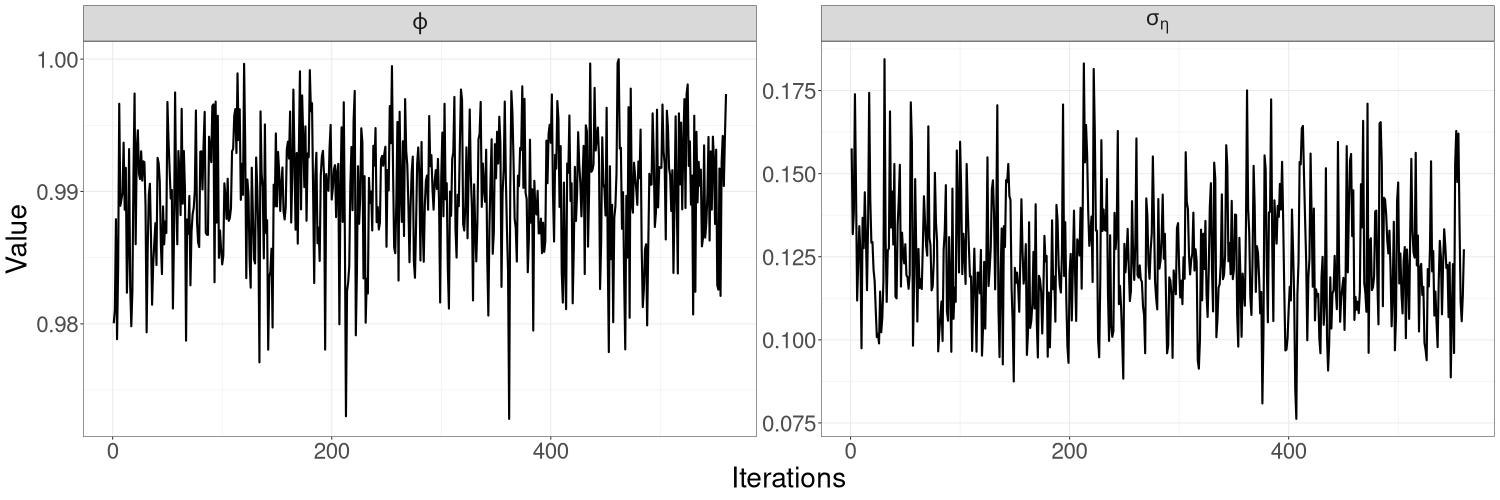}
        \caption{MCMC-stochvol trace plots (samples thinned by a factor of 50).}
        \label{fig:sv_real_stv_traceplot} 
     \end{subfigure}
     \caption{Trace plots for the univariate SV model applied to exchange rate data in Section~\ref{sec:real_data} of the main paper.}
     \label{fig:sv_real_traceplots}
 \end{figure}

\begin{figure}
     \centering
     \begin{subfigure}[b]{\linewidth}
         \centering
         \includegraphics[width = \linewidth]{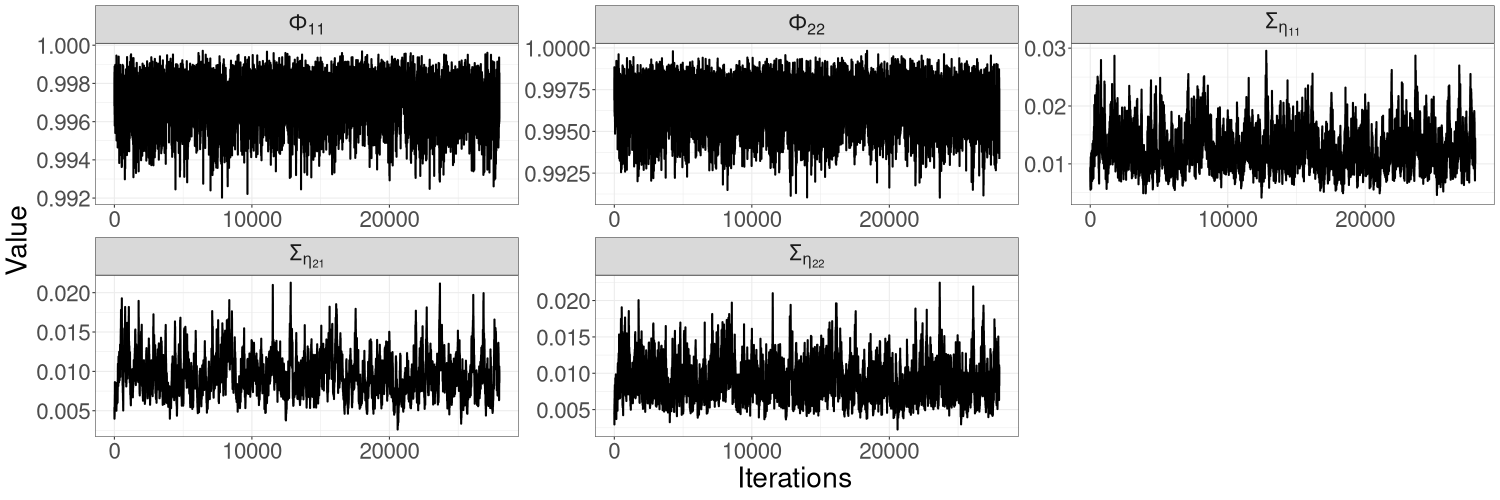}
        \caption{HMC-exact trace plots (samples not thinned).}
        \label{fig:multi_sv_real_hmc_traceplot} 
        \vspace{0.5cm}
     \end{subfigure}
    
     \begin{subfigure}[b]{\linewidth}
         \centering
         \includegraphics[width = \linewidth]{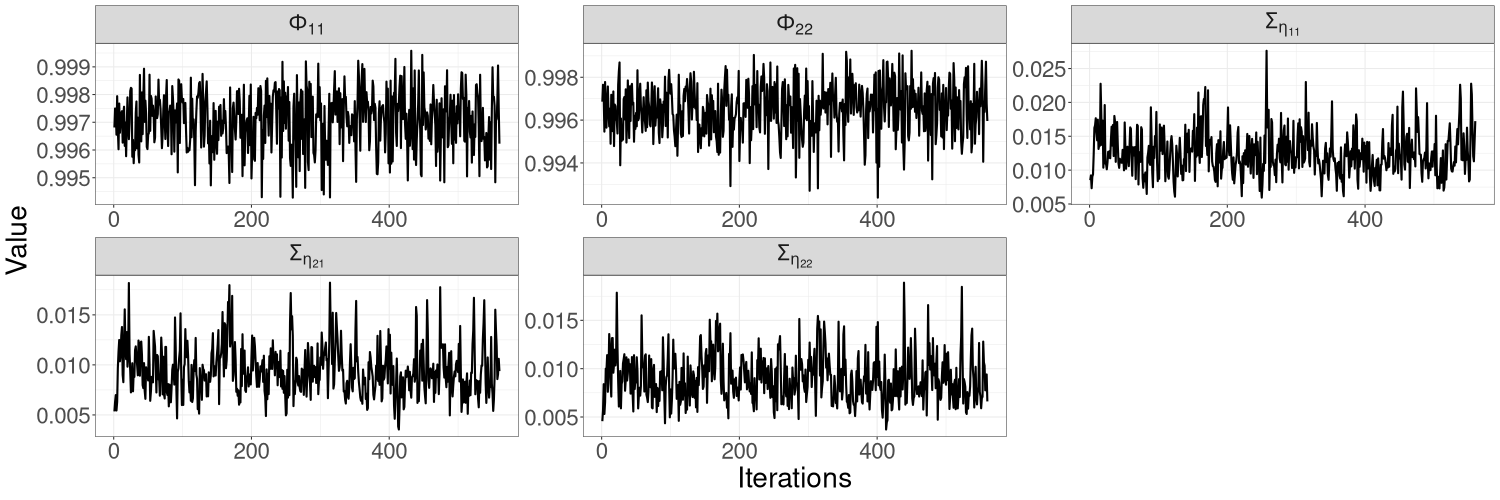}
        \caption{HMC-exact trace plots (samples thinned by a factor of 50).}
        \label{fig:multi_sv_real_hmc_traceplot_thin}
        \vspace{0.5cm}
     \end{subfigure}
    
     \begin{subfigure}[b]{\linewidth}
         \centering
         \includegraphics[width = \linewidth]{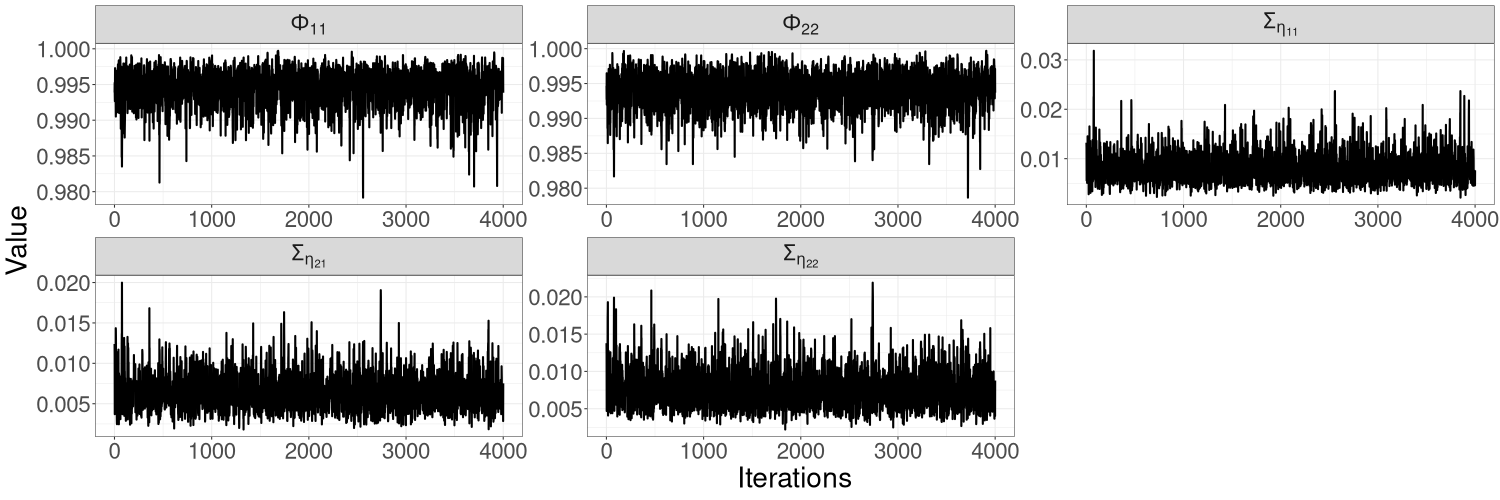}
        \caption{HMC-Whittle trace plots (samples not thinned).}
        \label{fig:multi_sv_real_hmcw_traceplot} 
     \end{subfigure}
     \caption{Trace plots for the bivariate SV model applied to exchange rate data in Section~\ref{sec:real_data} of the main paper.}
     \label{fig:multi_sv_real_traceplots}
 \end{figure}
    \clearpage
\section{Additional results for the univariate SV model in Section \ref{sec:sv_sim}}
\label{sec:other_param_vals}

This section provides additional figures for an experiment in which we simulate 100 datasets, each with length $T=2000$, from the univariate SV model in Section \ref{sec:sv_sim} for each value of $\phi \in \{0.7,0.8,0.9,0.99\}$ and $\sigma_\eta = 0.2$. The number of posterior draws for HMC-Whittle and HMC-exact are kept the same as in Section~\ref{sec:sv_sim} (see Table~\ref{tab:hmc_iters} in this online supplement for a summary). For each parameter setting, the $\hat{R}$ and ESS from HMC-Whittle and HMC-exact are reported in Table~\ref{tab:rep_ess_rhat}. In general, the ESS for HMC-Whittle are consistently above 1000 for all parameter settings, while those from HMC-exact tend to be much lower, especially when $\phi = 0.7$ and $\phi = 0.8$.

Figure \ref{fig:cis_phi09} in the main paper and Figures \ref{fig:cis_phi07}--\ref{fig:cis_phi099} show comparisons of the 95\% credible intervals from {R-VGA-Whittle} and HMC-exact (left) and R-VGA-Whittle and HMC-Whittle (right) for the first 25 simulated datasets for varying $\phi$. These figures show that the widths of the 95\% credible intervals obtained from R-VGA-Whittle and HMC-Whittle are similar, whereas those from HMC-exact are slightly different.

\begin{figure}[t]
     \centering
     \begin{subfigure}[b]{0.49\linewidth}
         \centering
         \centering
        \includegraphics[width=\linewidth]{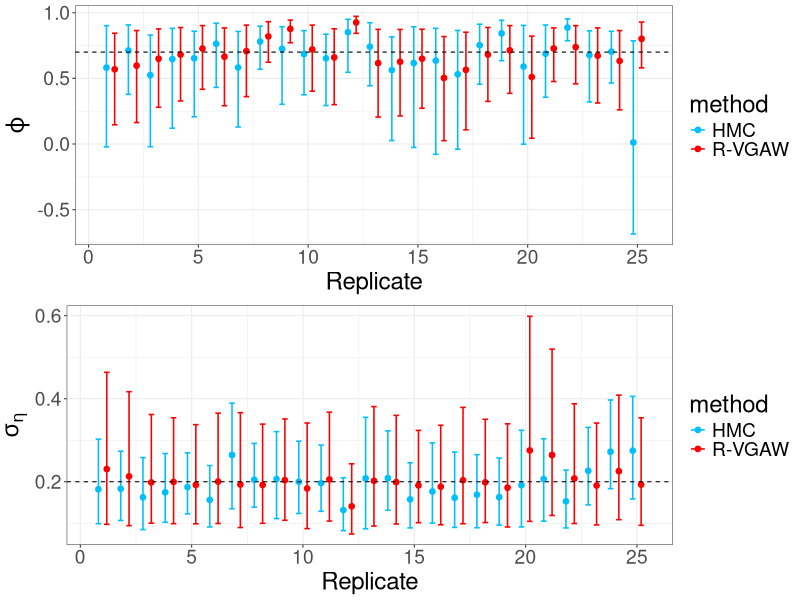}
         \caption{HMC-exact and R-VGA-Whittle.}
         \label{fig:ci_hmc_rvgaw_phi07}
     \end{subfigure}
     \hfill
     \begin{subfigure}[b]{0.49\linewidth}
         \centering
         \includegraphics[width=\linewidth]{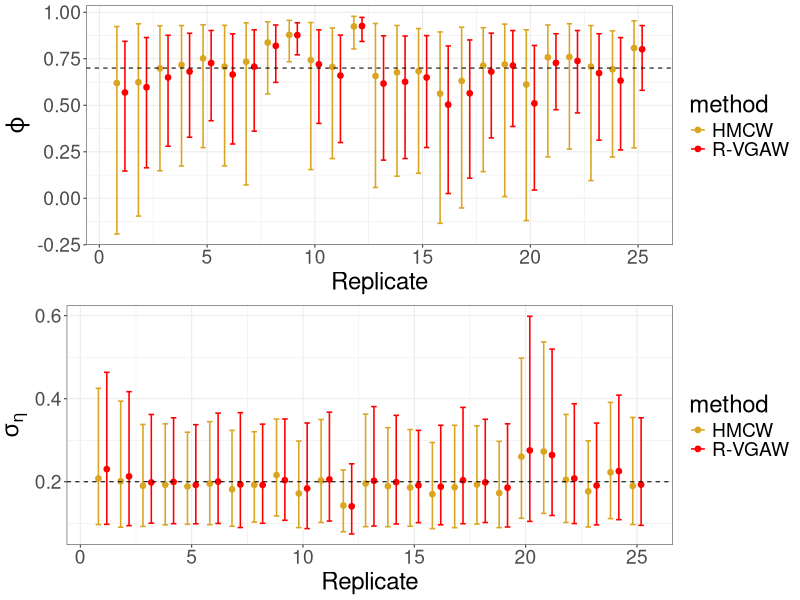}
         \caption{HMC-Whittle and R-VGA-Whittle.}
         \label{fig:ci_hmcw_rvgaw_phi07}
     \end{subfigure}
     \caption{Comparison of the 95\% credible intervals from R-VGA-Whittle and HMC-exact (left) and R-VGA-Whittle and HMC-Whittle (right) for the first 25 simulated datasets from the univariate SV model with $T=2000 $, $\phi = 0.7$, and $\sigma_\eta = 0.2$. Posterior means are marked with points.}
     \label{fig:cis_phi07}
\end{figure}

\begin{figure}[t]
     \centering
     \begin{subfigure}[b]{0.49\linewidth}
         \centering
         \centering
        \includegraphics[width=\linewidth]{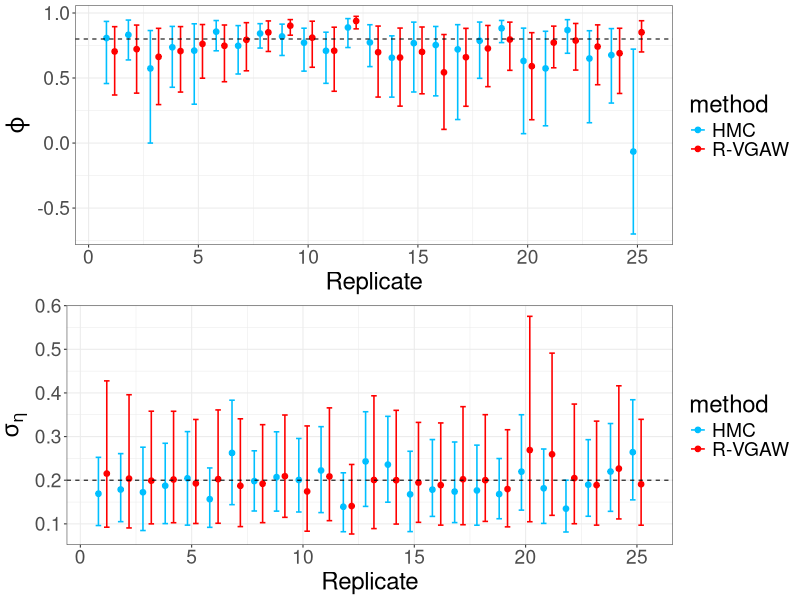}
         \caption{HMC-exact and R-VGA-Whittle.}
         \label{fig:ci_hmc_rvgaw_phi08}
     \end{subfigure}
     \hfill
     \begin{subfigure}[b]{0.49\linewidth}
         \centering
         \includegraphics[width=\linewidth]{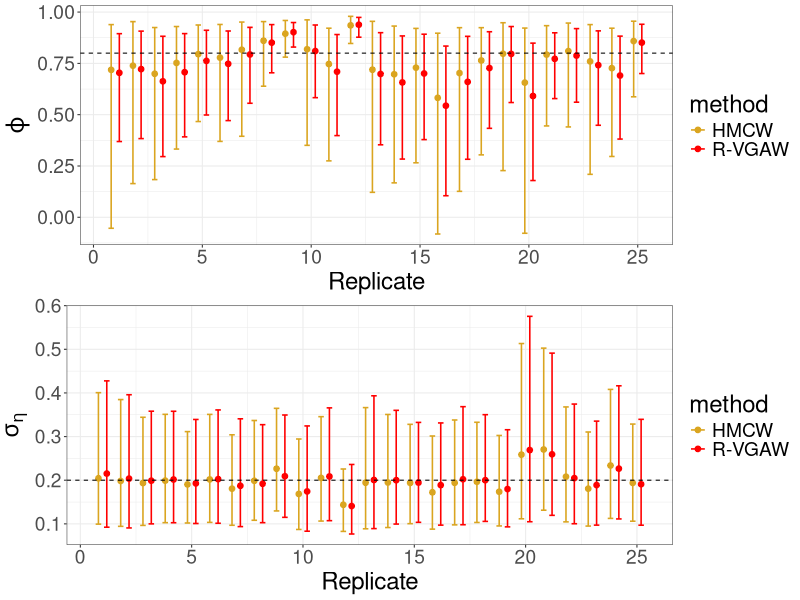}
         \caption{HMC-Whittle and R-VGA-Whittle.}
         \label{fig:ci_hmcw_rvgaw_phi08}
     \end{subfigure}
     \caption{Comparison of the 95\% credible intervals from R-VGA-Whittle and HMC-exact (left) and R-VGA-Whittle and HMC-Whittle (right) for the first 25 simulated datasets from the univariate SV model with $T=2000$, $\phi = 0.8$ and $\sigma_\eta = 0.2$. Posterior means are marked with points.}
     \label{fig:cis_phi08}
\end{figure}

\begin{figure}[t]
     \centering
     \begin{subfigure}[b]{0.49\linewidth}
         \centering
         \centering
        \includegraphics[width=\linewidth]{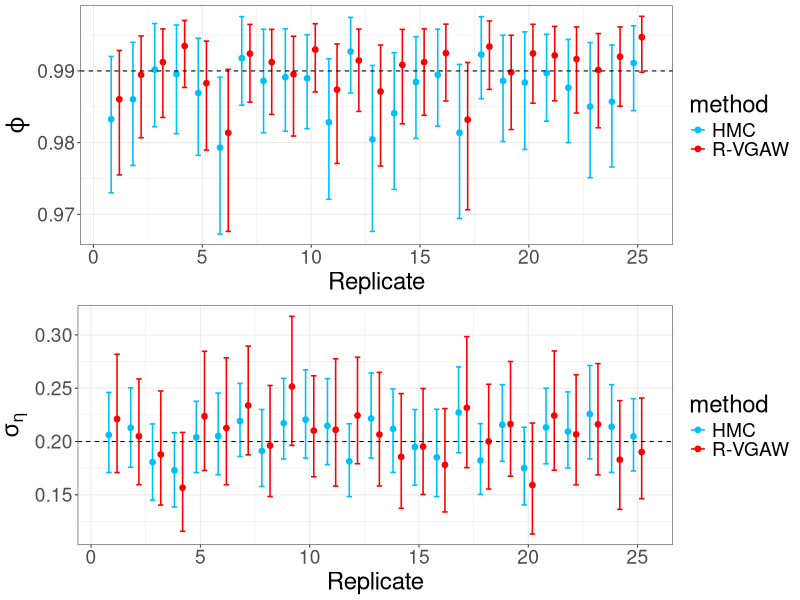}
         \caption{HMC-exact and R-VGA-Whittle.}
         \label{fig:ci_hmc_rvgaw_phi099}
     \end{subfigure}
     \hfill
     \begin{subfigure}[b]{0.49\linewidth}
         \centering
         \includegraphics[width=\linewidth]{Plots/revised_plots/repeat_exp/phi07/ci_hmcw_rvgaw_n2000_temperfirst5_blocksize100_0indiv_20250508.png}
         \caption{HMC-Whittle and R-VGA-Whittle.}
         \label{fig:ci_hmcw_rvgaw_phi099}
     \end{subfigure}
     \caption{Comparison of the 95\% credible intervals from R-VGA-Whittle and HMC-exact (left) and R-VGA-Whittle and HMC-Whittle (right) for the first 25 simulated datasets from the univariate SV model with $T=2000$, $\phi = 0.99$ and $\sigma_\eta = 0.2$. Posterior means are marked with points.}
     \label{fig:cis_phi099}
\end{figure}

\begin{table}[t]
\centering
\caption{$\hat{R}$ and ESS for HMC-Whittle and HMC-exact, averaged over 100 simulations of length $T=2000$, where the true values of $\phi$ are in the set $\{0.7, 0.8, 0.9, 0.99\}$, and $\sigma_\eta = 0.2$. \label{tab:rhatrepeated}}
\begin{tabular}{|l|rr|rr|rr|rr|}
\hline
\multicolumn{1}{|l|}{\multirow{3}{*}{True $\phi$}} & \multicolumn{4}{c|}{HMC-Whittle}                                                                                                                                              & \multicolumn{4}{c|}{HMC-exact}                                                                                                                                                \\ \cline{2-9} 
\multicolumn{1}{|l|}{}                                     & \multicolumn{2}{c|}{$\hat{R}$}                                                             & \multicolumn{2}{c|}{ESS}                                                              & \multicolumn{2}{c|}{$\hat{R}$}                                                             & \multicolumn{2}{c|}{ESS}                                                              \\ \cline{2-9} 
\multicolumn{1}{|l|}{}                                     & \multicolumn{1}{c|}{$\phi$} & \multicolumn{1}{c|}{$\sigma_\eta$} & \multicolumn{1}{c|}{$\phi$} & \multicolumn{1}{c|}{$\sigma_\eta$} & \multicolumn{1}{c|}{$\phi$} & \multicolumn{1}{c|}{$\sigma_\eta$} & \multicolumn{1}{c|}{$\phi$} & \multicolumn{1}{c|}{$\sigma_\eta$} \\ \hline
0.7                                                        & 1.0009                                   & 1.0008                                     & 1328                                     & 1525                                       & 1.0200                                   & 1.0294                                     & 130                                      & 64                                         \\
0.8                                                        & 1.0009                                   & 1.0008                                     & 1271                                     & 1475                                       & 1.0149                                   & 1.0226                                     & 134                                      & 67                                         \\
0.9                                                        & 1.0007                                   & 1.0007                                     & 3263                                     & 3584                                       & 1.0064                                   & 1.0100                                     & 199                                      & 116                                        \\
0.99                                                       & 1.0009                                   & 1.0006                                     & 1407                                     & 1405                                       & 1.0007                                   & 1.0028                                     & 447                                      & 447                                        \\
\hline
\end{tabular}
\label{tab:rep_ess_rhat}
\end{table}





\fi
\end{document}